\documentclass{elsartL}
\usepackage{graphicx}
\usepackage{amssymb}
\usepackage{amsmath}
\usepackage{mathrsfs}
\usepackage{mathtools}
\newcommand{\fsl}[1]{\ensuremath{\mathrlap{\!\not{\phantom{#1}}}#1}}
\begin{document}

\begin{frontmatter}
\title{Aspects of the ETH model of the pion-nucleon interaction}
\author[EM]{E. Matsinos{$^*$}},
\author[GR]{G. Rasche},
\address[EM]{Institute of Mechatronic Systems, Zurich University of Applied Sciences, Technikumstrasse 5, CH-8401 Winterthur, Switzerland}
\address[GR]{Physik-Institut der Universit\"at Z\"urich, Winterthurerstrasse 190, CH-8057 Z\"urich, Switzerland}

\begin{abstract}
The ETH model of the pion-nucleon ($\pi N$) interaction contains $t$-channel $\sigma$- and $\rho$-exchange graphs, as well as the $s$- and $u$-channel contributions with the well-established $s$ and $p$ baryon states with masses below $2$ GeV as virtual 
particles; the model amplitudes obey crossing symmetry and isospin invariance. In the present work, we give the analytical expressions for the model contributions to the $K$-matrix elements up to (and including) the $f$ waves. We also extract a new 
phase-shift solution after performing a partial-wave analysis of meson-factory $\pi^\pm p$ elastic-scattering data below $100$ MeV; included in our results now are also the effects of the variation of the $\sigma$-meson mass in the interval which is 
currently recommended by the Particle-Data Group. Finally, we revisit the subject of the $\pi N$ $\Sigma$ term and, using the model amplitudes, obtain $\Sigma = 72.4 \pm 3.1$ MeV. Our prediction agrees well with the result extracted with Olsson's method, 
after a few flaws in his paper [Phys. Lett. B 482 (2000) 50] were corrected.\\
\noindent {\it PACS:} 13.75.Gx; 25.80.Dj; 11.30.-j
\end{abstract}
\begin{keyword} $\pi^\pm p$ elastic scattering; $\pi N$ phase shifts; $\pi N$ coupling constant; $\pi N$ low-energy constants; $\pi N$ $\Sigma$ term
\end{keyword}
{$^*$}{Corresponding author. E-mail: evangelos[DOT]matsinos[AT]zhaw[DOT]ch, evangelos[DOT]matsinos[AT]sunrise[DOT]ch}
\end{frontmatter}

\section{\label{sec:Introduction}Introduction}

Quantum Chromodynamics (QCD) is the accepted Theory of the strong interaction; unfortunately, it yields approximate solutions only in selected cases (asymptotic freedom). As a result, phenomenological/empirical models are generally used in order to 
cover the dynamics in systems of hadrons. One such model, the ETH model, was put forth in the early 1990s to account for the pion-nucleon ($\pi N$) interaction at low energies.

The ETH model may be considered to be the product of the long-term study of the properties of pionic-atom data of isoscalar nuclei (i.e., of nuclei containing equal numbers of protons and neutrons) within the framework of the relativistic mean-field theory 
of the 1980s; the principal aim at those times was the explanation of the long-standing problem of the $s$-wave repulsion in the $\pi$-nucleus interaction (save for pionic hydrogen and $^3{\rm He}$, the strong shifts of the $1s$ levels in pionic atoms are 
repulsive) \cite{glm1}. In its original form, the model did not contain the spin-$\frac{1}{2}$ contributions of the $\Delta(1232)$; in the early 1990s, the emphasis was placed on the reproduction of the experimental results obtained at pion laboratory kinetic 
energy $T=0$ MeV, i.e., at the $\pi N$ threshold. The first attempts to reproduce the energy dependence of the then-available (and nowadays outdated) $\pi N$ phase shifts, even above the $\Delta(1232)$ resonance, turned out to be successful after the inclusion 
in the model of the spin-$\frac{1}{2}$ contributions of the $\Delta(1232)$ graphs \cite{glm2}. Essential in the foundation of the model was Ref.~\cite{glmbg}; analytical expressions for the contributions of the main Feynman graphs (simply `graphs' hereafter) 
of the model to the $s$- and $p$-wave $K$-matrix elements appeared in a compact form in Appendix A of that paper.

In a series of subsequent papers, two $\pi N$-related issues were mainly addressed: a) the reproduction of the low-energy ($T \leq 100$ MeV) $\pi^\pm p$ elastic-scattering and charge-exchange $\pi^- p \rightarrow \pi^0 n$ (CX) data \cite{mworg,mr1,mr2,mr3}, 
including the extraction of the values of the low-energy constants (LECs) of the $\pi N$ system, and b) the violation of isospin invariance in the hadronic part of the $\pi N$ interaction~\footnote{The first evidence for the violation of the isospin invariance 
in the hadronic part of the $\pi N$ interaction was presented in Ref.~\cite{glk} and appeared two years prior to Ref.~\cite{m}; the two research programmes were independent.} \cite{mworg,mr3,m}. The model was also involved in an iterative procedure which 
resulted in the determination of the electromagnetic (em) corrections \cite{gmorw1,gmorw2}, i.e., of the corrections which must be applied to the $\pi N$ phase shifts and to the partial-wave amplitudes on the way to the evaluation of the low-energy observables, 
namely of the differential cross section (DCS) and of the analysing power (AP). The long-term use of this model has demonstrated that it can account for the experimental information available at low energies almost as successfully as simple parameterisations of 
the $K$-matrix elements \cite{mworg,mr1,mr2,mr3,fm}, which do not contain theoretical constraints other than the expected low-energy behaviour of these elements. One may thus conclude that the model constitutes a firm basis for the parameterisation of the 
dynamics of the $\pi N$ system at low energies.

There are a number of reasons why a paper, replete with technical details, is expected to be useful.
\begin{itemize}
\item Limited information had been given in Ref.~\cite{glmbg} regarding the treatment of the main graphs of the model and the procedure leading to the analytical expressions of the $s$- and $p$-wave $K$-matrix elements appearing in Appendix A of that 
paper. Given the smallness of the effects which the higher baryon resonances (HBRs) induce at low energies, the details on their treatment and the expressions for their contributions to the model amplitudes had been omitted altogether; also omitted were 
all $d$- and $f$-wave contributions.
\item The complete publication of the model amplitudes is expected to facilitate its use in other works. The expressions may (occasionally) appear long, yet they can be implemented in a numerical analysis without much effort.
\item Due to its importance in QCD tests, the evaluation of the $\pi N$ $\Sigma$ term is of high significance in Hadronic Physics. There has been only one occasion in the past, in which we treated the $\Sigma$ term in the context of the model \cite{glmbg}. 
The expression given therein was a tree-level approximation; as a result, we had been reluctant, throughout our programme, to give the model predictions for the $\Sigma$ term alongside our results for other LECs. In the present work, we revisit the subject 
and show that the $\Sigma$ result, extracted with a method featuring a few LECs of the $\pi N$ system, is in good agreement with the value obtained with the simple expression of Ref.~\cite{glmbg}.
\item To make our analysis self-contained (i.e., independent of extraneous information on the $d$ and $f$ waves), it is needed to include in the model the $s$- and $u$-channel contributions with the well-established $d$ and $f$ HBRs as intermediate states. 
However, the theoretical treatment of (four out of six of) these fields is either too intricate (propagation of spin-$\frac{5}{2}$ particles) or not at all existing (propagation of spin-$\frac{7}{2}$ particles). The hope is that the present work will stimulate 
interest in the theoretical treatment of these fields.
\end{itemize}

The outline of the paper is as follows. In Section \ref{sec:General}, we deal with the description of the kinematics of the $\pi N$ system and with the definitions of the various quantities used in the present study; these details are usually `assumed known' 
in the scientific literature, urging the interested reader to search in books or to derive the expressions. Section \ref{sec:ModelAmplitudes} provides details on the treatment of the main graphs of the model (Subsections \ref{sec:Sigma}-\ref{sec:Delta}), as 
well as of those corresponding to the well-established $s$ and $p$ HBRs with masses below $2$ GeV (Subsection \ref{sec:HR}). Section \ref{sec:Results} is split into two parts: in the first part, the energy dependence of the model amplitudes is discussed 
and the contributions of each graph of the model to the $s$-wave scattering lengths and $p$-wave scattering volumes are given; the second part is dedicated to the treatment of the $\Sigma$ term within the context of the model. Section \ref{sec:Conclusions} 
contains our conclusions.

\section{\label{sec:General}Kinematics of the $\pi N$ system and definition of the amplitudes}

\subsection{\label{sec:Notation}Notation}

In the present work, we will make use of the following notation and conventions.
\begin{itemize}
\item The speed of light in vacuum $c$ is equal to $1$.
\item Einstein's summation convention is used.
\item $I_n$ denotes the $n \times n$ identity matrix.
\item $g^{\mu \nu}$ denotes the Minkowski metric with signature `$+ \, - \, - \, -$'.
\item $\epsilon_{ijk}$ is the Levi-Civita symbol; $\delta_{ij}$ is the Kronecker delta.
\item $\vec{\sigma}$ are the standard $2 \times 2$ Pauli matrices obeying the relations $[\sigma_i,\sigma_j] = 2 i \, \sum_k \epsilon_{ijk} \sigma_k$ and $\{\sigma_i,\sigma_j\} = 2 \delta_{ij} I_2$.
\item The isospin operators of the nucleon and of the pion are denoted by $\frac{1}{2} \vec{\tau}$ and $\vec{t}$.
\item $\gamma^{\mu}$ ($\mu = 0, 1, 2, 3$) are the standard Dirac $4 \times 4$ matrices, satisfying the relation $\{ \gamma^{\mu}, \gamma^{\nu} \} = 2 g^{\mu \nu} I_4$.
\item $m_p$ and $m_c$ denote the masses of the proton and of the charged pion.
\item $s$, $u$, and $t$ are the standard Mandelstam variables; two additional variables, $\nu$ and $\nu_B$, will be introduced in Subsection \ref{sec:ManPln}.
\item $P_l (\xi)$ denotes the standard Legendre polynomials.
\item For a $4$-vector $a$, $\fsl{a}=\gamma^{\mu} a_{\mu}$; the corresponding $3$-vector (i.e., the vector of the spatial components of the $4$-vector $a$) is denoted by $\vec{a}$.
\item $p_L$ and $q_L$ are the $4$-momenta of the nucleon (assumed to be a proton) and of the incident pion in the laboratory frame; in the description of the kinematics in Subsection \ref{sec:Kin}, the nucleon is assumed to be initially at rest in the laboratory 
frame ($\vec{p}_L = \vec{0}$). The pion laboratory kinetic energy $T$ satisfies the relation: $q_{L0}=T + m_c$.
\item $p_L^\prime$ and $q_L^\prime$ are the $4$-momenta of the scattered nucleon and of the scattered pion in the laboratory frame.
\item $\theta_L$ denotes the laboratory scattering angle of the pion.
\item CM stands for the centre of mass.
\item $p$ and $q$ are the $4$-momenta of the nucleon and of the incident pion in the CM frame, which is defined by $\vec{p} + \vec{q} = \vec{0}$.
\item $p^\prime$ and $q^\prime$ are the $4$-momenta of the scattered nucleon and of the scattered pion in the CM frame; of course, $\vec{p} \, ^{\prime} + \vec{q} \, ^{\prime} = \vec{0}$.
\item $\theta$ denotes the scattering angle in the CM frame.
\end{itemize}

Energy-momentum conservation enforces the relations $p_L + q_L = p^\prime_L + q^\prime_L$ and $p + q = p^\prime + q^\prime$. For elastic scattering, $q_0 = q^\prime_0$ (consequently, $p_0 = p^\prime_0$).

One combination of $4$-momenta, entering the form of the hadronic part of the scattering amplitude, is
\begin{equation} \label{eq:EQ01}
Q = \frac{1}{2} (q+q^\prime) \, \, \, .
\end{equation}

\subsection{\label{sec:Kin}Kinematics of the $\pi N$ system}

The standard Mandelstam variables $s$, $u$, and $t$ are defined as follows.
\begin{equation} \label{eq:EQ02}
s = (p + q)^2
\end{equation}
\begin{equation*}
u = (p - q^\prime)^2
\end{equation*}
\begin{equation*}
t = (q - q^\prime)^2
\end{equation*}
Being inner products of $4$-vectors, these quantities are invariant under Lorentz transformations. From Eq.~(\ref{eq:EQ02}), one obtains $s = (p_0 + q_0)^2 = W^2$, i.e., $s$ is simply equal to the square of the total energy $W$ in the CM frame. For the 
two-body scattering process $A+B \rightarrow C+D$, the sum $s+u+t$ is constant, equal to the sum of the squares of the masses of the incoming and outgoing particles $m_A^2+m_B^2+m_C^2+m_D^2$, to be denoted~\footnote{To avoid confusion with the $K$-matrix 
elements of subsequent sections, we will not follow Kibble's choice \cite{kbl} of using $K$ to represent the sum $s+u+t$. (The reader must bear in mind that Ref.~\cite{kbl} adheres to an older notation for the Mandelstam variables.) H\"ohler \cite{h} uses 
$\Sigma$ to denote the same quantity; however, we will reserve the symbol $\Sigma$ for another quantity, namely for the $\pi N$ $\Sigma$ term.} in the following as $\Lambda$; for $\pi^\pm p$ elastic scattering, $\Lambda = 2 m_p^2 + 2 m_c^2$. The constancy 
of $s+u+t$ implies that the hadronic part of the scattering amplitude depends on only two Mandelstam variables.

\subsubsection{\label{sec:LabToCM}Transformations between the laboratory and the CM frames}

The velocity of the CM in the laboratory frame (expressed as a fraction of $c$) is given by
\begin{equation} \label{eq:EQ03}
v = \frac{\lvert \vec{q}_L \rvert}{q_{L0} + m_p} \, \, \, .
\end{equation}
The Lorentz factor $\gamma$ is given by
\begin{equation} \label{eq:EQ04}
\gamma = \frac{1}{\sqrt{1 - v^2}} = \frac{q_{L0} + m_p}{\sqrt{s}} \, \, \, ;
\end{equation}
for the derivation of this equation, use has been made of the relation $s=m_p^2+m_c^2+2 m_p q_{L0}$ (evaluation of $s$ from quantities pertaining to the laboratory frame). Using $\lvert \vec{q} \, \rvert = \gamma (\lvert \vec{q}_L \rvert - v q_{L0})$, along 
with Eqs.~(\ref{eq:EQ03}) and (\ref{eq:EQ04}), one obtains
\begin{equation} \label{eq:EQ05}
\lvert \vec{q} \, \rvert = \frac{m_p \, \lvert \vec{q}_L \rvert}{\sqrt{s}} \, \, \, .
\end{equation}

\begin{figure}
\begin{center}
\includegraphics [width=15.5cm] {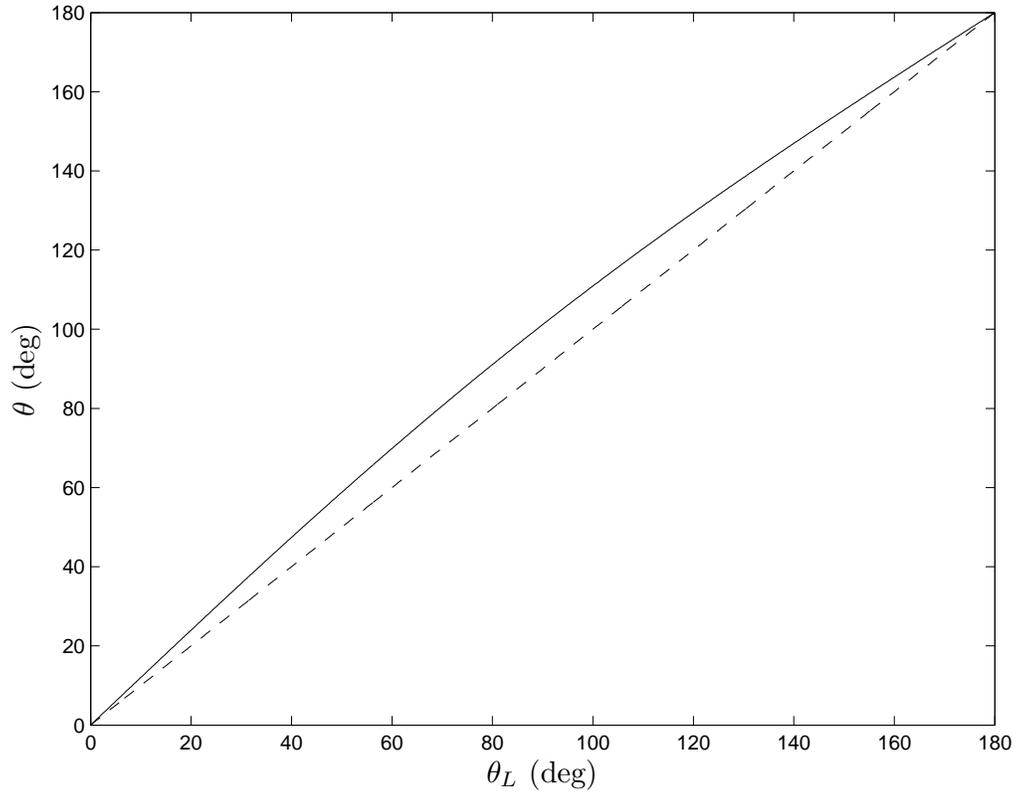}
\caption{\label{fig:Theta}A representative plot (solid line) of the centre-of-mass scattering angle $\theta$ as a function of the pion laboratory scattering angle $\theta_L$ for $\pi^\pm p$ elastic scattering at pion laboratory 
kinetic energy $T=60$ MeV; the $\theta$ values have been obtained using Eq.~(\ref{eq:EQ08}). The dashed line, representing the function $\theta=\theta_L$, has been added in order to provide an impression of the difference between 
the two angles.}
\end{center}
\end{figure}

We will now examine the dependence of the CM scattering angle $\theta$ on $\theta_L$. The transverse component of the pion momentum is not changed in the transformation from the laboratory to the CM frame.
\begin{equation*}
\lvert \vec{q}_L^{\, \prime} \rvert \sin \theta_L = \lvert \vec{q} \, ^{\prime} \rvert \sin \theta
\end{equation*}
The longitudinal component obeys the relation
\begin{equation*}
\lvert \vec{q}_L^{\, \prime} \rvert \cos \theta_L = \gamma ( \lvert \vec{q} \, ^{\prime} \rvert \cos \theta + v q^\prime_0 ) \, \, \, .
\end{equation*}
From these two equations, one obtains
\begin{equation} \label{eq:EQ06}
\tan \theta_L = \frac{\lvert \vec{q} \, ^{\prime} \rvert \sin \theta}{\gamma ( \lvert \vec{q} \, ^{\prime} \rvert \cos \theta + v q^\prime_0 )} \, \, \, .
\end{equation}
After introducing $\zeta = \frac{v q^\prime_0}{\lvert \vec{q} \, ^{\prime} \rvert}$, Eq.~(\ref{eq:EQ06}) is put in the form
\begin{equation} \label{eq:EQ07}
\gamma \tan \theta_L = \frac{\sin \theta}{\cos \theta + \zeta} \, \, \, .
\end{equation}
(Values of the variable $\zeta$, typical for low-energy $\pi^\pm p$ scattering, do not exceed a few $10^{-1}$. For instance, for elastic scattering at $T=20$ MeV, $\zeta \approx 0.164$; at $T=100$ MeV, $\zeta \approx 0.221$. The corresponding $\zeta$ values for 
the CX reaction are: $0.150$ and $0.217$.) For each $\theta_L$ value, two solutions for $\theta$ are obtained.
\begin{equation} \label{eq:EQ08}
\theta_\pm = 2 \, \arctan \big( \frac{-1 \pm \sqrt{1 + \gamma^2 (1-\zeta^2) \tan^2 \theta_L}}{\gamma (1-\zeta) \tan \theta_L} \big)
\end{equation}
Given that $\theta \rightarrow 0$ when $\theta_L \rightarrow 0$ and $\theta \rightarrow \pi$ when $\theta_L \rightarrow \pi$, the solution $\theta_+$ is the appropriate choice when $\theta_L < \frac{\pi}{2}$, whereas $\theta_-$ must be used for $\theta_L > \frac{\pi}{2}$. 
These two functions have the same limit for $\theta_L \to \frac{\pi}{2}$, namely the value $2 \, \arctan \sqrt{\frac{1+\zeta}{1-\zeta}}$. A representative (in the low-energy region) plot of $\theta$ for $\theta_L \in [0,\pi]$ is given in Fig.~\ref{fig:Theta}.

\begin{figure}
\begin{center}
\includegraphics [width=15.5cm] {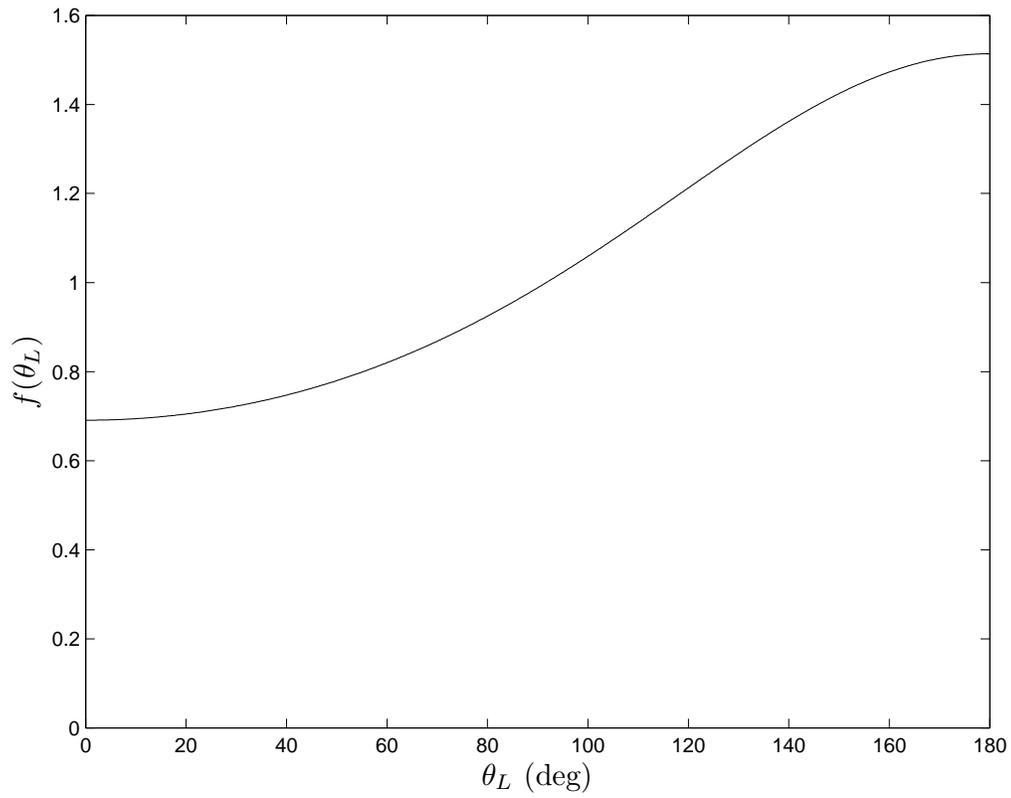}
\caption{\label{fig:f}A representative plot of the factor $f(\theta_L)$ of Eq.~(\ref{eq:EQ09}) for $\pi^\pm p$ elastic scattering at pion laboratory kinetic energy $T=60$ MeV.}
\end{center}
\end{figure}

We will next determine the relation between the $\pi N$ DCS in the laboratory and in the CM frames. Obviously,
\begin{equation*}
\frac{d\sigma}{d\Omega} = \frac{d\sigma}{d\Omega} \Big|_L \, \, \big| \frac{d\cos\theta_L}{d\cos\theta} \big| = \frac{d\sigma}{d\Omega} \Big|_L \, \, f(\theta_L) \, \, \, .
\end{equation*}
Using Eq.~(\ref{eq:EQ07}), one obtains
\begin{equation} \label{eq:EQ09}
f(\theta_L) = \gamma \, (1+\zeta \cos\theta) \, \big| \frac{\sin\theta_L}{\sin\theta} \big|^3 \, \, \, .
\end{equation}
A representative (in the low-energy region) plot of the function $f(\theta_L)$ for $\theta_L \in [0,\pi]$ is shown in Fig.~\ref{fig:f}. The importance of this `correction' in the extraction of the DCS values in the CM frame is obvious.

\subsubsection{\label{sec:ManPln}$\pi N$ scattering on the Mandelstam plane}

The equation $s = (p_0+q_0)^2$ leads to
\begin{equation*}
\vec{q}^{\, 2} = \frac{s^2 - \Lambda s + (m_p^2 - m_c^2)^2}{4 s} = \frac{\big( s - (m_p + m_c)^2 \big) \big( s - (m_p - m_c)^2 \big)}{4 s} \, \, \, .
\end{equation*}
As expected, $\vec{q}^{\, 2}$ depends solely on $s$. By setting $\vec{q}_L^{\, 2} < 0$ in Eq.~(\ref{eq:EQ05}), one determines the range of $s$ which cannot represent physical scattering: $s_1 \equiv (m_p - m_c)^2 < s < (m_p + m_c)^2 \equiv s_2$.

The procedure for obtaining the boundaries of the physical region on the Mandelstam plane has been put forth in Ref.~\cite{kbl}. The requirement that the scattering angle be real enforces the condition
\begin{equation*}
\begin{vmatrix}
m_p^2 & \frac{s-m_p^2-m_c^2}{2} & \frac{t-2 m_p^2}{2} \\
\frac{s-m_p^2-m_c^2}{2} & m_c^2 & \frac{u-m_p^2-m_c^2}{2} \\
\frac{t-2 m_p^2}{2} & \frac{u-m_p^2-m_c^2}{2} & m_p^2
\end{vmatrix}
\geq 0 \, \, \, .
\end{equation*}
This condition appears in-between Eqs.~(7) and (8) of Ref.~\cite{kbl}. Only the strict inequality appeared in Ref.~\cite{kbl}; however, there is no reason to avoid including the $\theta=0$ and $\theta=\pi$ cases in the physical scattering. After trivial algebraic 
operations, one may put this inequality in the form
\begin{equation} \label{eq:EQ10}
\left( s u - (m_p^2 - m_c^2) ^2 \right) t \geq 0 \, \, \, .
\end{equation}
The function on the left-hand side (lhs) of this inequality is the `Kibble function' of Ref.~\cite{h}. Before advancing to the delineation of the physical regions on the Mandelstam plane, two additional variables will be introduced, $\nu$ and $\nu_B$.
\begin{equation*}
\nu = \frac{s-u}{4 m_p} = \frac{4p_L \cdot q_L + t}{4 m_p}=q_{L0} + \frac{t}{4 m_p}
\end{equation*}
\begin{equation} \label{eq:EQ11}
\nu_B = -\frac{q \cdot q^\prime}{2 m_p} = \frac{t - 2 m_c^2}{4 m_p}
\end{equation}

To describe the energy and the angular dependence of the hadronic part of the scattering amplitude, one may choose any two (independent) of the aforementioned quantities, e.g., $s$ and $t$, or $\nu$ and $\nu_B$. Expressed in terms of $\nu$ and $\nu_B$, 
the standard Mandelstam variables are given by the expressions below.
\begin{equation} \label{eq:EQ12}
s = m_p \big( m_p + 2 (\nu - \nu_B) \big)
\end{equation}
\begin{equation} \label{eq:EQ13}
u = m_p \big( m_p - 2 (\nu + \nu_B) \big)
\end{equation}
\begin{equation*}
t = 2 m_c^2 + 4 m_p \nu_B
\end{equation*}
One of the popular choices in the study of the $\pi N$ system is the Mandelstam representation featuring the variables $\nu$ and $t$. As $\nu=\frac{s-u}{4 m_p}$,
\begin{equation} \label{eq:EQ14}
u = s - 4 m_p \nu
\end{equation}
and
\begin{equation} \label{eq:EQ15}
s + u + t = \Lambda \Rightarrow 2 s - 4 m_p \nu = \Lambda - t \Rightarrow s = \frac{\Lambda - t}{2} + 2 m_p \nu \, \, \, .
\end{equation}
The substitution of $s$ in Eq.~(\ref{eq:EQ14}) with the right-hand side (rhs) of the last of Eqs.~(\ref{eq:EQ15}) yields
\begin{equation} \label{eq:EQ16}
u =\frac{\Lambda - t}{2} - 2 m_p \nu \, \, \, .
\end{equation}
Equations (\ref{eq:EQ15}) and (\ref{eq:EQ16}) express $s$ and $u$ in terms of the variables $\nu$ and $t$. The value of $\nu_B$ is obtained from the last of Eqs.~(\ref{eq:EQ11}).

\begin{figure}
\begin{center}
\includegraphics [width=15.5cm] {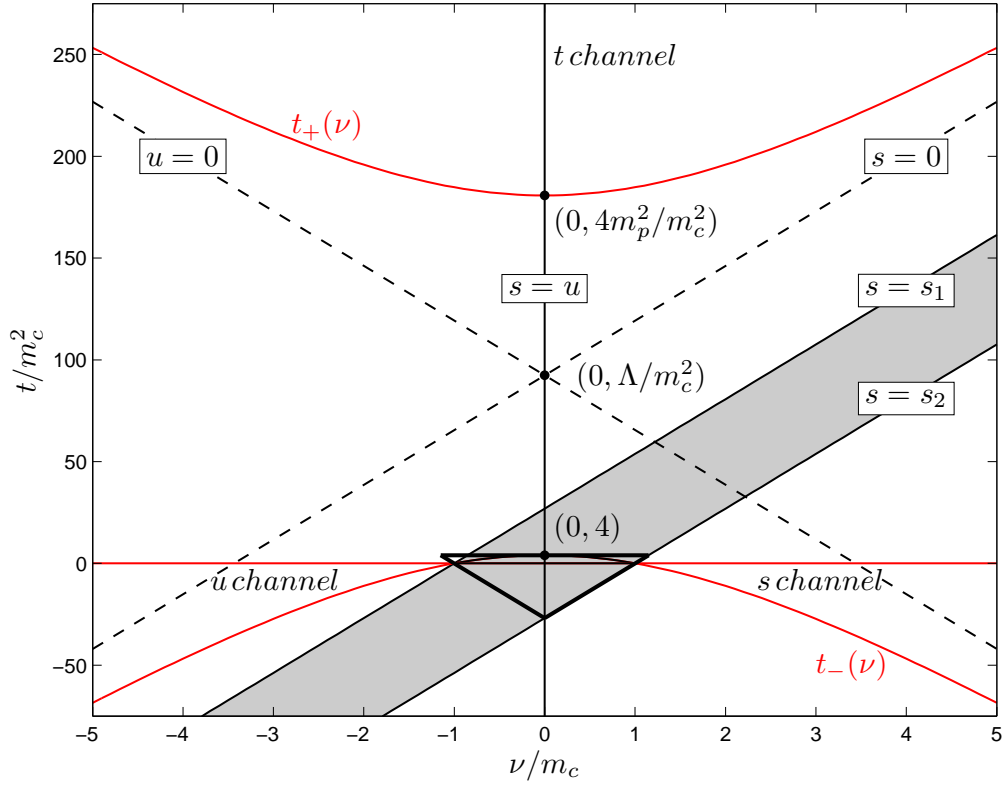}
\caption{\label{fig:Mp1}The regions of the Mandelstam plane. The gray band marks the region in which $\vec{q}_L^{\, 2} < 0$. The boundaries of the three regions, representing the physical processes involving two nucleons and two 
pions, are shown in red; these regions have been obtained from the solution of inequality (\ref{eq:EQ17}). The sides of the Mandelstam triangle have been drawn with increased line thickness.}
\end{center}
\end{figure}

The different regions on the $(\nu,t)$ plane are shown in Fig.~\ref{fig:Mp1}. The loci of constant $s$ are straight lines with slope equal to $4 m_p$, whereas the straight lines of constant $u$ correspond to slope $-4 m_p$. The straight lines with 
$s=0$ and $u=0$ intersect one another at $(\nu,t)=(0,\Lambda)$. The $s=u$ locus coincides with the $\nu=0$ straight line. As earlier mentioned, $\vec{q}_L^{\, 2} < 0$ when $s_1 < s < s_2$; this region, shown in gray in Fig.~\ref{fig:Mp1}, is 
delimited by the two straight lines with $s=s_1$ and $s=s_2$.

We now return to the determination of the boundaries of the physical region on the $(\nu,t)$ plane. Inserting $s$ and $u$ of Eqs.~(\ref{eq:EQ15}) and (\ref{eq:EQ16}) into expression (\ref{eq:EQ10}), one obtains
\begin{equation} \label{eq:EQ17}
\left( \big( \frac{\Lambda-t}{2} \big)^2 - 4 m_p^2 \nu^2 - (m_p^2 - m_c^2) ^2 \right) t \geq 0 \, \, \, .
\end{equation}
The equality is satisfied when $t=0$ or $t=t_\pm$, where
\begin{equation*}
t_\pm = \Lambda \pm 2 \sqrt{4 m_p^2 \nu^2 +(m_p^2 - m_c^2)^2} \, \, \, .
\end{equation*}
The curves representing the functions $t_\pm(\nu)$ are branches of the hyperbola admitting the $s=0$ and $u=0$ straight lines as asymptotes (see Fig.~\ref{fig:Mp1}). The upper branch (i.e., the $t_+$ solution) intersects the $\nu=0$ line at $t=4m_p^2$, 
whereas the lower one (i.e., the $t_-$ solution) at $t=4m_c^2$. As a result, for $t \leq 0$, the physical region is confined to $t_- \leq t \leq 0$; $t=0$ corresponds to $\theta=0$, whereas $t=t_-$ to $\theta=\pi$. 
For $t>0$, the physical region is characterised by $t \geq t_+$.

One point in the unphysical region, the so-called `Cheng-Dashen (CD) point' \cite{cd}, has received substantial attention during the past decades. The coordinates of the CD point are: $(\nu,t)=(0,2 m_c^2)$; the corresponding values of the other variables 
are: $\nu_B=0$ and $s=u=m_p^2$. The Mandelstam triangle is delimited by the straight lines $t=4 m_c^2$, $s=s_2$, and $u=s_2$ (see Fig.~\ref{fig:Mp1}).

\subsection{\label{sec:DefAmp}Definitions of the hadronic part of the scattering amplitude}

The most general Lorentz-invariant and parity-conserving $T$-matrix element in the CM frame is given by
\begin{equation} \label{eq:EQ18}
\mathscr{T} = \bar{u}_f (p^\prime) \, ( A + B \fsl{Q} ) \, u_i (p) \, \, \, ,
\end{equation}
where $u (p)$ is the Dirac spinor associated with the plane-wave of a nucleon with $4$-momentum $p$; $\bar{u} (p) = u^\dagger (p) \gamma^0$ is the conjugate spinor, whereas $u^\dagger (p)$ is the conjugate transpose of $u (p)$. The subscripts $i$ and 
$f$ stand for the nucleon spin and isospin in the initial and final state, respectively. The $4$-momentum $Q$ has already been introduced by Eq.~(\ref{eq:EQ01}). The following combination of the invariant amplitudes $A$ and $B$ is used in the extraction 
of the $\pi N$ $\Sigma$ term in schemes employing dispersion relations \cite{h,cd}:
\begin{equation*}
D = A + \nu B
\end{equation*}

To simplify the expressions and rid the various physical quantities (e.g., the $K$-matrix elements, the amplitudes, etc.) of obvious arguments, explicit reference to the dependence on the Mandelstam variables will be given in the rest of the paper only 
if necessary.

The Dirac spinor is of the form
\begin{equation*}
u (p) = \sqrt{\frac{p_0+m_p}{2 m_p}} \left( 
\begin{array}{c}
\phi \\
\frac{\vec{\sigma} \cdot \vec{p}}{p_0 + m_p} \phi \\
\end{array} \right) \, \, \, ,
\end{equation*}
where $\phi$ is an arbitrary two-spinor, satisfying $\phi^\dagger \phi=1$.

It is easy to show that Eq.~(\ref{eq:EQ18}) is rewritten (for $p_0 = p^\prime_0$) as
\begin{align*}
\mathscr{T} &= \frac{1}{2 m_p} \phi_f^\dagger \Big( \big( A + B (W - m_p) \big) (p_0+m_p) \nonumber \\
& \qquad \qquad \quad + \big( -A + B (W + m_p) \big) (p_0-m_p) \xi \nonumber \\
& \qquad \qquad \quad - \big( -A + B (W + m_p) \big) (p_0-m_p) i \vec{\sigma} \cdot \vec{n} \Big) \phi_i \, \, \, ,
\end{align*}
where $\xi=\cos\theta$. The vector $\vec{n}$, which is normal to the scattering plane, is defined as
\begin{equation*}
\vec{n}=\frac{\vec{p} \times \vec{p} \, ^{\prime}}{\vec{p} \, ^2}=\frac{\vec{q} \times \vec{q} \, ^{\prime}}{\vec{q} \, ^2} \, \, \, .
\end{equation*}
Of course, $\lvert \vec{n} \rvert = \sin \theta$.

The isospin states of the $\pi N$ system can be expressed in terms of eigenstates of (the square of) the total isospin $\vec{I}=\frac{1}{2} \vec{\tau}+\vec{t}$ and of its third component $I_3$. Within a formalism obeying isospin invariance, 
when modelling the hadronic part of the scattering amplitude as a sum of contributions associated with Feynman graphs, two quantities ($\alpha_k$ and $\beta_k$, where $k$ identifies the specific graph) are introduced to fix the relation between 
the $I=\frac{3}{2}$ and $I=\frac{1}{2}$ components. The final expression for the individual contributions to the $T$-matrix element, including the isospin structure, reads as
\begin{align} \label{eq:EQ19}
\mathscr{T}_k &= \frac{1}{2 m_p} \phi_f^\dagger \, (\alpha_k + \beta_k \vec{\tau} \cdot \vec{t}) \, \Big( \big( A_k + B_k (W - m_p) \big) (p_0+m_p) \nonumber \\
& \qquad \qquad \qquad + \big( -A_k + B_k (W + m_p) \big) (p_0-m_p) \xi \nonumber \\
& \qquad \qquad \qquad - \big( -A_k + B_k (W + m_p) \big) (p_0-m_p) i \vec{\sigma} \cdot \vec{n} \Big) \phi_i \, \, \, .
\end{align}

The projection operators for the states of total isospin $I=\frac{3}{2}$ and $I=\frac{1}{2}$ have the form
\begin{equation*}
\mathscr{P}_3=\frac{1}{3} (2 + \vec{\tau} \cdot \vec{t})
\end{equation*}
and
\begin{equation*}
\mathscr{P}_1=\frac{1}{3} (1 - \vec{\tau} \cdot \vec{t}) \, \, \, .
\end{equation*}

One may thus rewrite Eq.~(\ref{eq:EQ19}) as
\begin{align} \label{eq:EQ20}
\mathscr{T}_k &= \frac{1}{2 m_p} \phi_f^\dagger \, \big( (\alpha_k+\beta_k) \mathscr{P}_3 + (\alpha_k-2\beta_k) \mathscr{P}_1 \big) \, \Big( \big( A_k + B_k (W - m_p) \big) (p_0+m_p) \nonumber \\
& \qquad \qquad \qquad + \big( -A_k + B_k (W + m_p) \big) (p_0-m_p) \xi \nonumber \\
& \qquad \qquad \qquad - \big( -A_k + B_k (W + m_p) \big) (p_0-m_p) i \vec{\sigma} \cdot \vec{n} \Big) \phi_i \, \, \, .
\end{align}

The contributions to the hadronic part of the $\pi N$ scattering amplitude in the CM frame are defined as
\begin{equation*}
\mathscr{F}_k = \frac{m_p}{4 \pi W} \mathscr{T}_k
\end{equation*}
and will be put in the form
\begin{equation} \label{eq:EQ21}
\mathscr{F}_k = \phi_f^\dagger \, \big( (\alpha_k+\beta_k) \mathscr{P}_3 + (\alpha_k-2\beta_k) \mathscr{P}_1 \big) \, ( f_k + g_k \vec{\sigma} \cdot \vec{n} ) \phi_i \, \, \, .
\end{equation}
(In Section \ref{sec:ModelAmplitudes}, the subscripts $k$ will be omitted for simplicity.)

The quantities $f$ and $g \sin\theta$ (defined as the sums of the $f_k$ and $g_k \sin\theta$ contributions, respectively) are known as no-spin-flip and spin-flip amplitudes~\footnote{Our definition of the spin-flip amplitude contains the 
imaginary unit $i$ of Eq.~(\ref{eq:EQ20}).}. As mentioned earlier, the quantities $f$ and $g$ are functions of two Mandelstam variables; we choose $s$ and $\xi$ as independent variables. The expansions of the functions $f (s,\xi)$ and 
$g (s,\xi)$ in Legendre series lead to the extraction of the hadronic part of the partial-wave amplitudes.
\begin{equation} \label{eq:EQ22}
f (s,\xi) = f_{0+} (s) + \sum_{l>0} \big( (l+1) f_{l+} (s) + l f_{l-} (s) \big) P_l (\xi)
\end{equation}
\begin{equation} \label{eq:EQ23}
g (s,\xi) = i \sum_{l>0} \big( f_{l+} (s) - f_{l-} (s) \big) P_l^\prime (\xi) \, \, \, ,
\end{equation}
where the polynomials $P_l (\xi)$ satisfy the relations
\begin{equation} \label{eq:EQ24}
\int_{-1}^{1} P_l (\xi) P_m (\xi) d\xi = \frac{2}{2l+1} \delta_{lm}
\end{equation}
and
\begin{equation} \label{eq:EQ25}
\int_{-1}^{1} (1 - \xi^2) P_l^\prime (\xi) P_m^\prime (\xi) d\xi = \frac{2l(l+1)}{2l+1} \delta_{lm} \, \, \, .
\end{equation}
The subscripts $l\pm$ in Eqs.~(\ref{eq:EQ22}) and (\ref{eq:EQ23}) refer to the total angular momentum $l \pm \frac{1}{2}$. Taking Eqs.~(\ref{eq:EQ22})-(\ref{eq:EQ25}) into account, one obtains $f_{l+} (s)$ and $f_{l-} (s)$ for $l \neq 0$ 
by solving the following set of equations.
\begin{equation} \label{eq:EQ26}
(l+1) f_{l+} (s) + l f_{l-} (s) = \frac{2l+1}{2} \int_{-1}^{1} f (s,\xi) P_l (\xi) d\xi
\end{equation}
\begin{equation} \label{eq:EQ27}
f_{l+} (s) - f_{l-} (s) = - i \frac{2l+1}{2l(l+1)} \int_{-1}^{1} (1-\xi^2) g (s,\xi) P_l^\prime (\xi) d\xi
\end{equation}
The amplitude $f_{0+} (s)$ is obtained by applying Eq.~(\ref{eq:EQ26}) without the term $l f_{l-} (s)$ on the lhs of the equation. The expressions above refer to the hadronic part of each partial-wave amplitude, i.e., to the part which, 
in our papers, has been either associated with the graphs of the model or modelled via suitable parameterisations of the $K$-matrix elements. The expressions obtained after the inclusion of the em effects can be found in Section 2 of 
Ref.~\cite{mworg}.

For the sake of completeness, we give the Legendre polynomials up to $l=3$.
\begin{equation*}
P_0(\xi) = 1 \qquad \qquad \qquad P_1(\xi) = \xi
\end{equation*}
\begin{equation*}
P_2(\xi) = \frac{3 \xi^2-1}{2} \qquad \qquad \qquad P_3(\xi) = \frac{5\xi^3-3\xi}{2} 
\end{equation*}

Integrals of the form
\begin{equation*}
\phi_n (a,b) = \int_{-1}^{1} \frac{\xi^n \, d\xi}{a+b\xi} 
\end{equation*}
(where $\lvert a \rvert > \lvert b \rvert$) appear repeatedly in the partial-wave decomposition of the hadronic part of the scattering amplitude. In the present work, we need the $\phi_n (a,b)$ for $n \leq 5$. For $b \neq 0$,
\begin{align*}
\phi_0 (a,b) &= \frac{1}{b} \ln \frac{a+b}{a-b} \, \, \, , \qquad \phi_1 (a,b) = \frac{2 - a \, \phi_0 (a,b)}{b} \, \, \, , \nonumber \\
\phi_2 (a,b) &= - \frac{a}{b} \phi_1 (a,b) \, \, \, , \qquad \phi_3 (a,b) = \frac{2/3 - a \, \phi_2 (a,b)}{b} \, \, \, , \nonumber \\
\phi_4 (a,b) &= - \frac{a}{b} \phi_3 (a,b) \, \, \, , \qquad \phi_5 (a,b) = \frac{2/5 - a \, \phi_4 (a,b)}{b} \, \, \, .
\end{align*}
From now on, the symbol $\phi$ will denote these integrals; it must not be confused with the two-spinors associated with $u_{i,f} (p)$.

\section{\label{sec:ModelAmplitudes}Contributions of the graphs of the model to the partial-wave amplitudes}

\begin{figure}
\begin{center}
\includegraphics [width=15.5cm] {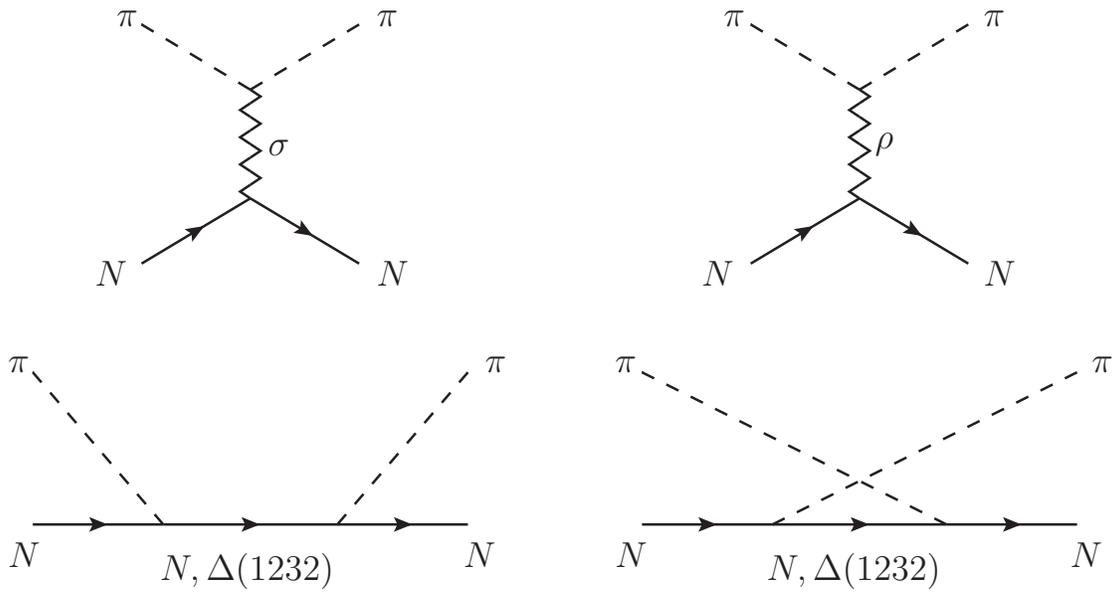}
\caption{\label{fig:FeynmanGraphsETHZ}The main graphs of the ETH model: scalar-isoscalar ($I=J=0$) and vector-isovector ($I=J=1$) t-channel graphs (upper part), and $N$ and $\Delta(1232)$ s- and u-channel graphs (lower part). 
The small contributions from the six well-established $s$ and $p$ higher baryon resonances with masses below $2$ GeV (not shown here) are also included analytically in the model.}
\end{center}
\end{figure}

The main graphs of the model comprise $t$-channel scalar-isoscalar ($I=J=0$) and vector-isovector ($I=J=1$) exchanges, as well as the $N$ and the $\Delta(1232)$ $s$- and $u$-channel contributions (see Fig.~\ref{fig:FeynmanGraphsETHZ}). 
The (small) contributions from six $s$ and $p$ HBRs with masses below $2$ GeV have also been included analytically \cite{glmbg}. The derivative coupling in the $t$-channel $I=J=0$ graph was added (for the sake of completeness) in Ref.~\cite{m}; 
after this (inessential) modification, no changes have been made to the model amplitudes.

The expressions for the contributions to the scattering amplitude of each of the graphs contained in the model have been obtained by applying the methodology of Ref.~\cite{bd}. The graphs of the model lead to real-valued $A(s,\xi)$ and 
$B(s,\xi)$ invariant amplitudes. Consequently, the solutions of the set of Eqs.~(\ref{eq:EQ26}) and (\ref{eq:EQ27}) are also real; we will identify these solutions with the standard $K$-matrix elements $K_{l\pm} (s)$. The partial-wave 
amplitudes $f_{l\pm} (s)$ will be obtained from $K_{l\pm} (s)$ via a unitarisation prescription which will be introduced in Subsection \ref{sec:P11}. Being evident, the explicit reference to the energy dependence of the $K$-matrix elements 
will be avoided hereafter. Of course, these remarks hold separately for each value of the total isospin, i.e., $I=\frac{3}{2}$ and $I=\frac{1}{2}$. The isospin amplitudes $A^I$ are defined by the relation $A=A^{3/2} \mathscr{P}_3 + A^{1/2} \mathscr{P}_1$. 
Using the isospin invariant operators $I_2$ and $\vec{\tau} \cdot \vec{t}$, instead of the projection operators $\mathscr{P}_3$ and $\mathscr{P}_1$, one may also decompose $A$ as $A^+ I_2 + A^- \vec{\tau} \cdot \vec{t}$. This relation defines 
the isospin-even (or isoscalar) invariant amplitude $A^+$ and the isospin-odd (or isovector) invariant amplitude $A^-$. Of course, the same remarks hold in the case of the invariant amplitude $B$.

Due to a sign-convention difference, the contributions to $A^-$ and $B^-$ of the present work are opposite to those of Ref.~\cite{h}. The amplitudes $A^+$ and $B^+$ have the same sign. Using the explicit forms of 
$\mathscr{P}_3$ and $\mathscr{P}_1$, one obtains
\begin{equation*}
A^+ = \frac{2 A^{3/2} + A^{1/2}}{3} \quad\quad\quad\quad A^- = \frac{A^{3/2} - A^{1/2}}{3}
\end{equation*}
and analogous relations for the invariant amplitude $B$.

The values of the relevant physical constants (see Table \ref{tab:Constants}) have been taken from the most recent compilation of the Particle-Data Group (PDG) \cite{pdg}. Exempting the definition of the spin-flip amplitude $g$, our conventions 
for the various physical quantities (i.e., for the amplitudes, for the $s$-wave scattering lengths and $p$-wave scattering volumes, etc.) follow Ref.~\cite{ew}.

\subsection{\label{sec:Sigma}$t$-channel $\sigma$ exchange}

Despite the fact that the $\sigma$ meson may be considered to be the `Higgs boson of QCD', necessary for the spontaneous breaking of the chiral symmetry and for understanding the spectrum of the masses of the hadrons \cite{oo,dp}, the debate 
on its very existence has been long. In several studies over a period of thirty years, the determination of the physical content of the state (i.e., whether it is a quark-antiquark or a diquark-antidiquark pair, e.g., see Ref.~\cite{ao} and the 
references therein) has received equal attention as the extraction of its mass and decay width. First experimental evidence of the existence of a light $\sigma$ particle came from the DM2 Collaboration, from fits to the $\pi \pi$ invariant 
mass in J/$\psi$ decays \cite{au}; a broad resonance-like structure around (the invariant mass of) $500$ MeV may be seen in Fig.~13 of that report. The PDG compilation \cite{pdg} lists this state under the identifier $f_0(500)$; it also discusses its 
peculiarities (see pp. 707-9) and provides a list of the relevant literature (see pp. 711-13). (Earlier versions of the PDG compilation referred to the state as $f_0(600)$; in the 1960s, the $\sigma$ meson had also appeared as $\epsilon$ 
meson in the literature.)

The $t$-channel $I=J=0$ contribution to the hadronic amplitude is approximated in the model by the exchange of a $\pi \pi$ resonance, identified as the light $\sigma$ meson ($I^G (J^{PC}) = 0^+(0^{++})$). Given the quantum numbers of the exchanged 
meson, the parameters $\alpha$ and $\beta$, entering the isospin decomposition of the scattering amplitude, have the values: $\alpha=1$ and $\beta=0$; therefore, the $\sigma$-exchange contributions to the $I=\frac{3}{2}$ and $I=\frac{1}{2}$ $K$-matrix 
elements are equal, see Eq.~(\ref{eq:EQ20}). This exchange introduces two coupling constants: $g_{\pi \pi \sigma}$ (in fact, this coupling constant is put equal to $g_{\pi \pi \sigma} m_c$), corresponding to the $\pi \pi \sigma$ vertex, and 
$g_{\sigma N N}$, associated with the $\sigma N N$ vertex.

The $\sigma$-meson mass $m_\sigma$ enters the expressions via its propagator. Up to now, the results of our partial-wave analysis (PWA) of the low-energy $\pi^\pm p$ scattering data have shown practical insensitivity to the choice of the $m_\sigma$ 
value. In our analyses after and including Ref.~\cite{m}, $m_\sigma$ had been fixed at $860$ MeV \cite{torn}. In the present work, $m_\sigma$ will be varied in the interval which is currently recommended by the PDG \cite{pdg}. We address this issue in 
the beginning of Section \ref{sec:Results}.

The $\pi \pi \sigma$ interaction Lagrangian density is of the form
\begin{equation} \label{eq:EQ28}
\Delta \mathscr{L}_{\pi \sigma} = - g_{\pi \pi \sigma} \left( m_c \vec{\pi} \, ^2 - \frac{\kappa_\sigma}{m_c} \partial_\mu \vec{\pi} \cdot \partial^\mu \vec{\pi} \right) \varphi \, \, \, ,
\end{equation}
where $\vec{\pi}$ and $\varphi$ stand for the quantum fields of the pion and of the exchanged scalar meson, respectively. The parameter $\kappa_\sigma$ fixes the relative contribution of the derivative term. The $\sigma N N$ interaction Lagrangian 
density is of the form
\begin{equation*}
\Delta \mathscr{L}_{\sigma N} = - g_{\sigma N N} \bar{\psi} \varphi \psi \, \, \, ,
\end{equation*}
where $\psi$ denotes the field of the nucleon.

The $\sigma$-exchange contribution to the invariant amplitude $B$ vanishes. The contribution to the invariant amplitude $A$ of the first term within the brackets on the rhs of Eq.~(\ref{eq:EQ28}) is given by
\begin{equation*}
A = \frac{2 g_{\pi \pi \sigma} g_{\sigma N N} m_c}{m_\sigma^2 - t} \, \, \, .
\end{equation*}
Upon inspection of this equation, it becomes evident that only the product of the two coupling constants can be determined from the fits to the experimental data. Inspired by the Fermi coupling constant of the weak interaction, the parameter 
$G_\sigma$ is introduced via the relation: $G_\sigma m_\sigma^2 = g_{\pi \pi \sigma} g_{\sigma N N}$. The previous expression for $A$ may thus be rewritten as
\begin{equation} \label{eq:EQ29}
A = \frac{2 G_\sigma m_\sigma^2 m_c}{m_\sigma^2 - t} \, \, \, .
\end{equation}

Inserting $A$ of Eq.~(\ref{eq:EQ29}), along with $B=0$, into Eq.~(\ref{eq:EQ20}), using Eq.~(\ref{eq:EQ21}) to obtain $f_k$ and $g_k$, and subsequently solving the set of Eqs.~(\ref{eq:EQ26}) and (\ref{eq:EQ27}) with 
$f(s,\xi)=f_k$ and $g(s,\xi)=g_k$ (given the linearity of the Legendre decomposition, Eqs.~(\ref{eq:EQ26}) and (\ref{eq:EQ27}) also hold for the contributions of each specific graph), one obtains the following expressions for the 
$K$-matrix elements up to the $f$ waves.
\begin{align*}
K_{0+} &= \frac{\lambda}{2} \big( (p_0 + m_p) \phi_0 - (p_0 - m_p) \phi_1 \big) \nonumber \\
K_{1+} &= \frac{\lambda}{4} \big( 2 (p_0 + m_p) \phi_1 + (p_0 - m_p) (\phi_0 - 3 \phi_2) \big) \nonumber \\
K_{1-} &= \frac{\lambda}{2} \big( (p_0 + m_p) \phi_1 - (p_0 - m_p) \phi_0 \big) \nonumber \\
K_{2+} &= \frac{\lambda}{4} \big( - (p_0 + m_p) (\phi_0 - 3 \phi_2) + (p_0 - m_p) (3 \phi_1 - 5 \phi_3) \big) \nonumber \\
K_{2-} &= - \frac{\lambda}{4} \big( (p_0 + m_p) (\phi_0 - 3 \phi_2) + 2 (p_0 - m_p) \phi_1 \big) \nonumber \\
K_{3+} &= - \frac{\lambda}{16} \big( 4 (p_0 + m_p) (3 \phi_1 - 5 \phi_3) + (p_0 - m_p) (3 \phi_0 - 30 \phi_2 + 35 \phi_4) \big) \nonumber \\
K_{3-} &= \frac{\lambda}{4} \big( - (p_0 + m_p) (3 \phi_1 - 5 \phi_3) + (p_0 - m_p) (\phi_0 - 3 \phi_2) \big) \, \, \, ,
\end{align*}
where the energy-dependent arguments ($a$,$b$)=($m_\sigma^2+2\vec{q} \, ^2$,$-2\vec{q} \, ^2$) are implied in all $\phi_n$. The quantity $\lambda$ is defined (in the present subsection) as
\begin{equation} \label{eq:EQ30}
\lambda = \frac{G_\sigma m_\sigma^2 m_c}{4 \pi W} \, \, \, .
\end{equation}

The contribution to the invariant amplitude $A$ of the second term within the brackets on the rhs of Eq.~(\ref{eq:EQ28}) (derivative $\pi \sigma$ coupling) is given by
\begin{equation*}
A = \frac{G_\sigma \kappa_\sigma m_\sigma^2}{m_c} \left( \frac{m_\sigma^2 - 2 m_c^2}{m_\sigma^2 - t} - 1 \right) \, \, \, .
\end{equation*}
Evidently, the contribution of the derivative $\pi \sigma$ coupling to the invariant amplitudes $A^+$ and $D^+$ at the CD point vanishes. As the derivative term in Eq.~(\ref{eq:EQ28}) was introduced in Ref.~\cite{m}, the following 
contributions to the $K$-matrix elements had not been listed in Ref.~\cite{glmbg}.
\begin{align*}
K_{0+} &= - \frac{\lambda^\prime}{2} \big( (p_0 + m_p) (2 - \mu \phi_0) + (p_0 - m_p) \mu \phi_1 \big) \nonumber \\
K_{1+} &= \frac{\lambda^\prime \mu}{4} \big( 2 (p_0 + m_p) \phi_1 + (p_0 - m_p) (\phi_0 - 3 \phi_2) \big) \nonumber \\
K_{1-} &= \frac{\lambda^\prime}{2} \big( (p_0 + m_p) \mu \phi_1 + (p_0 - m_p) (2 - \mu \phi_0) \big) \nonumber \\
K_{2+} &= \frac{\lambda^\prime \mu}{4} \big( - (p_0 + m_p) (\phi_0 - 3 \phi_2) + (p_0 - m_p) (3 \phi_1 - 5 \phi_3) \big) \nonumber \\
K_{2-} &= - \frac{\lambda^\prime \mu}{4} \big( (p_0 + m_p) (\phi_0 - 3 \phi_2) + 2 (p_0 - m_p) \phi_1 \big) \nonumber \\
K_{3+} &= - \frac{\lambda^\prime \mu}{16} \big( 4 (p_0 + m_p) (3 \phi_1 - 5 \phi_3) + (p_0 - m_p) (3 \phi_0 - 30 \phi_2 + 35 \phi_4) \big) \nonumber \\
K_{3-} &= \frac{\lambda^\prime \mu}{4} \big( - (p_0 + m_p) (3 \phi_1 - 5 \phi_3) + (p_0 - m_p) (\phi_0 - 3 \phi_2) \big) \, \, \, ,
\end{align*}
where $\mu = m_\sigma^2 - 2 m_c^2$ and
\begin{equation*}
\lambda^\prime = \frac{\lambda \kappa_\sigma}{2 m_c^2} \, \, \, ,
\end{equation*}
the quantity $\lambda$ having been defined in Eq.~(\ref{eq:EQ30}).

As earlier mentioned, given the isospin decomposition of the amplitude in the case of the $\sigma$ exchange, the $I=\frac{3}{2}$ and $I=\frac{1}{2}$ $K$-matrix elements come out equal for all $l$ values; therefore, 
$K_{l\pm}^{3/2}=K_{l\pm}^{1/2}=K_{l\pm}$.

\subsection{\label{sec:Rho}$t$-channel $\rho$ exchange}

The $t$-channel $I=J=1$ contribution is described in the model by the $\rho$ ($I^G (J^{PC}) = 1^+(1^{--})$) exchange. The $\pi \pi \rho$ interaction Lagrangian density is of the form
\begin{equation*}
\Delta \mathscr{L}_{\pi \rho} = g_{\pi \pi \rho} \vec{\rho} \, ^\mu \cdot (\partial_\mu \vec{\pi} \times \vec{\pi} ) \, \, \, ,
\end{equation*}
where $\vec{\rho} \, ^\mu$ denotes the field of the exchanged vector meson. The $\rho N N$ interaction Lagrangian density is of the form
\begin{equation} \label{eq:EQ31}
\Delta \mathscr{L}_{\rho N} = - g_{\rho N N} \bar{\psi} \frac{\vec{\tau}}{2} \cdot \left( \gamma^\mu \vec{\rho} \, _\mu + \frac{\kappa_\rho}{2 m_p} \sigma^{\mu \nu} \partial_\mu \vec{\rho}_\nu \right) \psi \, \, \, ,
\end{equation}
where the parameter $\kappa_\rho$ fixes the relative contribution of the tensor coupling to the $\rho N N$ vertex~\footnote{The sign of the tensor contribution in formula (A.8.28) of Ref.~\cite{h}, p.~565, is opposite to that of our 
Eq.~(\ref{eq:EQ31}). (This difference is not related to the sign convention in the isovector amplitudes between the present work and Ref.~\cite{h}.) In spite of this difference however, the expressions for the $\rho$-exchange 
contributions to the invariant amplitudes $A$ and $B$ in Ref.~\cite{h} agree with those given herein. We thus consider the sign of the tensor-coupling term in Eq.~(A.8.28) of Ref.~\cite{h} to be a typographical error.}. The matrices 
$\sigma^{\mu \nu}$ are defined by the relation: $\sigma^{\mu \nu} = \frac{i}{2} [\gamma^\mu , \gamma^\nu]$. As in the case of the $\sigma$ exchange, we will introduce the parameter $G_\rho$ via the relation 
$G_\rho m_\rho^2 = g_{\pi \pi \rho} g_{\rho N N}$.

The contribution of the first term within the brackets on the rhs of Eq.~(\ref{eq:EQ31}) (vector $\rho N$ coupling) to the invariant amplitude $A$ vanishes; the contribution to the invariant amplitude $B$ is given by
\begin{equation} \label{eq:EQ32}
B = - \frac{G_\rho m_\rho^2}{m_\rho^2 - t} \, \, \, .
\end{equation}

From Eq.~(\ref{eq:EQ32}), along with $A=0$, one obtains the following expressions.
\begin{align*}
K_{0+} &= \frac{\lambda}{2} \big( (W - m_p) (p_0 + m_p) \phi_0 + (W + m_p) (p_0 - m_p) \phi_1 \big) \nonumber \\
K_{1+} &= \frac{\lambda}{4} \big( 2 (W - m_p) (p_0 + m_p) \phi_1 - (W + m_p) (p_0 - m_p) (\phi_0 - 3 \phi_2) \big) \nonumber \\
K_{1-} &= \frac{\lambda}{2} \big( (W - m_p) (p_0 + m_p) \phi_1 + (W + m_p) (p_0 - m_p) \phi_0 \big) \nonumber \\
K_{2+} &= - \frac{\lambda}{4} \big( (W - m_p) (p_0 + m_p) (\phi_0 - 3 \phi_2) + (W + m_p) (p_0 - m_p) (3 \phi_1 - 5 \phi_3) \big) \nonumber \\
K_{2-} &= \frac{\lambda}{4} \big( - (W - m_p) (p_0 + m_p) (\phi_0 - 3 \phi_2) + 2 (W + m_p) (p_0 - m_p) \phi_1 \big) \nonumber \\
K_{3+} &= \frac{\lambda}{16} \big( - 4 (W - m_p) (p_0 + m_p) (3 \phi_1 - 5 \phi_3) + (W + m_p) (p_0 - m_p) (3 \phi_0 - 30 \phi_2 + 35 \phi_4) \big) \nonumber \\
K_{3-} &= - \frac{\lambda}{4} \big( (W - m_p) (p_0 + m_p) (3 \phi_1 - 5 \phi_3) + (W + m_p) (p_0 - m_p) (\phi_0 - 3 \phi_2) \big) \, \, \, ,
\end{align*}
where the energy-dependent arguments ($a$,$b$)=($m_\rho^2+2\vec{q} \, ^2$,$-2\vec{q} \, ^2$) are implied in all $\phi_n$. In the present subsection, the quantity $\lambda$ is defined as
\begin{equation*}
\lambda = - \frac{G_\rho m_\rho^2}{8 \pi W} \, \, \, .
\end{equation*}

The contribution of the second term within the brackets on the rhs of Eq.~(\ref{eq:EQ31}) (tensor $\rho N$ coupling) leads to the following expressions for the invariant amplitudes.
\begin{equation*}
A = \frac{G_\rho m_\rho^2 \kappa_\rho \nu}{m_\rho^2 - t}
\end{equation*}
\begin{equation*}
B = - \frac{G_\rho m_\rho^2 \kappa_\rho}{m_\rho^2 - t}
\end{equation*}

One finally obtains the expressions.
\begin{align*}
K_{0+} &= \frac{\lambda^\prime}{2} \big( 2 W (\phi_0-\phi_1) - p_0 (\phi_0-2\phi_1+\phi_2) - m_p (\phi_0-\phi_2) \big) \nonumber \\
K_{1+} &= \frac{\lambda^\prime}{4} \big( 2 W (\phi_0+2\phi_1-3\phi_2) - p_0 (\phi_0+\phi_1-5\phi_2+3\phi_3) + m_p (\phi_0-3\phi_1-\phi_2+3\phi_3) \big) \nonumber \\
K_{1-} &= \frac{\lambda^\prime}{2} \big( - 2 W (\phi_0-\phi_1) + p_0 (\phi_0-2\phi_1+\phi_2) - m_p (\phi_0-\phi_2) \big) \nonumber \\
K_{2+} &= \frac{\lambda^\prime}{4} \big( - 2 W (\phi_0-3\phi_1-3\phi_2+5\phi_3) + p_0 (\phi_0-4\phi_1+8\phi_3-5\phi_4) \nonumber \\
&\qquad \quad + m_p (\phi_0+2\phi_1-6\phi_2-2\phi_3+5\phi_4) \big) \nonumber \\
K_{2-} &= \frac{\lambda^\prime}{4} \big( - 2 W (\phi_0+2\phi_1-3\phi_2) + p_0 (\phi_0+\phi_1-5\phi_2+3\phi_3) + m_p (\phi_0-3\phi_1-\phi_2+3\phi_3) \big) \nonumber \\
K_{3+} &= \frac{\lambda^\prime}{16} \big( - 2 W (3\phi_0+12\phi_1-30\phi_2-20\phi_3+35\phi_4) \nonumber \\
&\qquad \quad + p_0 (3\phi_0+9\phi_1-42\phi_2+10\phi_3+55\phi_4-35\phi_5) \nonumber \\
&\qquad \quad - m_p (3\phi_0-15\phi_1-18\phi_2+50\phi_3+15\phi_4-35\phi_5) \big) \nonumber \\
K_{3-} &= \frac{\lambda^\prime}{4} \big( 2 W (\phi_0-3\phi_1-3\phi_2+5\phi_3) - p_0 (\phi_0-4\phi_1+8\phi_3-5\phi_4) \nonumber \\
&\qquad \quad + m_p (\phi_0+2\phi_1-6\phi_2-2\phi_3+5\phi_4) \big) \, \, \, ,
\end{align*}
where
\begin{equation*}
\lambda^\prime = \frac{G_\rho m_\rho^2 \kappa_\rho \vec{q} \, ^2}{16 \pi W m_p} \, \, \, .
\end{equation*}
Evidently, the contribution of the tensor $\rho N$ coupling to the isovector $s$-wave scattering length $b_1$ vanishes.

The parameters $\alpha$ and $\beta$, entering the isospin decomposition of the scattering amplitude in the case of the $\rho$ exchange, have the values: $\alpha=0$ and $\beta=1$; therefore, $K_{l\pm}^{3/2}=K_{l\pm}$ and 
$K_{l\pm}^{1/2}=-2 K_{l\pm}$.

\subsection{\label{sec:Nucleon}$s$- and $u$-channel $N$ graphs}

The most general form of the $\pi N N$ interaction Lagrangian density is
\begin{equation*}
\Delta \mathscr{L}_{\pi N} = - \frac{g_{\pi N N}}{1+x} \bar{\psi} \gamma^5 \vec{\tau} \cdot \left( i x \vec{\pi} + \frac{1}{2 m_p} \gamma^\mu \partial_\mu \vec{\pi} \right) \psi \, \, \, ,
\end{equation*}
where $\gamma^5 = i \gamma^0 \gamma^1 \gamma^2 \gamma^3$. 
The quantity $g_{\pi N N}$ is the standard $\pi N$ coupling constant; the parameter $x$ describes the strength of the pseudoscalar admixture in the $\pi N N$ vertex~\footnote{A similar admixture of the two couplings was used 
in Ref.~\cite{gs}. Their relevant quantity $\lambda$ is related to our $x$ via the expression $\lambda=\frac{x}{1+x}$. The two descriptions of the $\pi N N$ vertex are equivalent.}. Both pure pseudovector ($x=0$) and pure 
pseudoscalar ($x \to \infty$) $\pi N$ couplings have been used in the past. It is known that these two couplings are equivalent for \emph{free} nucleons; the equivalence is expected to break in case of bound or off-shell nucleons. 
A short discussion on this subject may be found in Refs.~\cite{ew} (p.~13) and \cite{gs}. Despite the fact that the recent fits of the model have been performed using a pure pseudovector $\pi N$ coupling (see beginning of Section 
\ref{sec:Results}), the general expressions will be given here, i.e., those for arbitrary real $x$.

The contributions of the $s$-channel graph to the invariant amplitudes $A$ and $B$ are as follows.
\begin{equation*}
A = \frac{g_{\pi N N}^2}{2 m_p} \left( 1 - \frac{x^2}{(1+x)^2} \right)
\end{equation*}
\begin{equation*}
B = - g_{\pi N N}^2 \left( \frac{1}{s-m_p^2} + \frac{1}{4 m_p^2 (1+x)^2} \right)
\end{equation*}

The $s$-channel graph leads to the following expressions.
\begin{align*}
K_{0+} &= - \frac{g_{\pi N N}^2}{8 \pi W} \big( \frac{W-(3+4x)m_p}{4 m_p^2(1+x)^2} + \frac{1}{W+m_p} \big) (p_0 + m_p) \nonumber \\
K_{1-} &= - \frac{g_{\pi N N}^2}{8 \pi W} \big( \frac{W+(3+4x)m_p}{4 m_p^2(1+x)^2} + \frac{1}{W-m_p} \big) (p_0 - m_p)
\end{align*}
The contributions to all other partial waves vanish.

For the $s$-channel $N$ graph, the parameters $\alpha$ and $\beta$, entering the isospin decomposition of the scattering amplitude, have the values: $\alpha=1$ and $\beta=-1$; therefore, $K_{l\pm}^{3/2}=0$ and 
$K_{l\pm}^{1/2}=3 K_{l\pm}$.

The contributions of the $u$-channel graph to the invariant amplitudes are as follows.
\begin{equation*}
A = \frac{g_{\pi N N}^2}{2 m_p} \left( 1 - \frac{x^2}{(1+x)^2} \right)
\end{equation*}
\begin{equation*}
B = g_{\pi N N}^2 \left( \frac{1}{u-m_p^2} + \frac{1}{4 m_p^2 (1+x)^2} \right)
\end{equation*}

The $u$-channel graph leads to the following expressions.
\begin{align*}
K_{0+} &= \lambda \Big( \big( \frac{W+(1+4x)m_p}{2 m_p^2(1+x)^2} - (W - m_p) \phi_0 \big) (p_0 + m_p) - (W + m_p) (p_0 - m_p) \phi_1 \Big) \nonumber \\
K_{1+} &= \frac{\lambda}{2} \big( - 2 (W - m_p) (p_0 + m_p) \phi_1 + (W + m_p) (p_0 - m_p) (\phi_0 - 3 \phi_2) \big) \nonumber \\
K_{1-} &= \lambda \Big( - (W - m_p) (p_0 + m_p) \phi_1 + \big( \frac{W-(1+4x)m_p}{2 m_p^2(1+x)^2} - (W + m_p) \phi_0 \big) (p_0 - m_p) \Big) \nonumber \\
K_{2+} &= \frac{\lambda}{2} \big( (W - m_p) (p_0 + m_p) (\phi_0 - 3 \phi_2) + (W + m_p) (p_0 - m_p) (3 \phi_1 - 5 \phi_3) \big) \nonumber \\
K_{2-} &= \frac{\lambda}{2} \big( (W - m_p) (p_0 + m_p) (\phi_0 - 3 \phi_2) - 2 (W + m_p) (p_0 - m_p) \phi_1 \big) \nonumber \\
K_{3+} &= \frac{\lambda}{8} \big( 4 (W - m_p) (p_0 + m_p) (3 \phi_1 - 5 \phi_3) - (W + m_p) (p_0 - m_p) (3 \phi_0 -30 \phi_2 + 35 \phi_4) \big) \nonumber \\
K_{3-} &= \frac{\lambda}{2} \big( (W - m_p) (p_0 + m_p) (3 \phi_1 - 5 \phi_3) + (W + m_p) (p_0 - m_p) (\phi_0 - 3 \phi_2) \big) \, \, \, ,
\end{align*}
where
\begin{equation*}
\lambda=\frac{g_{\pi N N}^2}{16 \pi W} \, \, \, .
\end{equation*}
In the expressions for the $u$-channel graph, the energy-dependent arguments ($a$,$b$)=($2 p_0 q_0 - m_c^2$,$2\vec{q} \, ^2$) are implied in all $\phi_n$.

For the $u$-channel $N$ graph, the parameters $\alpha$ and $\beta$, entering the isospin decomposition of the scattering amplitude, have the values: $\alpha=1$ and $\beta=1$; therefore, 
$K_{l\pm}^{3/2}=2 K_{l\pm}$ and $K_{l\pm}^{1/2}=- K_{l\pm}$.

We will now elaborate further on the invariant amplitudes obtained from the graphs of the present subsection. Taking into account the isospin decomposition of the scattering amplitude (i.e., the quantities $\alpha$ and $\beta$), 
one obtains the following contributions to $A^\pm$ and $B^\pm$.
\begin{equation*}
A^+ = \frac{g_{\pi N N}^2}{m_p} \left( 1 - \frac{x^2}{(1+x)^2} \right)
\end{equation*}
\begin{equation*}
B^+ = - g_{\pi N N}^2 \left( \frac{1}{s-m_p^2} - \frac{1}{u-m_p^2} \right)
\end{equation*}
\begin{equation*}
A^- = 0
\end{equation*}
\begin{equation*}
B^- = g_{\pi N N}^2 \left( \frac{1}{s-m_p^2} + \frac{1}{u-m_p^2} + \frac{1}{2 m_p^2 (1+x)^2} \right)
\end{equation*}
The isoscalar invariant amplitude $B^+$ is independent of $x$, whereas the isovector invariant amplitude $A^-$ vanishes. Both invariant amplitudes $B^\pm$ are singular along the straight lines 
$s=m_p^2$ and $u=m_p^2$ on the $(\nu,t)$ plane (corresponding to $t = 2 m_c^2 \pm 4 m_p \nu$). Using Eqs.~(\ref{eq:EQ12}) and (\ref{eq:EQ13}), one may put the invariant amplitudes $B^\pm$ into the forms:
\begin{equation*}
B^+ = \frac{g_{\pi N N}^2}{m_p} \frac{\nu}{\nu_B^2-\nu^2}
\end{equation*}
and
\begin{equation*}
B^- = - \frac{g_{\pi N N}^2}{m_p} \left( \frac{\nu_B}{\nu_B^2-\nu^2} - \frac{1}{2 m_p (1+x)^2} \right) \, \, \, .
\end{equation*}

As a result, the nucleon contribution to the isoscalar amplitude $D^+$ reads as
\begin{equation} \label{eq:EQ33}
D^+_N = \frac{g_{\pi N N}^2}{m_p} \left( 1 - \frac{x^2}{(1+x)^2} + \frac{\nu^2}{\nu_B^2-\nu^2} \right) = \frac{g_{\pi N N}^2}{m_p} \left( \frac{\nu_B^2}{\nu_B^2-\nu^2} - \frac{x^2}{(1+x)^2} \right) \, \, \, .
\end{equation}
When extrapolating the $D^+$ amplitude into the unphysical region, toward the CD point, the first term within the brackets in the last part of the previous equation (pseudovector Born-term contribution) is omitted.

\subsection{\label{sec:Delta}$s$- and $u$-channel $\Delta(1232)$ graphs}

The treatment of graphs, involving the exchange of a virtual fermion with spin $J>\frac{1}{2}$, is rather intricate; as a result of the off-shellness of the intermediate state, the propagator is expected to contain, apart from 
the `nominal' contributions of the spin-$J$ state, contributions from the states of lower spin $J-n$, where $1 \leq n \leq J-\frac{1}{2}$. (Only $n=1$ is possible for $J=\frac{3}{2}$.) Fixing the admixture of each of these 
states can hardly be made on the basis of theoretical arguments; as a result, each such state introduces one additional free parameter.

The first attempts to construct the propagator of the $\Delta(1232)$ date back to the late 1930s \cite{fp} and early 1940s \cite{rs}. The propagator, then obtained, has been known in the literature as the `Rarita-Schwinger propagator'; 
it comprises spin-$\frac{3}{2}$ and spin-$\frac{1}{2}$ contributions and contains an arbitrary complex parameter $A \neq -\frac{1}{2}$. The expression for the Rarita-Schwinger propagator may be found in Ref.~\cite{h}, p.~562, Eq.~(A.8.9). 
In the mid 1980s, Williams proposed a propagator which did not contain spin-$\frac{1}{2}$ contributions \cite{wl}, but shortly afterwards Benmerrouche, Davidson, and Mukhopadhyay \cite{bdm} demonstrated that the `Williams propagator' has 
no inverse, hence it cannot be correct. Other propagators appeared in the late 1990s, without \cite{pasca1,pasca2} or with \cite{haber} spin-$\frac{1}{2}$ contributions. In our analyses after (and including) Ref.~\cite{glm2}, we have 
followed the Rarita-Schwinger formalism as given in Ref.~\cite{h}, with $A=-1$.

The interaction Lagrangian density
\begin{equation} \label{eq:EQ34}
\Delta \mathscr{L}_{\pi N \Delta} = \frac{g_{\pi N \Delta}}{2 m_p} \bar{\Psi}_\mu \vec{T} \cdot \Theta^{\mu \nu} \partial_\nu \vec{\pi} \psi + h.c.
\end{equation}
introduces two parameters: the coupling constant~\footnote{When comparing values of the coupling constant $g_{\pi N \Delta}$, the reader must bear in mind that $g_{\pi N \Delta}$ of Ref.~\cite{h} contains the factor $2 m_p$ 
appearing as denominator in our Eq.~(\ref{eq:EQ34}). One possibility of fixing the coupling constant $g_{\pi N \Delta}$ from the decay width of the $\Delta(1232)$ will be given in Subsection \ref{sec:P11} (see footnote \ref{ftn:FTN1}).} 
$g_{\pi N \Delta}$ and the parameter $Z$ associated with the vertex factor
\begin{equation*}
\Theta^{\mu \nu} = g^{\mu \nu} - \left( Z + \frac{1}{2} \right) \gamma^\mu \gamma^\nu \, \, \, .
\end{equation*}
In Eq.~(\ref{eq:EQ34}), $\Psi_\mu$ stands for the vector-spinor field of the $\Delta(1232)$; the spinor index is suppressed. $\vec{T}$ is the transition operator between $I=\frac{3}{2}$ and $I=\frac{1}{2}$ states.

Some authors have argued \cite{pasca2,te,kem} that the spin-$\frac{1}{2}$ contributions to the $\Delta(1232)$ field are redundant in the framework of an Effective Field Theory (EFT), as such off-shell effects can be absorbed in other 
terms of the effective Lagrangian. Nevertheless, it is not clear how this claim impairs the Rarita-Schwinger formalism, at least in terms of its use in phenomenology. We will return to this point in Ref.~\cite{mr4}.

The fixation of parameter $Z$ from theoretical principles has been explored in a number of studies. To start with, using the subsidiary condition $\gamma^\mu \Theta_{\mu \nu} = 0$, Peccei suggested the use of $Z = -\frac{1}{4}$ \cite{pe}. 
Nath, Etemadi, and Kimel \cite{nek} considered Peccei's condition restrictive and, invoking the Lorentz invariance of the resulting $S$-matrix, recommended $Z=\frac{1}{2}$. However, it was argued in Ref.~\cite{bdm} that the $Z=\frac{1}{2}$ 
choice leads to unexpected properties of the $\Delta(1232)$ radiative decay. H\"ohler \cite{h} assumed a cautious attitude regarding the arguments in favour of such `theoretical preferences', thus hinting at the extraction of the value of the 
parameter $Z$ from measurements; we have followed this approach in our PWAs of the low-energy $\pi N$ measurements and obtained results which favour $Z=-\frac{1}{2}$. This observation may be helpful in the treatment of the spin-$\frac{5}{2}$ 
and spin-$\frac{7}{2}$ propagators; if true, the inclusion in the model of the graphs with the $d$ and $f$ HBRs as intermediate states could become possible without the introduction of additional free parameters.

The detailed expressions for the pole and non-pole contributions to the invariant amplitudes $A$ and $B$ may be found in Ref.~\cite{h}, pp.~562 and 564. These contributions had appeared earlier in Ref.~\cite{nek}, but the concise 
formulae of Ref.~\cite{h} are more attractive for a compact implementation~\footnote{One of us (E.M.) has performed the lengthy calculation of the contributions of the $\Delta(1232)$ graphs and confirms the validity of the results 
of Refs.~\cite{h} and \cite{nek}. Regarding the propagation of a massive spin-$\frac{3}{2}$ field, our findings will appear elsewhere \cite{mr4}.}. (The reader must also bear in mind that, compared to Ref.~\cite{h}, the invariant 
amplitudes $B$ are defined with an opposite sign in Ref.~\cite{nek}.) To keep the present work self-contained, the expressions of Ref.~\cite{h} (after they have been modified accordingly, to comply with the notations and conventions 
of the present work) will also appear here.

We now comment on the separation of the contributions of the $\Delta(1232)$ graphs into pole and non-pole parts. The $s$-channel contributions to the invariant amplitudes $A$ and $B$ are functions of the Mandelstam variables 
$s$ and $t$, and may be put in the form of a sum of three terms, each containing a different power of $s-m_\Delta^2$, from $-1$ (inversely proportional) to $1$ (linear); the pole contributions comprise only the terms containing 
$(s-m_\Delta^2)^{-1}$. (Regarding the $u$-channel contributions, the previous comment applies after substituting $s$ with $u$.) Of course, it is true that the separation of the contributions into pole and non-pole parts rests on 
the choice of the independent variables in a study. For instance, if one chooses to use $s$ and $\xi$ (instead of $s$ and $t$), some terms currently categorised in the pole part will move to the non-pole part. To avoid needless 
complication and to facilitate the comparison of our results with the literature, we refrain from redefining the pole and non-pole contributions of the $\Delta(1232)$ graphs, thus following the separation method of Refs.~\cite{h} 
and \cite{nek}.

\subsubsection{\label{sec:PoleDelta}Pole contributions}

The contributions of the $s$-channel graph to the invariant amplitudes $A$ and $B$ are as follows.
\begin{equation} \label{eq:EQ35}
A = \frac{g_{\pi N \Delta}^2}{36 m_p^2} \frac{\alpha_1 + \alpha_2 t}{m_{\Delta}^2-s}
\end{equation}
\begin{equation} \label{eq:EQ36}
B = \frac{g_{\pi N \Delta}^2}{36 m_p^2} \frac{\beta_1 + \beta_2 t}{m_{\Delta}^2-s} \, \, \, ,
\end{equation}
where
\begin{align} \label{eq:EQ37}
\alpha_1 &= 3 (m_\Delta + m_p) \vec{q}_\Delta \, ^2 + (m_\Delta - m_p) (p_{0\Delta} + m_p)^2 \, \, \, , \nonumber \\
\alpha_2 &= \frac{3}{2} (m_\Delta + m_p) \, \, \, , \nonumber \\
\beta_1 &= 3 \vec{q}_\Delta \, ^2 - (p_{0\Delta} + m_p)^2 \, \, \, , \nonumber \\
\beta_2 &= \frac{3}{2} \, \, \, .
\end{align}
In these expressions, $m_{\Delta}$ denotes the mass of the $\Delta(1232)$, $\lvert \vec{q}_\Delta \rvert$ is the CM momentum at $s=m_{\Delta}^2$, and $p_{0\Delta}$ the total CM energy of the nucleon at $s=m_{\Delta}^2$. The 
expressions for the $K$-matrix elements are somewhat simplified if a few additional quantities are introduced.
\begin{align} \label{eq:EQ38}
\alpha_0 &= \alpha_1 - 2\alpha_2 \vec{q} \, ^2 \nonumber \\
\beta_+ &= (\beta_1 - 2\beta_2 \vec{q} \, ^2) (W+m_p) \nonumber \\
\beta_- &= (\beta_1 - 2\beta_2 \vec{q} \, ^2) (W-m_p) \nonumber \\
\beta_+^\prime &= \beta_2 (W+m_p) \nonumber \\
\beta_-^\prime &= \beta_2 (W-m_p)
\end{align}

The $s$-channel graph leads to the following expressions.
\begin{align*}
K_{0+} &= \lambda \big( (\alpha_0 + \beta_-) (p_0 + m_p) + \frac{2 \vec{q} \, ^2}{3} (-\alpha_2 + \beta_+^\prime) (p_0 - m_p) \big) \nonumber \\
K_{1+} &= \frac{2 \lambda \vec{q} \, ^2}{3} (\alpha_2 + \beta_-^\prime) (p_0 + m_p) \nonumber \\
K_{1-} &= K_{1+} + \lambda ( - \alpha_0 + \beta_+ ) (p_0 - m_p) \nonumber \\
K_{2-} &= \frac{2 \lambda \vec{q} \, ^2}{3} (-\alpha_2 + \beta_+^\prime) (p_0 - m_p) \, \, \, ,
\end{align*}
where
\begin{equation} \label{eq:EQ39}
\lambda=\frac{g_{\pi N \Delta}^2}{288 \pi W m_p^2 (m_\Delta^2-W^2)} \, \, \, .
\end{equation}
The contributions to all other partial waves vanish. The $(m_\Delta^2-W^2)$ factor in the denominator of $\lambda$ introduces a pole only in $K_{1+}$, not in the other elements.

For the $s$-channel $\Delta(1232)$ graph, the parameters $\alpha$ and $\beta$ (not to be confused with the quantities defined in Eqs.~(\ref{eq:EQ37}) and (\ref{eq:EQ38}), all of which carry subscripts), 
entering the isospin decomposition of the scattering amplitude, have the values: $\alpha=2$ and $\beta=1$; therefore, $K_{l\pm}^{3/2}=3 K_{l\pm}$ and $K_{l\pm}^{1/2}=0$.

The contributions of the $u$-channel graph to the invariant amplitudes $A$ and $B$ are as follows.
\begin{equation} \label{eq:EQ40}
A = \frac{g_{\pi N \Delta}^2}{36 m_p^2} \frac{\alpha_1 + \alpha_2 t}{m_{\Delta}^2-u}
\end{equation}
\begin{equation} \label{eq:EQ41}
B = - \frac{g_{\pi N \Delta}^2}{36 m_p^2} \frac{\beta_1 + \beta_2 t}{m_{\Delta}^2-u} \, \, \, ,
\end{equation}
where the quantities $\alpha_1$, $\alpha_2$, $\beta_1$, and $\beta_2$ have been defined in Eqs.~(\ref{eq:EQ37}). In the following, we will also make use of the quantities $\alpha_0$, $\beta_+$, $\beta_-$, $\beta_+^\prime$, 
and $\beta_-^\prime$ of Eqs.~(\ref{eq:EQ38}).

The $u$-channel graph leads to the following expressions.
\begin{align} \label{eq:EQ42}
K_{0+} &= \frac{\lambda^\prime}{2} \Big( \big( (\alpha_0 - \beta_-) \phi_0 + 2 (\alpha_2 - \beta_-^\prime) \vec{q} \, ^2 \phi_1 \big) (p_0 + m_p) \nonumber \\
&\qquad\quad - \big( (\alpha_0 + \beta_+) \phi_1 + 2 (\alpha_2 + \beta_+^\prime) \vec{q} \, ^2 \phi_2 \big) (p_0 - m_p) \Big) \nonumber \\
K_{1+} &= \frac{\lambda^\prime}{4} \Big( 2 \big( (\alpha_0 - \beta_-) \phi_1 + 2 (\alpha_2 - \beta_-^\prime) \vec{q} \, ^2 \phi_2 \big) (p_0 + m_p) \nonumber \\
&\qquad\quad + \big( (\alpha_0 + \beta_+) (\phi_0 - 3\phi_2) + 2 (\alpha_2 + \beta_+^\prime) \vec{q} \, ^2 (\phi_1 - 3\phi_3) \big) (p_0-m_p) \Big) \nonumber \\
K_{1-} &= \frac{\lambda^\prime}{2} \Big( \big( (\alpha_0 - \beta_-) \phi_1 + 2 (\alpha_2 - \beta_-^\prime) \vec{q} \, ^2 \phi_2 \big) (p_0 + m_p) \nonumber \\
&\qquad\quad - \big( (\alpha_0 + \beta_+) \phi_0 + 2 (\alpha_2 + \beta_+^\prime) \vec{q} \, ^2 \phi_1 \big) (p_0-m_p) \Big) \nonumber \\
K_{2+} &= -\frac{\lambda^\prime}{4} \Big( \big( (\alpha_0 - \beta_-) (\phi_0 - 3\phi_2) + 2 (\alpha_2 - \beta_-^\prime) \vec{q} \, ^2 (\phi_1-3\phi_3) \big) (p_0 + m_p) \nonumber \\
&\qquad\qquad - \big( (\alpha_0 + \beta_+) (3\phi_1 - 5\phi_3) + 2 (\alpha_2 + \beta_+^\prime) \vec{q} \, ^2 (3\phi_2-5\phi_4) \big) (p_0 - m_p) \Big) \nonumber \\
K_{2-} &= -\frac{\lambda^\prime}{4} \Big( \big( (\alpha_0 - \beta_-) (\phi_0 - 3\phi_2) + 2 (\alpha_2 - \beta_-^\prime) \vec{q} \, ^2 (\phi_1 - 3\phi_3) \big) (p_0 + m_p) \nonumber \\
&\qquad\qquad + 2 \big( (\alpha_0 + \beta_+) \phi_1 + 2 (\alpha_2 + \beta_+^\prime) \vec{q} \, ^2 \phi_2 \big) (p_0 - m_p) \Big) \nonumber \\
K_{3+} &= -\frac{\lambda^\prime}{16} \Big( 4 \big( (\alpha_0 - \beta_-) (3\phi_1 - 5\phi_3) + 2 (\alpha_2 - \beta_-^\prime) \vec{q} \, ^2 (3\phi_2 - 5\phi_4) \big) (p_0 + m_p) \nonumber \\
&\qquad\qquad + \big( (\alpha_0 + \beta_+) (3\phi_0 - 30\phi_2 + 35\phi_4) + 2 (\alpha_2 + \beta_+^\prime) \vec{q} \, ^2 (3\phi_1 - 30\phi_3 + 35\phi_5) \big) (p_0 - m_p) \Big) \nonumber \\
K_{3-} &= \frac{\lambda^\prime}{4} \Big( - \big( (\alpha_0 - \beta_-) (3\phi_1 - 5\phi_3) + 2 (\alpha_2 - \beta_-^\prime) \vec{q} \, ^2 (3\phi_2 - 5\phi_4) \big) (p_0 + m_p) \nonumber \\
&\qquad\quad + \big( (\alpha_0 + \beta_+) (\phi_0 - 3\phi_2) + 2 (\alpha_2 + \beta_+^\prime) \vec{q} \, ^2 (\phi_1 - 3\phi_3) \big) (p_0 - m_p) \Big) \, \, \, ,
\end{align}
where
\begin{equation} \label{eq:EQ43}
\lambda^\prime=\frac{g_{\pi N \Delta}^2}{288 \pi W m_p^2} \, \, \, .
\end{equation}
In Eqs.~(\ref{eq:EQ42}), the energy-dependent arguments ($a$,$b$)=($m_\Delta^2 - m_p^2 -m_c^2 + 2 p_0 q_0$,$2\vec{q} \, ^2$) are implied in all $\phi_n$.

For the $u$-channel $\Delta(1232)$ graph, the parameters $\alpha$ and $\beta$, entering the isospin decomposition of the scattering amplitude, have the values: $\alpha=2$ and $\beta=-1$; therefore, 
$K_{l\pm}^{3/2}=K_{l\pm}$ and $K_{l\pm}^{1/2}=4 K_{l\pm}$.

\subsubsection{\label{sec:NonPoleDelta}Non-pole contributions}

The non-pole contributions are split into isoscalar and isovector parts. These contributions affect only the $s$, $p$, and $d$ waves~\footnote{The claim of Ref.~\cite{h}, p.~564, that the non-pole contributions affect 
only the $s$ and $p$ waves is not correct. The invariant amplitudes $A$ and $B$ are indeed linear in $t$, yet a quadratic term in $t$, contributing to the $d$ waves in the Legendre expansion of $f(s,\xi)$, is 
introduced via the second of the three terms contained in the last factor within brackets in Eq.~(\ref{eq:EQ20}).}.

The contribution of the isoscalar part to the invariant amplitudes $A$ and $B$ reads as follows.
\begin{equation} \label{eq:EQ44}
A^+ = - \frac{g_{\pi N \Delta}^2}{9 m_p^2 m_\Delta} \big( (p_{0\Delta}+m_p) (2 m_\Delta-m_p) + (2 + \frac{m_p}{2 m_\Delta}) m_c^2 + (t - 2 m_c^2) Y \big)
\end{equation}
\begin{equation} \label{eq:EQ45}
B^+ = - \frac{4 g_{\pi N \Delta}^2 Z^2 \nu}{9 m_p m_\Delta^2} \, \, \, ,
\end{equation}
where
\begin{equation*}
Y=(2+\frac{m_p}{m_\Delta}) Z^2 + (1+\frac{m_p}{m_\Delta}) Z \, \, \, .
\end{equation*}

To somewhat simplify the expressions for the $K$-matrix elements, we will define four quantities:
\begin{align*}
\alpha_1 &= (p_{0\Delta}+ m_p) (2 m_\Delta-m_p) + (2+\frac{m_p}{2 m_\Delta}) m_c^2 - 2 q_0^2 Y \, \, \, , \nonumber \\
\alpha_2 &= 2 \vec{q} \, ^2 Y \, \, \, , \nonumber \\
\beta_1 &= 2 p_0 q_0 + \vec{q} \, ^2 \, \, \, , \nonumber \\
\beta_2 &= \vec{q} \, ^2 \, \, \, .
\end{align*}

The non-pole isoscalar contributions to the partial waves for $l \leq 3$ are as follows.
\begin{align*}
K_{0+} &= -\lambda \Big( \big( \alpha_1+\frac{2Z^2\beta_1}{m_\Delta} (W-m_p) \big) (p_0+m_p)+\frac{1}{3} \big( -\alpha_2+\frac{2Z^2\beta_2}{m_\Delta} (W+m_p) \big)(p_0-m_p) \Big) \nonumber \\
K_{1+} &= -\frac{\lambda}{3} \big( \alpha_2+\frac{2Z^2\beta_2}{m_\Delta} (W-m_p) \big) (p_0+m_p) \nonumber \\
K_{1-} &= K_{1+}+\lambda \big( \alpha_1-\frac{2Z^2\beta_1}{m_\Delta} (W+m_p) \big) (p_0-m_p) \nonumber \\
K_{2-} &= \frac{\lambda}{3} \big( \alpha_2 - \frac{2 Z^2 \beta_2}{m_\Delta} (W+m_p) \big) (p_0-m_p) \, \, \, , \nonumber \\
\end{align*}
where
\begin{equation} \label{eq:EQ46}
\lambda=\frac{g_{\pi N \Delta}^2}{72 \pi W m_p^2 m_\Delta} \, \, \, .
\end{equation}
The contributions to all other partial waves vanish. For the isoscalar part of the non-pole contributions, $K_{l\pm}^{3/2}=K_{l\pm}$ and $K_{l\pm}^{1/2}=K_{l\pm}$.

The contributions of the isovector part to the invariant amplitudes $A$ and $B$ are as follows.
\begin{equation} \label{eq:EQ47}
A^- = \frac{2 g_{\pi N \Delta}^2 Y \nu}{9 m_p m_\Delta}
\end{equation}
\begin{equation} \label{eq:EQ48}
B^- = - \frac{g_{\pi N \Delta}^2}{36 m_p^2} \Big( (1+\frac{m_p}{m_\Delta})^2 +\frac{8m_pY}{m_\Delta} +\frac{4}{m_\Delta^2} \big( (m_c^2 -\frac{t}{2}) Z^2 - m_c^2 Z \big) \Big)
\end{equation}

We will now redefine the auxiliary quantities $\alpha_1$, $\alpha_2$, $\beta_1$, and $\beta_2$ for the remaining part of the present subsection.
\begin{align*}
\alpha_1 &= 2 p_0 q_0 + \vec{q} \, ^2 \nonumber \\
\alpha_2 &= \vec{q} \, ^2 \nonumber \\
\beta_1 &= (1+\frac{m_p}{m_\Delta})^2 + \frac{8m_pY}{m_\Delta} +\frac{4}{m_\Delta^2} ( q_0^2 Z^2 - m_c^2 Z ) \nonumber \\
\beta_2 &= -\frac{4 Z^2 \vec{q} \, ^2}{m_\Delta^2}
\end{align*}

The non-pole isovector contributions to the partial waves for $l \leq 3$ are as follows.
\begin{align*}
K_{0+} &= \lambda^\prime \Big( \big( \frac{4Y\alpha_1}{m_\Delta}-\beta_1(W-m_p) \big) (p_0+m_p)- \frac{1}{3} \big( \frac{4Y\alpha_2}{m_\Delta}+\beta_2 (W+m_p) \big) (p_0-m_p) \Big) \nonumber \\
K_{1+} &= \frac{\lambda^\prime}{3} \big( \frac{4Y\alpha_2}{m_\Delta}-\beta_2(W-m_p) \big) (p_0+m_p) \nonumber \\
K_{1-} &= K_{1+}-\lambda^\prime \big( \frac{4Y\alpha_1}{m_\Delta}+\beta_1(W+m_p) \big) (p_0-m_p) \nonumber \\
K_{2-} &= -\frac{\lambda^\prime}{3} \big( \frac{4Y\alpha_2}{m_\Delta}+\beta_2(W+m_p) \big) (p_0-m_p) \, \, \, , \nonumber \\
\end{align*}
where
\begin{equation} \label{eq:EQ49}
\lambda^\prime=\frac{g_{\pi N \Delta}^2}{288 \pi W m_p^2} \, \, \, .
\end{equation}
The contributions to all other partial waves vanish. For the isovector part of the non-pole contributions, $K_{l\pm}^{3/2}=K_{l\pm}$ and $K_{l\pm}^{1/2}=-2 K_{l\pm}$.

\subsection{\label{sec:HR}$s$- and $u$-channel graphs with the $s$ and $p$ HBRs as intermediate states}

Given in Subsections \ref{sec:Sigma}-\ref{sec:Delta} were the detailed contributions of the main graphs of the model to the $K$-matrix elements up to the $f$ waves. However, the model contains additional $s$- and $u$-channel 
graphs which, owing to the relative weakness of the corresponding coupling constants and to the larger masses of the intermediate states, contribute significantly less to the invariant amplitudes at low energies. Up to the 
present time, the states which have been treated analytically are those of the well-established (four-star, in the PDG listings \cite{pdg}) $s$ and $p$ HBRs with masses~\footnote{Of course, the upper bound of $2$ GeV, which 
had been set in Ref.~\cite{glmbg} regarding the mass of the accepted baryon states, is arbitrary. The selected value enabled the analytical inclusion in the model of all the significant contributions, whereas the states with 
larger masses were considered to be too distant to affect the results obtained at low energies to a `detectable' degree.} below $2$ GeV. In spite of the smallness of these contributions at low energies (to the extent that they 
are frequently ignored in low-energy studies of the $\pi N$ system), they have been part of the model after (and including) Ref.~\cite{glmbg}; the expressions for these contributions will be given in the present subsection. 
The contributions of the four $N$-type states will be discussed in Subsections \ref{sec:P11}-\ref{sec:P13}, those of the two $\Delta$-type states in Subsections \ref{sec:S31} and \ref{sec:P31}.

To avoid complexity in the notation, one variable will denote the masses of all these states $R$, namely $M_R$. The coupling constants $g_{\pi N R}$ will be fixed from the partial widths for the decay processes 
$R \rightarrow \pi N$ (explicit expressions will be given in each case); each partial width $\Gamma$ is defined as the product of the total width $\Gamma_T$ of the specific resonance and of the branching ratio $\eta$ for its 
$\pi N$ decay mode. The relevant values of the masses, of the total decay widths, and of the branching ratios are given in Table \ref{tab:Constants}.

All $N$-type $s$-channel contributions are characterised by an isospin decomposition with $\alpha=1$ and $\beta=-1$ ($K_{l\pm}^{3/2}=0$ and $K_{l\pm}^{1/2}=3 K_{l\pm}$); all $u$-channel contributions (of the $N$-type resonances) 
follow the $\alpha=1$ and $\beta=1$ decomposition ($K_{l\pm}^{3/2}=2 K_{l\pm}$ and $K_{l\pm}^{1/2}=-K_{l\pm}$). All $s$-channel $\Delta$-type isospin decompositions correspond to $\alpha=2$ and $\beta=1$ ($K_{l\pm}^{3/2}=3 K_{l\pm}$ 
and $K_{l\pm}^{1/2}=0$); finally, all $u$-channel contributions (of the $\Delta$-type resonances) may be obtained after using $\alpha=2$ and $\beta=-1$ ($K_{l\pm}^{3/2}=K_{l\pm}$ and $K_{l\pm}^{1/2}=4 K_{l\pm}$).

In the expressions of the $u$-channel graphs, the energy-dependent arguments ($a$,$b$)=($M_R^2 -m_p^2 - m_c^2 + 2 p_0 q_0$,$2\vec{q} \, ^2$) will be implied in all $\phi_n$. The quantities $\lvert \vec{q}_R \rvert$ and $p_{0R}$ 
will denote the CM momentum and the total CM energy of the nucleon at the specific resonance position ($s=M_R^2$).

\subsubsection{\label{sec:P11}$N(1440)$}

The $N(1440)$, also known as `Roper resonance', is an $I(J^P) = \frac{1}{2}(\frac{1}{2}^+)$ state (like the nucleon). As a result, the only modifications in the expressions of Subsection \ref{sec:Nucleon} are due to the mass 
of the intermediate state. For the sake of completeness, the general expressions for the invariant amplitudes $A$ and $B$ will be given, i.e., treating the $\pi N R$ coupling in the most general manner. The 
expressions for the contributions of the $s$-channel graphs read as:
\begin{equation} \label{eq:EQ50}
A = \frac{g_{\pi N R}^2}{2 m_p} \frac{s - m_p^2}{s - M_R^2} \left( \frac{2 m_p^2 (r-1) x^2}{(s-m_p^2) (1+x)^2} + 1 - \frac{x^2 - \frac{r-1}{2}}{(1+x)^2} \right)
\end{equation}
and
\begin{equation} \label{eq:EQ51}
B = - g_{\pi N R}^2 \frac{s - m_p^2}{s - M_R^2} \left( \frac{1}{s - m_p^2} + \frac{(r-1)(x+\frac{1}{2})}{(s - m_p^2) (1+x)^2} + \frac{1}{4 m_p^2 (1+x)^2} \right) \, \, \, ,
\end{equation}
where $r$ denotes the ratio $M_R / m_p$. The expressions for the contributions of the $u$-channel graphs are:
\begin{equation} \label{eq:EQ52}
A = \frac{g_{\pi N R}^2}{2 m_p} \frac{u - m_p^2}{u - M_R^2} \left( \frac{2 m_p^2 (r-1) x^2}{(u - m_p^2) (1+x)^2} + 1 - \frac{x^2-\frac{r-1}{2}}{(1+x)^2} \right)
\end{equation}
and
\begin{equation} \label{eq:EQ53}
B = g_{\pi N R}^2 \frac{u - m_p^2}{u - M_R^2} \left( \frac{1}{u - m_p^2} + \frac{(r-1)(x+\frac{1}{2})}{(u - m_p^2) (1+x)^2} + \frac{1}{4 m_p^2 (1+x)^2} \right) \, \, \, .
\end{equation}

In the beginning of Section \ref{sec:Results}, we will explain why our recent fits to the data have been performed using a pure pseudovector coupling in the graphs of Subsection \ref{sec:Nucleon}. As we will treat the 
(significantly smaller) contributions of the graphs with an $N(1440)$ intermediate state similarly, we restrict ourselves to $x=0$ in the remaining part of the present subsection. For a pure pseudovector coupling, Eqs.~(\ref{eq:EQ50}) 
and (\ref{eq:EQ51}) lead to the following $s$-channel contributions.
\begin{align} \label{eq:EQ54}
K_{0+} &= - \frac{g_{\pi N R}^2}{32 \pi W m_p^2} \frac{(W-m_p)^2}{W+M_R} (p_0 + m_p) \nonumber \\
K_{1-} &= - \frac{g_{\pi N R}^2}{32 \pi W m_p^2} \frac{(W+m_p)^2}{W-M_R} (p_0 - m_p)
\end{align}
The contributions to all other partial waves vanish.

The $u$-channel contributions, obtained from Eqs.~(\ref{eq:EQ52}) and (\ref{eq:EQ53}), read as:
\begin{align} \label{eq:EQ55}
K_{0+} &= \frac{\lambda}{2} (2 \alpha_1 - \beta_1 \phi_0 - \beta_2 \phi_1 ) \nonumber \\
K_{1+} &= \frac{\lambda}{4} \big( - 2 \beta_1 \phi_1 + \beta_2 (\phi_0 - 3 \phi_2) \big) \nonumber \\
K_{1-} &= \frac{\lambda}{2} (2 \alpha_2 - \beta_2 \phi_0 - \beta_1 \phi_1 ) \nonumber \\
K_{2+} &= \frac{\lambda}{4} \big( \beta_1 (\phi_0 - 3 \phi_2) + \beta_2 (3 \phi_1 - 5 \phi_3) \big) \nonumber \\
K_{2-} &= \frac{\lambda}{4} \big( \beta_1 (\phi_0 - 3 \phi_2) - 2 \beta_2 \phi_1 \big) \nonumber \\
K_{3+} &= \frac{\lambda}{16} \big( 4 \beta_1 (3 \phi_1 - 5 \phi_3) - \beta_2 (3 \phi_0 - 30 \phi_2 + 35 \phi_4) \big) \nonumber \\
K_{3-} &= \frac{\lambda}{4} \big( \beta_1 (3 \phi_1 - 5 \phi_3) + \beta_2 (\phi_0 - 3 \phi_2) \big) \, \, \, ,
\end{align}
where
\begin{equation*}
\lambda=\frac{g_{\pi N R}^2}{32 \pi W m_p^2}
\end{equation*}
and
\begin{align} \label{eq:EQ56}
\alpha_1 &= (W + M_R) (p_0 + m_p) \, \, \, , \nonumber \\
\alpha_2 &= (W - M_R) (p_0 - m_p) \, \, \, , \nonumber \\
\beta_1 &= (M_R + m_p)^2 (W + M_R - 2 m_p) (p_0 + m_p) \, \, \, , \nonumber \\
\beta_2 &= (M_R + m_p)^2 (W - M_R + 2 m_p) (p_0 - m_p) \, \, \, .
\end{align}

We will next explain the method followed in Ref.~\cite{glmbg} for fixing the coupling constants $g_{\pi N R}$. Close to a resonance pole, the dominant contribution to the cross section follows the relativistic Breit-Wigner 
distribution. The shape of this distribution is determined by the form of the propagator, which contains the denominator $s-M_R^2 + i \sqrt{s} \, \Gamma(s)$; $\Gamma(s)$ will be identified here as the energy-dependent partial 
width for the decay mode $R \rightarrow \pi N$.

The $N(1440)$ creates a pole in the P11 partial wave, see the second of Eqs.~(\ref{eq:EQ54}). Within our unitarisation prescription \cite{glmbg}, the scattering amplitudes $f_{l\pm}^I$ are related to the corresponding 
$K$-matrix elements via the expression
\begin{equation} \label{eq:EQ57}
f_{l\pm}^I = \frac{K_{l\pm}^I}{1-i \lvert \vec{q} \, \rvert K_{l\pm}^I} \, \, \, .
\end{equation}
Restricting ourselves to the P11 partial wave and inserting $K_{1-}^{1/2}$ from the second of Eqs.~(\ref{eq:EQ54}) into Eq.~(\ref{eq:EQ57}) (also including a factor of $3$ from the isospin structure), we obtain a complex 
denominator, the imaginary part of which may be directly associated with the partial decay width of the resonance (in fact, with the term $\sqrt{s} \, \Gamma(s)$). If one defines the (constant) partial decay width of the 
resonance as the value of the energy-dependent partial decay width $\Gamma(s)$ at the resonance position ($s=M_R^2$), one derives the following expression.
\begin{equation*}
\Gamma = \frac{3 g_{\pi N R}^2 \lvert \vec{q}_R \rvert ^3 (M_R+m_p)^2}{16 \pi M_R m_p^2 (p_{0R} + m_p) }
\end{equation*}
Using the values of Table \ref{tab:Constants}, one obtains for $N(1440)$: $g_{\pi N R} \approx 4.8$.

The same procedure will be followed for fixing the coupling constants for all the HBRs~\footnote{The same scheme could have been used to fix $g_{\pi N \Delta}$. The application of the method in the case of the $\Delta(1232)$ 
leads to the relation:
\begin{equation*}
\Gamma = \frac{g_{\pi N \Delta}^2 \lvert \vec{q}_\Delta \rvert ^3 (p_{0\Delta} + m_p)}{48 \pi m_\Delta m_p^2} \, \, \, ,
\end{equation*}
which yields $g_{\pi N \Delta}=29.28 \pm 0.38$; the $g_{\pi N \Delta}$ value of Table \ref{tab:ModelParameters} ($29.81 \pm 0.27$) is in good agreement with this result.\label{ftn:FTN1}}:
\begin{itemize}
\item identification of the $K$-matrix element containing the specific pole,
\item application of the unitarisation prescription (\ref{eq:EQ57}) to that element,
\item extraction of the (energy-dependent) partial decay width,
\item evaluation of the partial decay width at the resonance position ($s=M_R^2$), and
\item solution of the resulting equation with respect to $g_{\pi N R}$.
\end{itemize}

\subsubsection{\label{sec:S11}$N(1535)$ and $N(1650)$}

There are two neighbouring S11 ($I(J^P) = \frac{1}{2}(\frac{1}{2}^-)$) states in the mass range considered here for the contributions of the HBRs: $N(1535)$ and $N(1650)$. The interaction Lagrangian density for a pure vector 
coupling~\footnote{In principle, a scalar coupling could also be considered. However, given the smallness of the effects which the HBRs induce at low energies, the vector coupling suffices for our purposes.} is of the form:
\begin{equation*}
\Delta \mathscr{L}_{\pi N R} = - i \frac{g_{\pi N R}}{2 m_p} \bar{\Psi} \vec{\tau} \cdot \gamma^\mu \partial_\mu \vec{\pi} \psi \, \, \, ,
\end{equation*}
where $\Psi$ stands for the spinor field of the HBR.

The contributions of the $s$-channel graph to the invariant amplitudes read as:
\begin{equation*}
A = - \frac{g_{\pi N R}^2}{4 m_p^2} \frac{s - m_p^2}{s - M_R^2} (M_R - m_p)
\end{equation*}
and
\begin{equation*}
B = - \frac{g_{\pi N R}^2}{4 m_p^2} \frac{s + m_p^2 - 2 M_R m_p}{s - M_R^2} \, \, \, ,
\end{equation*}
whereas those of the $u$-channel graph are:
\begin{equation*}
A = - \frac{g_{\pi N R}^2}{4 m_p^2} \left( 1 + \frac{(M_R + m_p)^2}{u - M_R^2} \right) (M_R - m_p) 
\end{equation*}
and
\begin{equation*}
B = \frac{g_{\pi N R}^2}{4 m_p^2} \left( 1 + \frac{(M_R - m_p)^2}{u - M_R^2} \right) \, \, \, .
\end{equation*}

For the $s$-channel graph, one obtains
\begin{align*}
K_{0+} &= - \frac{g_{\pi N R}^2}{32 \pi W m_p^2} \frac{(W-m_p)^2}{W-M_R} (p_0 + m_p) \, \, \, , \nonumber \\
K_{1-} &= - \frac{g_{\pi N R}^2}{32 \pi W m_p^2} \frac{(W+m_p)^2}{W+M_R} (p_0 - m_p) \, \, \, .
\end{align*}
The contributions to all other partial waves vanish.

For the $u$-channel graph, one may use Eqs.~(\ref{eq:EQ55}) with
\begin{equation*}
\lambda=\frac{g_{\pi N R}^2}{32 \pi W m_p^2}
\end{equation*}
and
\begin{align} \label{eq:EQ58}
\alpha_1 &= (W - M_R) (p_0 + m_p) \, \, \, , \nonumber \\
\alpha_2 &= (W + M_R) (p_0 - m_p) \, \, \, , \nonumber \\
\beta_1 &= (M_R - m_p)^2 (W - M_R - 2 m_p) (p_0 + m_p) \, \, \, , \nonumber \\
\beta_2 &= (M_R - m_p)^2 (W + M_R + 2 m_p) (p_0 - m_p) \, \, \, .
\end{align}
Evidently, the quantities $\alpha_1$, $\alpha_2$, $\beta_1$, and $\beta_2$ of Eqs.~(\ref{eq:EQ58}) may be obtained from Eqs.~(\ref{eq:EQ56}) via the substitution $M_R \rightarrow -M_R$, which is a speedy way to determine 
the contributions to the $K$-matrix elements of an intermediate $s$ baryon state, once those of the corresponding $p$ state are known, and vice versa.

From the $s$-channel $K_{0+}^{1/2}$, one obtains the following relation between the partial decay widths and the coupling constants $g_{\pi N R}$.
\begin{equation*}
\Gamma = \frac{3 g_{\pi N R}^2 \lvert \vec{q}_R \rvert (M_R-m_p)^2 (p_{0R} + m_p)}{16 \pi M_R m_p^2 }
\end{equation*}
Using the entries of Table \ref{tab:Constants}, one obtains similar $g_{\pi N R}$ values for the $N(1535)$ and $N(1650)$, between $2.1$ to $2.2$.

\subsubsection{\label{sec:P13}$N(1720)$}

Being an $I(J^P) = \frac{1}{2}(\frac{3}{2}^+)$ state, the contributions of the graph with an $N(1720)$ intermediate state to the $K$-matrix elements and to the invariant amplitudes are similar to those of Subsection 
\ref{sec:Delta} for the $\Delta(1232)$ graphs. The only difference relates to the isospin decomposition of the $K$-matrix elements, which follows that of the nucleon ($\alpha=1$ and $\beta=-1$ for the $s$-channel graph, 
$\alpha=1$ and $\beta=1$ for the $u$-channel graph). The contributions are (again) split into pole and non-pole parts. To obtain the expressions for the graphs with an $N(1720)$ intermediate state from those of Subsection 
\ref{sec:Delta}, one must first substitute $m_\Delta$ with $M_R$. The pole expressions (Subsection \ref{sec:PoleDelta}) should then be multiplied by $3$ (typical for the transition from $I=\frac{3}{2}$ to $I=\frac{1}{2}$ 
states), the isoscalar part of the non-pole contributions (first part of Subsection \ref{sec:NonPoleDelta}) by $\frac{3}{2}$, and the isovector part of the non-pole contributions (second part of Subsection \ref{sec:NonPoleDelta}) 
by $-3$ \cite{mr4}. To avoid misunderstanding, the affected expressions (apart from the substitution of $m_\Delta$ with $M_R$) are: (\ref{eq:EQ35}), (\ref{eq:EQ36}), (\ref{eq:EQ39}), (\ref{eq:EQ40}), (\ref{eq:EQ41}), 
(\ref{eq:EQ43}), (\ref{eq:EQ44}), (\ref{eq:EQ45}), (\ref{eq:EQ46}), (\ref{eq:EQ47}), (\ref{eq:EQ48}), and (\ref{eq:EQ49}); the remaining equations of Subsection \ref{sec:Delta} are applicable as they stand.

From the $s$-channel $K_{1+}^{1/2}$, one obtains the following relation between the partial decay width and the coupling constant $g_{\pi N R}$.
\begin{equation*}
\Gamma = \frac{g_{\pi N R}^2 \lvert \vec{q}_R \rvert ^3 (p_{0R} + m_p)}{16 \pi M_R m_p^2 }
\end{equation*}
Using the corresponding entries of Table \ref{tab:Constants}, one obtains for $N(1720)$: $g_{\pi N R} \approx 2.2$.

\subsubsection{\label{sec:S31}$\Delta(1620)$}

The $\Delta(1620)$ is an S31 ($I(J^P) = \frac{3}{2}(\frac{1}{2}^-)$) state. Therefore, the relevant expressions may be obtained from those given for the graphs with an S11 intermediate state in Subsection \ref{sec:S11}, 
after the division of the S11 $K$-matrix elements by $3$ and the appropriate inclusion of the isospin structure.

For the $s$-channel graph, one finally obtains
\begin{align*}
K_{0+} &= - \frac{g_{\pi N R}^2}{96 \pi W m_p^2} \frac{(W-m_p)^2}{W-M_R} (p_0 + m_p) \, \, \, , \nonumber \\
K_{1-} &= - \frac{g_{\pi N R}^2}{96 \pi W m_p^2} \frac{(W+m_p)^2}{W+M_R} (p_0 - m_p) \, \, \, .
\end{align*}
The contributions to all other partial waves vanish.

For the $u$-channel graph, one may use Eqs.~(\ref{eq:EQ55}) with
\begin{equation*}
\lambda=\frac{g_{\pi N R}^2}{96 \pi W m_p^2} \, \, \, ,
\end{equation*}
along with the quantities $\alpha_1$, $\alpha_2$, $\beta_1$, and $\beta_2$ as defined in Eqs.~(\ref{eq:EQ58}).

From the $s$-channel $K_{0+}^{3/2}$, one obtains the following relation between the partial decay width and the coupling constant $g_{\pi N R}$.
\begin{equation*}
\Gamma = \frac{g_{\pi N R}^2 \lvert \vec{q}_R \rvert (M_R-m_p)^2 (p_{0R} + m_p)}{16 \pi M_R m_p^2 }
\end{equation*}
Using the corresponding entries of Table \ref{tab:Constants}, one obtains for $\Delta(1620)$: $g_{\pi N R} \approx 2.2$.

\subsubsection{\label{sec:P31}$\Delta(1910)$}

The $\Delta(1910)$ is a P31 ($I(J^P) = \frac{3}{2}(\frac{1}{2}^+)$) state. Therefore, the relevant expressions may be obtained from the formulae given for the graph with a P11 intermediate state in Subsection \ref{sec:P11}, 
after the division of the P11-related $K$-matrix elements by $3$ and the appropriate inclusion of the isospin structure.

For the $s$-channel graph, one finally obtains
\begin{align*}
K_{0+} &= - \frac{g_{\pi N R}^2}{96 \pi W m_p^2} \frac{(W-m_p)^2}{W+M_R} (p_0 + m_p) \, \, \, , \nonumber \\
K_{1-} &= - \frac{g_{\pi N R}^2}{96 \pi W m_p^2} \frac{(W+m_p)^2}{W-M_R} (p_0 - m_p) \, \, \, .
\end{align*}
The contributions to all other partial waves vanish.

For the $u$-channel graph, one may use Eqs.~(\ref{eq:EQ55}) with
\begin{equation*}
\lambda=\frac{g_{\pi N R}^2}{96 \pi W m_p^2} \, \, \, ,
\end{equation*}
along with the quantities $\alpha_1$, $\alpha_2$, $\beta_1$, and $\beta_2$ as defined in Eqs.~(\ref{eq:EQ56}).

From the $s$-channel $K_{1-}^{3/2}$, one obtains the following relation between the partial decay width and the coupling constant $g_{\pi N R}$.
\begin{equation*}
\Gamma = \frac{g_{\pi N R}^2 \lvert \vec{q}_R \rvert ^3 (M_R+m_p)^2}{16 \pi M_R m_p^2 (p_{0R} + m_p)}
\end{equation*}
Using the corresponding entries of Table \ref{tab:Constants}, one obtains for $\Delta(1910)$: $g_{\pi N R} \approx 2.0$.

\section{\label{sec:Results}Additional considerations}

In Section \ref{sec:ModelAmplitudes}, we gave the detailed contributions of the graphs of the model in the partial waves for $l \leq 3$. Within the context of the model, the $s$- and $p$-wave results may be thought of as complete, 
in that they contain all the well-established pole contributions in the mass range up to $2$ GeV. On the contrary, given that the corresponding $d$ and $f$ HBRs have not been included in the model, the $d$ and $f$ waves obtained 
so far must consequently be considered as incomplete.

According to the PDG compilation \cite{pdg}, the well-established $N$-type $d$ and $f$ HBRs below $2$ GeV are: D13$(1520)$, D15$(1675)$, and F15$(1680)$. Additionally, the $\Delta$-type $d$ and $f$ HBRs are: D33$(1700)$, F35$(1905)$, 
and F37$(1950)$. The first state in each category (i.e., the D13$(1520)$ and the D33$(1700)$) could be included in the model without much effort, after applying the formalism of Subsection \ref{sec:Delta} for the propagation of a 
spin-$\frac{3}{2}$ particle and standard transformations between the $p$ and $d$ states. The first report on the treatment of the propagation of spin-$\frac{5}{2}$ particles has recently appeared \cite{slm}. Given the complexity of the 
subject, the complete development of the formalism and the extraction of amplitudes which can be directly used in our scheme are bound to take some time. At present, it is also not clear how one could use the experience gained from the 
$\Delta(1232)$ graphs (i.e., concerning the preferred value of the parameter $Z$) and simplify the vertex factors accordingly in the case of spin-$\frac{5}{2}$ graphs. Regarding the propagation of massive spin-$\frac{7}{2}$ states, we 
are not aware of any attempt aiming at their treatment. Of course, as long as the aforementioned $d$ and $f$ states are not included in the model, the only possibility left to us to reliably account also for the $d$ and $f$ partial 
waves is to fix them from an external source; we have chosen to use the current solution of the SAID analysis \cite{abws} for $l=2$ and $l=3$. The inclusion in the model of the $d$ and $f$ HBRs is worth the effort, as it will remove the 
(tiny) dependence of our results on extraneous material, thus making our analysis self-contained.

As each of the main graphs of the model, detailed in Subsections \ref{sec:Sigma}-\ref{sec:Delta}, introduces two parameters, the model contains eight parameters in total. When a fit to the data is performed treating all these parameters 
as free, it turns out that $G_\sigma$, $G_\rho$, $\kappa_\rho$, and $x$ are strongly correlated; as a result, it is not possible to determine the values of all these quantities. In accordance with $\pi N$ models developed within an EFT framework, 
we fix $x$ at $0$. The fits of the model to the data are thus performed (since Ref.~\cite{m}) on the basis of the variation of seven parameters: $G_\sigma$, $\kappa_\sigma$, $G_\rho$, $\kappa_\rho$, $g_{\pi N N}$, $g_{\pi N \Delta}$, and $Z$. Future 
fits might involve fewer free parameters. For instance, the coupling constant $g_{\pi N \Delta}$ could be fixed from the decay width of the $\Delta(1232)$ (see footnote \ref{ftn:FTN1}) and/or $Z$ could be fixed at $-\frac{1}{2}$. Other 
options include the use of a formalism in which no spin-$\frac{1}{2}$ contributions are present in the spin-$\frac{3}{2}$ field.

We will assume that the physical quantities appearing in the present section (i.e., the fit parameters, the $s$-wave scattering lengths and $p$-wave scattering volumes, the $\pi N$ phase shifts, etc.) are not purely-hadronic as they still 
contain residual em effects. These effects relate in particular to the use of the physical masses (instead of the unknown hadronic ones) of the proton, of the neutron, and of the charged and neutral pion (in the hadronic part of the interaction) 
in the determination of the em corrections. Unfortunately, it is not possible at the present time to assess the importance of these residual contributions. As a result, we must retain the cautious attitude of considering all hadronic quantities 
of the present work `em-modified'~\footnote{In previous works, we emphasised that we are dealing with em-modified quantities in a framework of formal isospin invariance by using the symbol `\~{}' over the hadronic quantities. In fact, 
there is no such need; all hadronic quantities relating to or obtained from any analysis of the experimental data (be they parameters of hadronic models, predictions derived on their basis, extracted amplitudes, etc.) are unavoidably affected. 
Given the presence of these residual contributions, there is no purely-hadronic quantity in the present work (as well as in any other PWA of the $\pi N$ data).}. However, as the repetitive use of this term is tedious, we will omit it.

The results of our previous PWA of low-energy $\pi^\pm p$ elastic-scattering data may be found in Ref.~\cite{mr1}; that paper also contains the details of our procedure regarding the identification of the outliers in the database. Recent 
modifications in our analysis software and database structure enabled the inclusion in the database of the present work also of the MEIER04 AP measurements (Ref.~[31] in Ref.~\cite{mr1}), comprising 28 data points; as a result, the truncated 
combined $\pi^\pm p$ elastic-scattering database of the present work contains the 668 data points of the final database of Ref.~\cite{mr1} and the 28 MEIER04 measurements (none of which turned out to be an outlier), hence a total of 696 data 
points. The confidence level $\mathrm{p}_{min}$, used in the statistical tests, was set to the equivalent of a $2.5 \sigma$ effect in the normal distribution \cite{mr1}. Compared to Ref.~\cite{mr1}, there is one additional, more important 
change. The quantity $m_\sigma$ (associated with the graphs of Subsection \ref{sec:Sigma}), which in Ref.~\cite{mr1} had been fixed at $860$ MeV \cite{torn}, is varied herein in the interval which is currently recommended by the PDG \cite{pdg}. 
Fits were performed at seven $m_\sigma$ values, from $400$ to $550$ MeV with an increment of $25$ MeV. The corresponding variation in the final $\chi^2$ values in these fits did not exceed about $0.4$, the minimal $\chi^2$ value of $909.9$ 
corresponding to the low $m_\sigma$ end~\footnote{When fixing $m_\sigma$ at $860$ MeV \cite{torn}, the $\chi^2$ value of the fit to the same data is equal to $910.9$.}; as earlier mentioned, the sensitivity of our analysis in the physical 
region to the choice of $m_\sigma$ is very low. On the other hand, when shifting $m_\sigma$ from $860$ MeV to the range of values investigated herein, a change of about one standard deviation was seen in the $\Sigma$ value, where the 
$\sigma$-exchange contribution is dominant. In the Monte-Carlo simulation, yielding the various predictions (for the $\pi N$ phase shifts and the LECs, as well as for the $\pi N$ $\Sigma$ term), the results of each of these seven fits for the 
model parameters (fitted values and uncertainties), as well as the corresponding Hessian matrices, are taken into account (with equal weight). As a result, all our predictions also contain now the effects of the variation of $m_\sigma$. The 
average fitted values of the model parameters, as well as their uncertainties, corresponding to the variation of $m_\sigma$ as described above, are contained in Table \ref{tab:ModelParameters}.

In Subsection \ref{sec:PR}, we will discuss the energy dependence of the various contributions to the $s$, $p$, $d$, and $f$ $K$-matrix elements for $T \leq 100$ MeV. We will also discuss the various contributions at threshold. Subsection 
\ref{sec:SigmaTerm} will be dedicated to the $\pi N$ $\Sigma$ term.

\subsection{\label{sec:PR}Results in the physical region}

\subsubsection{\label{sec:PhaseShifts}$\pi N$ phase shifts}

The relation between the $\pi N$ (hadronic) partial-wave amplitudes $f_{l\pm}^I$ and the corresponding $K$-matrix elements $K_{l\pm}^I$ is given by Eq.~(\ref{eq:EQ57}); $K_{l\pm}^I$ denotes the sum of the contributions (within each specific 
partial wave) detailed in Subsections \ref{sec:Sigma}-\ref{sec:HR}. Assuming no inelasticity, the $\pi N$ phase shifts $\delta_{l\pm}^I$ are associated with the $\pi N$ partial-wave amplitudes via the relation
\begin{equation*}
f_{l\pm}^I = \frac{\exp(2 i \delta_{l\pm}^I) - 1}{2 i \lvert \vec{q} \, \rvert} = \frac{\sin \delta_{l\pm}^I \, \cos \delta_{l\pm}^I}{\lvert \vec{q} \, \rvert} + i \frac{\sin^2 \delta_{l\pm}^I}{\lvert \vec{q} \, \rvert} \, \, \, .
\end{equation*}
On the other hand, Eq.~(\ref{eq:EQ57}) may be rewritten as
\begin{equation*}
f_{l\pm}^I = \frac{K_{l\pm}^I \, (1 + i \lvert \vec{q} \, \rvert K_{l\pm}^I)}{1 + \vec{q} \, ^2 (K_{l\pm}^I)^2} \, \, \, .
\end{equation*}
The direct comparison of the last two expressions leads to the relation
\begin{equation*}
\lvert \vec{q} \, \rvert K_{l\pm}^I = \frac{\Im [ f_{l\pm}^I ]}{\Re [ f_{l\pm}^I ]} = \tan \delta_{l\pm}^I \, \, \, .
\end{equation*}
In our PWAs of the low-energy $\pi^\pm p$ scattering data, this equation yields $\delta_{l\pm}^I$ from $K_{l\pm}^I$ in the $s$ and $p$ waves. Our current phase-shift solution is given in Table \ref{tab:HPS}. The $\pi N$ phase shifts 
$\delta_{l\pm}^I$ are subsequently corrected for em effects and lead to the observables, following the long chain of equations given in Section 2 of Ref.~\cite{mworg}. Regarding the em corrections in our PWAs, we make use of the values 
of Refs.~\cite{gmorw1,gmorw2}, whereas the (small) inelasticity corrections (which had not been re-assessed in Refs.~\cite{gmorw1,gmorw2}) are taken from the NORDITA work \cite{two1,two2,two3}; the same inelasticity corrections had been 
employed when determining the em corrections in Refs.~\cite{gmorw1,gmorw2} from modern (meson-factory) low-energy $\pi^\pm p$ elastic-scattering data.

In the physical region, the unitarisation prescription introduced by Eq.~(\ref{eq:EQ57}) creates complex partial-waves amplitudes $f_{l\pm}^I$ from real $K$-matrix elements $K_{l\pm}^I$, the latter being directly 
linked to the graphs of the model (or, in general, being suitably parameterised). In the unphysical region, $\lvert \vec{q} \, \rvert$ is imaginary; as a result, the partial-waves amplitudes $f_{l\pm}^I$ become real.

\subsubsection{\label{sec:ED}Energy dependence of the partial-wave $K$-matrix elements}

\begin{figure}
\begin{center}
\includegraphics [width=15.5cm] {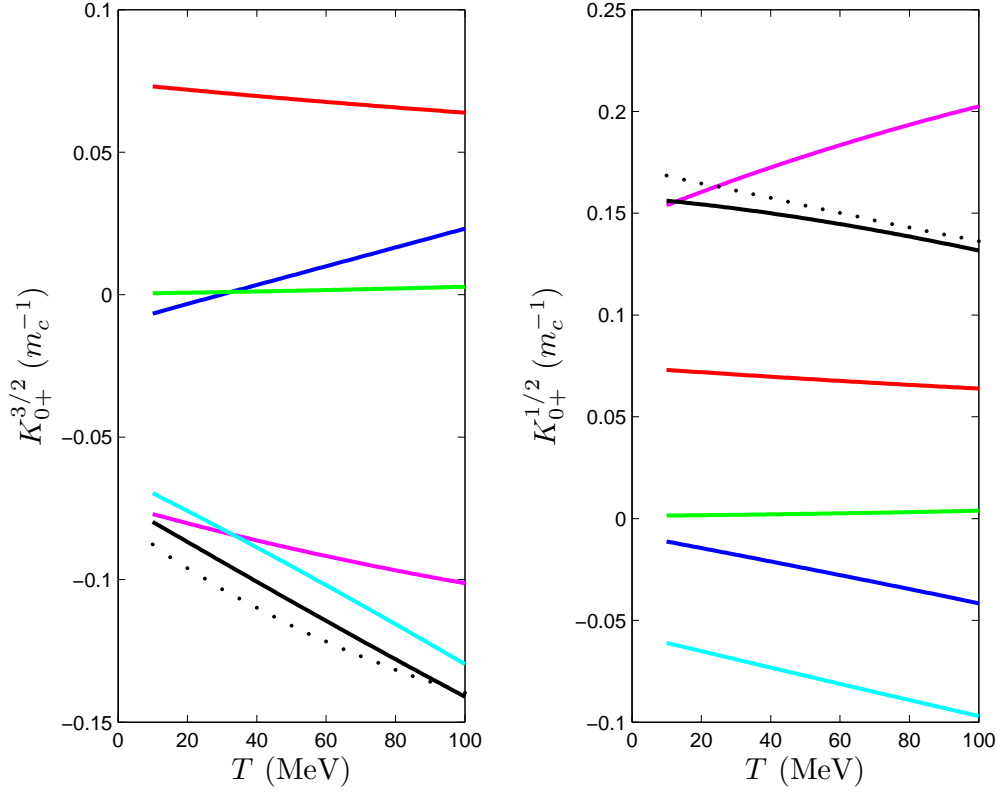}
\caption{\label{fig:sWaves}The contributions of the graphs of the ETH model to the $s$-wave ($l=0$) $K$-matrix elements in the low-energy region. $T$ denotes the pion laboratory kinetic energy. The $t$-channel $\sigma$-exchange 
contributions (Subsection \ref{sec:Sigma}) are shown in red; the $t$-channel $\rho$-exchange contributions (Subsection \ref{sec:Rho}) in magenta; the $s$- and $u$-channel contributions of the $N$ graphs (Subsection \ref{sec:Nucleon}) 
in dark blue; the $s$- and $u$-channel contributions of the $\Delta(1232)$ graphs (Subsection \ref{sec:Delta}) in cyan; finally, the $s$- and $u$-channel contributions from the graphs with the well-established $s$ and $p$ higher baryon 
resonances with masses below $2$ GeV as intermediate states (Subsection \ref{sec:HR}) in green; the black curves correspond to the sum of these contributions. These results have been obtained with a pure pseudovector coupling ($x=0$) 
in the graphs of Subsection \ref{sec:Nucleon} (and in those of Subsection \ref{sec:P11}); regarding the contributions of the $\sigma$-exchange graphs of Subsection \ref{sec:Sigma}, $m_\sigma$ was fixed at $475$ MeV, the central value 
of the recommended $m_\sigma$ range by the PDG \cite{pdg}. The single points represent the current solution of the SAID analysis \cite{abws} sampled with a step of $5$ MeV.}
\end{center}
\end{figure}

\begin{figure}
\begin{center}
\includegraphics [width=15.5cm] {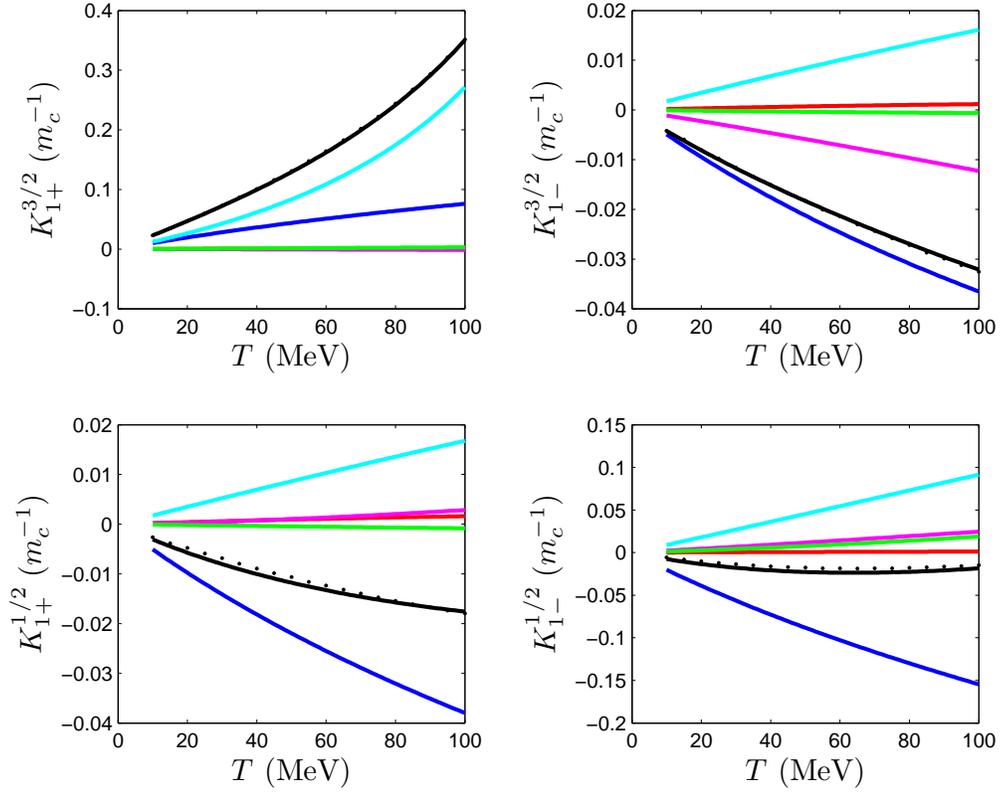}
\caption{\label{fig:pWaves}The contributions of the graphs of the ETH model to the $p$-wave ($l=1$) $K$-matrix elements in the low-energy region. $T$ denotes the pion laboratory kinetic energy. The colours of the curves are 
explained in the caption of Fig.~\ref{fig:sWaves}.}
\end{center}
\end{figure}

\begin{figure}
\begin{center}
\includegraphics [width=15.5cm] {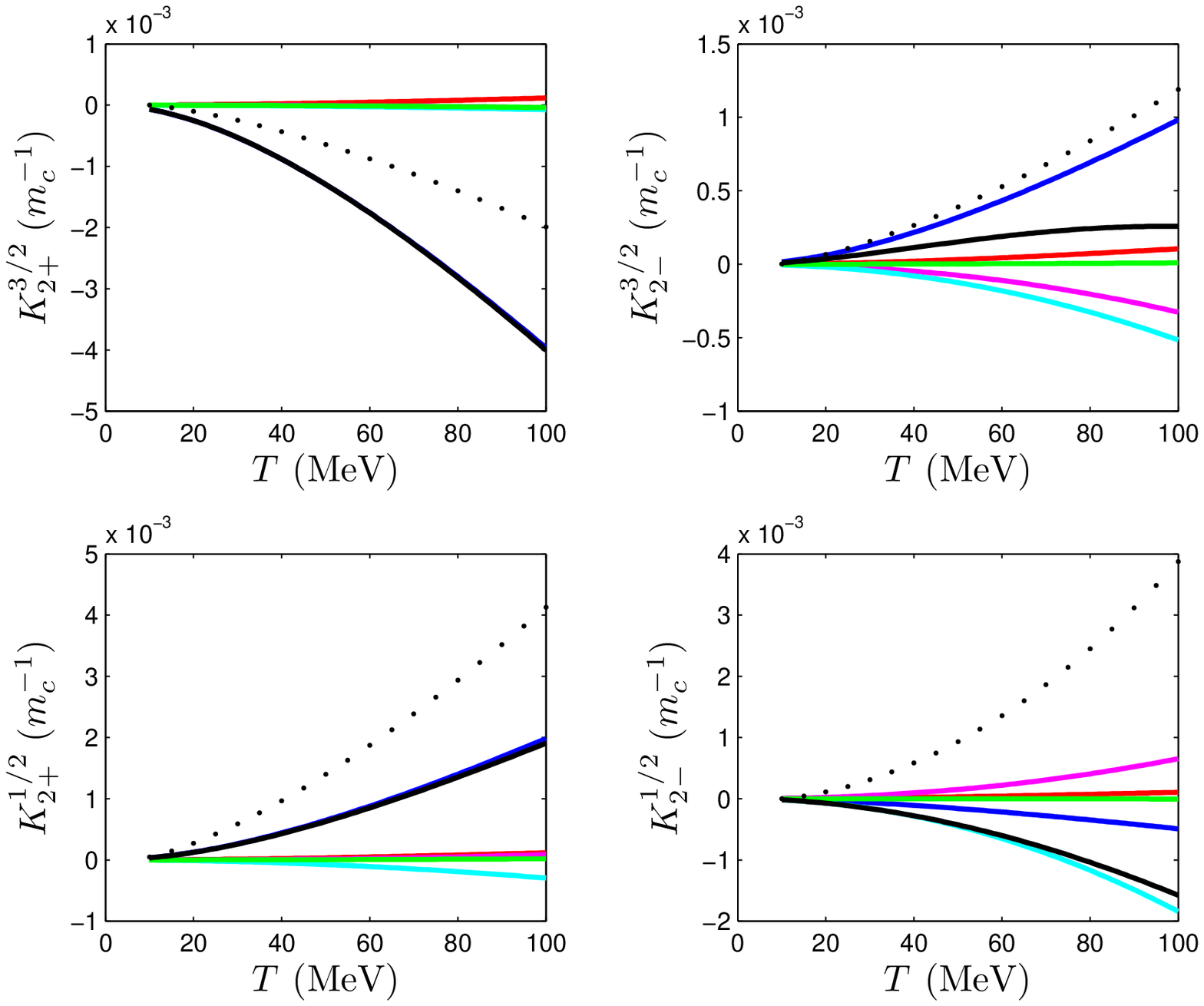}
\caption{\label{fig:dWaves}The contributions of the graphs of the ETH model to the $d$-wave ($l=2$) $K$-matrix elements in the low-energy region. $T$ denotes the pion laboratory kinetic energy. The colours of the curves are 
explained in the caption of Fig.~\ref{fig:sWaves}. To suppress artefacts which are due to the truncation of small values, simple polynomials have been fitted to the $d$-wave phase shifts of the current solution of the 
SAID analysis \cite{abws}.}
\end{center}
\end{figure}

\begin{figure}
\begin{center}
\includegraphics [width=15.5cm] {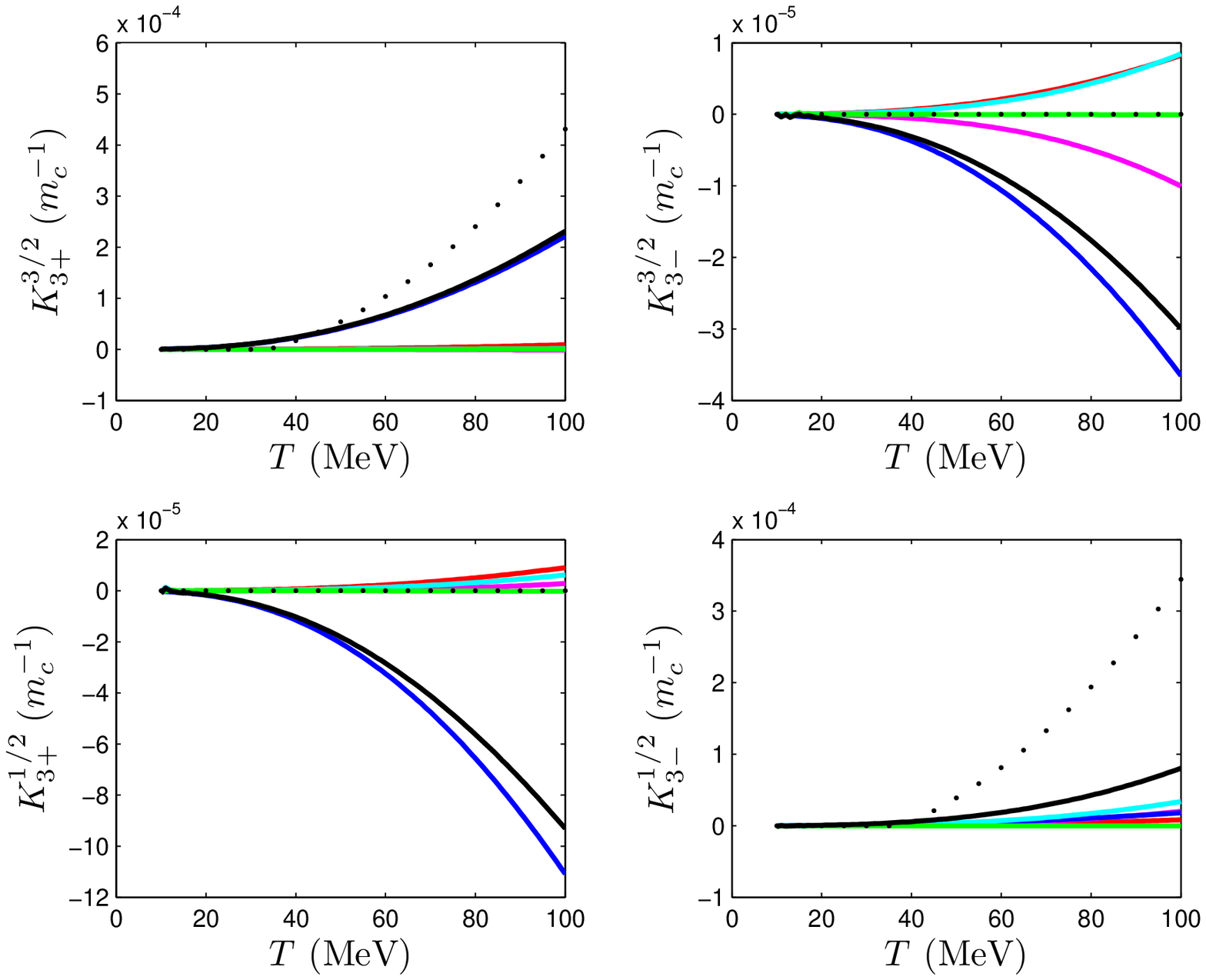}
\caption{\label{fig:fWaves}The contributions of the graphs of the ETH model to the $f$-wave ($l=3$) $K$-matrix elements in the low-energy region. $T$ denotes the pion laboratory kinetic energy. The colours of the curves are 
explained in the caption of Fig.~\ref{fig:sWaves}. To suppress artefacts which are due to the truncation of small values, simple polynomials have been fitted to the $f$-wave phase shifts of the current solution of the 
SAID analysis \cite{abws}.}
\end{center}
\end{figure}

The energy dependence of the $K$-matrix elements $K_{l\pm}^I$ of the model for $T \leq 100$ MeV is shown in Figs.~\ref{fig:sWaves}-\ref{fig:fWaves}, along with the current solution (as of June 12, 2013) of the SAID analysis \cite{abws}; 
we have already commented on the differences in the $s$-wave part of the $\pi N$ interaction between these two solutions \cite{mworg,mr1}. Regarding the model amplitudes, the expressions of Subsections \ref{sec:Sigma}-\ref{sec:HR} have been 
used, along with the optimal model-parameter vector corresponding to $m_\sigma=475$ MeV (the central value of the interval which is currently recommended by the PDG \cite{pdg}); the fitted uncertainties have not been used. In most cases, we 
observe near cancellations of large contributions, which may explain the largeness of the correlations among the model parameters observed during the optimisation phase \cite{mworg,mr1,mr3}.

Shown in Tables \ref{tab:SLII} and \ref{tab:SLSI} are the contributions of the graphs of the model to the $\pi N$ $s$-wave scattering lengths and $p$-wave scattering volumes in two standard formats (isoscalar-isovector and spin-isospin); 
in spin-isospin format, these quantities are defined by the equations:
\begin{equation*}
a_{0+}^{I} = \lim_{\lvert \vec{q} \, \rvert \to 0} K_{0+}^{I} \, \, \, , a_{1\pm}^{I} = \lim_{\lvert \vec{q} \, \rvert \to 0} \frac{K_{1\pm}^{I}}{\vec{q} \, ^2} \, \, \, .
\end{equation*}
The isoscalar $b_0$ (also denoted as $a^+_{0+}$) and isovector $b_1$ $s$-wave scattering lengths are expressed in terms of $a_{0+}^{3/2}$ and $a_{0+}^{1/2}$ via the relations:
\begin{equation} \label{eq:EQ59}
b_0 \equiv a^+_{0+} = \frac{2 a_{0+}^{3/2} + a_{0+}^{1/2}}{3} \, \, \, , b_1 = \frac{a_{0+}^{3/2} - a_{0+}^{1/2}}{3} \, \, \, .
\end{equation}
The relations between the two forms of the $p$-wave scattering volumes read as:
\begin{align*}
c_0 &= +\frac{4}{3}\:a_{1+}^{3/2}+\frac{2}{3}\:a_{1-}^{3/2}+\frac{2}{3}\:a_{1+}^{1/2}+\frac{1}{3}\:a_{1-}^{1/2} \, \, \, , \\
c_1 &= +\frac{2}{3}\:a_{1+}^{3/2}+\frac{1}{3}\:a_{1-}^{3/2}-\frac{2}{3}\:a_{1+}^{1/2}-\frac{1}{3}\:a_{1-}^{1/2} \, \, \, , \\
d_0 &= -\frac{2}{3}\:a_{1+}^{3/2}+\frac{2}{3}\:a_{1-}^{3/2}-\frac{1}{3}\:a_{1+}^{1/2}+\frac{1}{3}\:a_{1-}^{1/2} \, \, \, , \\
d_1 &= -\frac{1}{3}\:a_{1+}^{3/2}+\frac{1}{3}\:a_{1-}^{3/2}+\frac{1}{3}\:a_{1+}^{1/2}-\frac{1}{3}\:a_{1-}^{1/2} \, \, \, . \\
\end{align*}

Three remarks may be made upon inspection of Tables \ref{tab:SLII} and \ref{tab:SLSI}.
\begin{itemize}
\item The isovector $s$-wave scattering length $b_1$ is accounted for, almost entirely, by the $\rho$-exchange graph.
\item Two large contributions to the isoscalar $s$-wave scattering length $b_0$, namely those of the $\sigma$-exchange and $\Delta(1232)$ graphs, nearly cancel one another; the resulting $b_0$ value almost vanishes.
\item The smallness of the contributions of the $s$ and $p$ HBRs in the entirety of the low-energy region is noticeable. Even in the case of $K_{1-}^{1/2}$, which contains the nearest (to the upper bound of the energy considered in our PWAs) 
of these states (i.e., the $N(1440)$), the effects are not significant. However, this remark must not be interpreted as discouragement for including in the model the missing $d$ and $f$ HBRs. On one hand, self-consistency dictates that the 
model contain the contributions of all the well-established states (up to $l=3$) with masses below $2$ GeV, not only those of the $s$ and $p$ HBRs. Additionally, the inclusion of the $d$ and $f$ HBRs will enable us to use the model $d$ and 
$f$ waves in our analyses.
\end{itemize}

In our PWAs of the low-energy $\pi^\pm p$ scattering data in Refs.~\cite{mworg,mr1,mr2,mr3,fm}, simple parameterisations of the $K$-matrix elements, which are devoid of theoretical constraints (e.g., those imposed by crossing symmetry 
which the model amplitudes obey), had been used alongside the model amplitudes. The success of the chosen forms in accounting for the experimental information may be understood on the basis of Figs.~\ref{fig:sWaves} and \ref{fig:pWaves}. 
The $K$-matrix elements $K_{0+}^{3/2}$ and $K_{0+}^{1/2}$ of the model are almost linear functions of $T$ in the low-energy region; as a result, the quadratic forms chosen in the simple parameterisations of these $K$-matrix elements more 
than suffice. A similar remark applies to the $p$-wave $K$-matrix elements.

Regarding our result for the $s$-wave scattering length $a_{0+}^{3/2}$, our values have been stable since we first dealt with the low-energy $\pi^+ p$ data of the modern experiments \cite{fm}. Irrespective of the analysis method (i.e., of the use 
of the model or of simple parameterisations of the $K$-matrix elements, of the choice of the minimisation function, etc.), the extracted $a_{0+}^{3/2}$ values have always exceeded about $-0.077 \, m_c^{-1}$. We have extensively commented 
\cite{mworg,mr1,mr3} on the mismatch between our prediction for the $\pi^- p$ elastic-scattering length
\begin{equation*}
a^{cc}=\frac{a_{0+}^{3/2}+2a_{0+}^{1/2}}{3}=0.0809 \pm 0.0012 \, m_c^{-1}
\end{equation*}
and the result obtained directly at threshold from the strong shift of the $1s$ level in pionic hydrogen (the application of the corrections of Ref.~\cite{orwmg} to the experimental result of Ref.~\cite{ss} leads to 
$a^{cc}=0.0859 \pm 0.0006 \, m_c^{-1}$); this discrepancy, which amounts to an effect at the level of $3.8 \sigma$ in the normal distribution, is currently not understood.

An approach for extracting the $s$-wave scattering lengths from experimental information obtained at threshold, also including the strong shift of the $1s$ level ($\epsilon_{1s}$) in pionic deuterium (which is related to the isoscalar $s$-wave 
scattering length $b_0$), appeared in Ref.~\cite{bhhknp}. We will first compare the corrected values for the $\pi^- p$ elastic-scattering length $a^{cc}$, obtained in Refs.~\cite{orwmg} and \cite{bhhknp}. In Ref.~\cite{orwmg}, the input value 
for $\epsilon_{1s}$ in pionic hydrogen was equal to $-7.116 \pm 0.013$ eV, matching well the subsequent result $-7.120 \pm 0.012$ eV, which Ref.~\cite{bhhknp} used~\footnote{The final results of the Pionic-Hydrogen Collaboration are expected in 
the near future \cite{gott}.}. The corrected $a^{cc}$ value of Ref.~\cite{orwmg} $0.0859 \pm 0.0006 \, m_c^{-1}$ agrees very well with the result $0.0861 \pm 0.0009 \, m_c^{-1}$, extracted in Ref.~\cite{bhhknp} (our quantity $a^{cc}$ is denoted 
therein as $a_{\pi^- p}$), see the second of their Eqs.~(17). Considering the differences in the methodology between the two approaches, this agreement is very satisfactory.

The further comparison between the results of Ref.~\cite{bhhknp} and the material of the present work is difficult. A meaningful comparison between the results of any two approaches rests on the similarity (better, compatibility) of the input 
and the removal of the (same) unwanted contributions from the important physical quantities. As mentioned earlier, our $s$-wave scattering lengths are expected to contain additional em effects, which have not been removed by the em corrections 
of Refs.~\cite{gmorw1,gmorw2}; owing to the fact that the physical masses of the interacting particles (instead of the unknown hadronic ones) have been used in Ref.~\cite{bhhknp}, the same remark also applies in their case. However, one 
additional problem is lurking. There is no guarantee that the em effects, removed in our analysis of the scattering data and in Ref.~\cite{bhhknp} (at threshold), are `matching'; it is certainly re-assuring that the two corrected $a^{cc}$ 
results \cite{orwmg,bhhknp} agree, yet additional tests are needed in order to establish the similarity of the removed em contributions for $T \leq 100$ MeV and for all three available reactions at low energies (i.e., for the two elastic-scattering 
processes and for the CX reaction). As our em corrections (both for the scattering data and at threshold) have long been available \cite{gmorw1,gmorw2,orwmg}, it may be easier for the authors of Ref.~\cite{bhhknp} to 
investigate the compatibility of the results obtained in the two schemes. As we have already pointed out \cite{mr3}, a consistent scheme for removing reliably the em effects at all energies is needed.

One additional point needs to be stressed. Isospin invariance is fulfilled by the ETH model; no isospin-breaking graphs (e.g., involving $\rho-\omega$ or $\eta-\pi^0$ mixing) have ever been included in it. If isospin-breaking effects are present 
in the experimental data, they disguise themselves in our approach as changes in the fitted values of the model parameters. As the partial-wave amplitudes of the model are functions of the parameter vector, they come out different in the fits 
involving different combinations of the three available reactions at low energies, failing to fulfill the triangle identity. To a large extent, our approach is data-driven, in that it is left to the input data to decide on the possibility and 
on the level of the isospin-breaking effects. Assuming that the model comprises a firm basis for the analysis of the low-energy $\pi N$ data, that the missing em corrections are not sizeable, and that there are no major problems with the 
absolute normalisation of the bulk of the modern low-energy $\pi N$ database, any significant discrepancies in our results can only be attributed to the violation of the isospin invariance in the hadronic interaction. Evidently, our $s$-wave 
scattering lengths contain all isospin-breaking contributions which are removed in Ref.~\cite{bhhknp}. As a result, it is not clear how we could possibly compare further our results with theirs.

A few $\pi N$ models \cite{la,pt,mo}, of variable similarity (and increased complexity compared) to ours, have surfaced since Ref.~\cite{glmbg} appeared. Earlier attempts to account for the $\pi N$ data on the basis of hadronic models had been 
cited in Ref.~\cite{glmbg}; comments on those earlier attempts may be found in Subsection 6.4 therein.
\begin{itemize}
\item Lahiff and Afnan \cite{la} presented a description of the $\pi N$ phase shifts in terms of solutions of the Bethe-Salpeter (BS) equation directly in four dimensions. The potentials, which the authors used in their BS equation, were 
derived from $t$-channel graphs with $\sigma$ and $\rho$ exchanges, as well as $s$- and $u$-channel contributions with $N$ and $\Delta(1232)$ intermediate states; their hadronic model uses only the derivative $\pi \sigma$ coupling and does not 
include any HBRs effects. Regarding the $\Delta(1232)$ graphs, two approaches have been followed in Ref.~\cite{la}: the Rarita-Schwinger formalism and Pascalutsa's method \cite{pasca1}. In their work, the authors start from bare vertices and 
propagators, which become dressed as their potential is iterated in the BS equation. To tackle convergence issues, the authors introduce cut-off functions, associated with each vertex which their model contains. The paper provides helpful 
and interesting insight into the contributions of the graphs of their model to the $\pi N$ phase shifts (e.g., see their Fig.~6 and the corresponding text). Among the interesting conclusions of that work is the remark that the phase shift 
$\delta^{1/2}_{1-}$ (P11) is better reproduced in the Rarita-Schwinger formalism for the $\pi N \Delta$ interaction. Furthermore, similarly to ours, their analysis favours the solution $Z=-\frac{1}{2}$; their corresponding $Z$ value (in 
the paper, they make use of the parameter $x_\Delta \equiv -(Z+\frac{1}{2})$) is around $-0.4$ (see their Table II). Finally, as far as the coupling constant $g_{\pi N \Delta}$ is concerned, their result with the Rarita-Schwinger formalism 
is similar to ours, whereas the value obtained with Pascalutsa's method is (perhaps, owing to the use of only the derivative $\pi \sigma$ coupling in Ref.~\cite{la}) unreasonably large.
\item Pascalutsa and Tjon \cite{pt} developed a relativistic, covariant, and unitary model along the general lines of Ref.~\cite{la}, also taking account of the P11$(1440)$, S11$(1535)$, and D13$(1520)$ contributions. In their model, the 
spin-$\frac{3}{2}$ fields have been treated in a variety of ways, also including the Rarita-Schwinger formalism. The authors finally investigated the reproduction of the $\pi N$ phase shifts up to $T=600$ MeV (see their Fig.~8). At present, 
it is not clear to us why the results of their Table II for $g_{\pi N \Delta}$, columns labelled as $N \rho \Delta$ (WT) and $N \rho \Delta$ (VMD), differ so drastically from ours.
\item Mei\ss ner and Oller \cite{mo} developed a chiral, unitary, relativistic approach for the general description of the meson-baryon interaction, and applied it to $\pi N$ elastic scattering. Their model is built on the basis of the 
tree-level contributions of the lowest-order meson-baryon Lagrangian obtained within the framework of Chiral-Perturbation Theory (ChPT), onto which the effects of the $\Delta(1232)$ and $N(1440)$ graphs, as well as those pertaining to 
the meson resonances in the $t$ channel, are added. Similarly to us, the authors follow the Rarita-Schwinger formalism in the description of the $\pi N \Delta$ interaction. Owing to their method, featuring subtracted dispersion relations, 
the authors do not need to introduce any form factors and cut-off functions in their approach. Their $\pi N$ scattering amplitude is subsequently matched to the one obtained within the framework of the Heavy-Baryon ChPT (HBChPT) at third 
order, close to (and slightly below) threshold, where the HBChPT amplitude is expected to converge. The authors finally fit their model to $\pi N$ phase shifts, which they successfully reproduce (see their Fig.~9), at least up to $T=150$ 
MeV. Exempting the isoscalar $s$-wave scattering length $b_0$ (the $b_0$ results in their three fits are considerably larger than the value obtained in the present work, see our Table \ref{tab:SLII}), the reproduction of the LECs of the 
$\pi N$ system with their model appears to be reasonable. The result for the coupling constant $g_{\pi N \Delta}$ of Ref.~\cite{mo} (see their Table 1) is compatible with our value, whereas their values for the parameter $Z$ (ranging between 
$-0.16$ and $-0.05$) are significantly larger than our result of Table \ref{tab:ModelParameters}.
\end{itemize}
The common characteristic of Refs.~\cite{la,pt,mo} is that they fit their models to $\pi N$ phase shifts, rather than to genuine $\pi N$ measurements. In all three works, the authors chose two phase-shift solutions as their `input data': their 
first solution was taken to be one of the `standard' Karlsruhe-Helsinki (KH) or Karlsruhe (KA) analyses of the 1980s, the other was one of the `popular' phase-shift solutions of the SAID group during the 1990s (SM95). It would be interesting to 
investigate the changes in the main results of Refs.~\cite{la,pt,mo}, if they used as input the phase-shift solution of the present work (Table \ref{tab:HPS}), which has been obtained on the exclusive basis of low-energy information and its 
uncertainties reflect directly those of the experimental data.

As explained earlier, the $d$ and $f$ waves are fixed in our PWAs from the current solution of the SAID analysis \cite{abws}. For the time being, we are compelled to do so, given that the contributions of the HBRs in six (out of eight) of these 
partial waves are not contained in the model. The most striking difference in the $d$ waves between the $K$-matrix elements of the model and the current solution of the SAID analysis occurs in D13 ($K_{2-}^{1/2}$); the two contributions are of 
opposite sign. The D33 ($K_{2-}^{3/2}$) of the model remains small (compared to the current solution of the SAID analysis), whereas the ratios of the values of the two solutions in D35 ($K_{2+}^{3/2}$) and D15 ($K_{2+}^{1/2}$) involve a factor 
of about $2$. It remains to be seen how the inclusion of the well-established $d$ HBRs with masses below $2$ GeV (i.e., of D13$(1520)$, D15$(1675)$, and D33$(1700)$) affects the $d$-wave $K$-matrix elements. There is general mismatch also in the 
$f$ waves, albeit of lesser significance given the size of these contributions. To conclude, the use in a PWA of the model $d$ and $f$ waves, as they currently stand, is not recommended; the fixation of the corresponding $\pi N$ phase shifts 
from an external source is thus mandatory.

\subsection{\label{sec:SigmaTerm}The $\pi N$ $\Sigma$ term}

Almost all theoretical (and, surprisingly, several experimental) papers on the $\pi N$ system have been written in a way which communicates to the reader the message that the extrapolation of the hadronic part of the scattering amplitude into 
the unphysical region (to obtain an estimate of the $\pi N$ $\Sigma$ term) is the principal motivation for conducting experiments in the physical one. This fact alone emphasises the importance of the $\Sigma$ term in QCD tests. In the first 
part of the present subsection, we will describe why this quantity is of interest in Hadronic Physics. In the second part, we will obtain a new prediction for the $\Sigma$ term.

\subsubsection{\label{sec:SigmaTermHistory}History}

The introduction of the $\Sigma$ term dates back to the early 1970s, when Cheng and Dashen \cite{cd} were set on investigating whether the SU(2)$\bigotimes$SU(2) or the SU(3) symmetry is better obeyed by the strong interaction. The $\Sigma$ term 
was then proposed as a measure of goodness of the former symmetry. To obtain a reliable estimate of this quantity, Cheng and Dashen recommended the use of fixed-$t$ dispersion relations in the extrapolation of the real part of the isoscalar 
amplitude $\bar{D}^+ (s,\xi) = \bar{A}^+ (s,\xi) + \nu \bar{B}^+ (s,\xi)$ into the unphysical region~\footnote{Regarding the developments in the methodology of schemes involving dispersion relations, an approach featuring the Roy-Steiner 
equations for $\pi N$ scattering was proposed in the recent past as the means for the reliable extrapolation of the $\pi N$ scattering amplitude into the unphysical region \cite{dhkm}.}; the bar above the amplitudes indicates the removal of the 
nucleon Born-term contribution (see end of Subsection \ref{sec:Nucleon}). The $\Sigma$ term is defined via the relation
\begin{equation} \label{eq:EQ60}
\Sigma = F_\pi^2 \Re [ \bar{D}^+ (s_{CD},\xi_{CD}) ] \, \, \, ,
\end{equation}
where $s_{CD}=m_p^2$ and $\xi_{CD}=-\frac{m_c^2}{4 m_p^2-m_c^2}$. The quantity $F_\pi$ is the pion-decay constant (see Table \ref{tab:Constants}).

We denote the masses of the $u$- and of the $d$-quark by $m_u$ and $m_d$, respectively. The scalar form factor $\sigma(t)$ is then defined as the matrix element of the $u$- and $d$-quark QCD Hamiltonian mass term between two proton states 
with $4$-momenta $p$ and $p^\prime$ \cite{gls1}:
\begin{equation*}
\bar{u}(p^\prime) \sigma(t) u(p) = \frac{m_u + m_d}{4 m_p} < p^\prime | \bar{u} u + \bar{d} d | p > \, \, \, ,
\end{equation*}
where $t=(p^\prime-p)^2$.

The $\pi N$ $\sigma$ term is defined as
\begin{equation*}
\sigma(0) = \frac{m_u + m_d}{4 m_p} < p | \bar{u} u + \bar{d} d | p > \, \, \, .
\end{equation*}
Defining furthermore
\begin{equation*}
\hat{\sigma}=\frac{m_u + m_d}{4 m_p} < p | \bar{u} u + \bar{d} d - 2 \bar{s} s | p >
\end{equation*}
and the strange-quark content of the proton as
\begin{equation*}
y = \frac{2 < p | \bar{s} s | p >}{< p | \bar{u} u + \bar{d} d | p >} \, \, \, ,
\end{equation*}
one obtains:
\begin{equation*}
\sigma(0) = \frac{\hat{\sigma}}{1-y} \, \, \, .
\end{equation*}
The quantity $\hat{\sigma}$ may be found in the literature also as $\sigma_0$, not to be confused with $\sigma(0)$.

The extraction of a value for the parameter $y$ relies on the evaluation of $\sigma(0)$ and $\hat{\sigma}$.
\begin{itemize}
\item Estimates of the quantity $\sigma(0)$ may be obtained from the $\pi N$ amplitude (via the $\Sigma$ value) or theoretically (e.g., from Lattice-QCD calculations). We will return to the first option shortly. As far as the 
second option is concerned, a list of relevant works, as well as the results of an analysis of available $\sigma(0)$ estimates, may be found in Ref.~\cite{bali}.
\item The quantity $\hat{\sigma}$ was first evaluated in the early 1980s, within the framework of the ChPT: $\hat{\sigma} = 35 \pm 5$ MeV \cite{g,gl}. A subsequent value, obtained in the context of the HBChPT \cite{bor}, corroborated the 
$\hat{\sigma}$ result of Refs.~\cite{g,gl}. However, the authors of a recent paper \cite{agco} reckon that the analyses of Refs.~\cite{g,gl,bor} were afflicted by large systematic effects, pertaining to relativistic corrections and to 
the omission of the decuplet resonances, and that more reliable results may be obtained after employing Lorentz covariant Baryon ChPT with the explicit inclusion of the $\Delta(1232)$ field; their $\hat{\sigma}$ result turned out to be 
substantially larger than the values of Refs.~\cite{g,gl,bor}, namely in the vicinity of $60$ MeV.
\end{itemize}

The association between the quantities $\Sigma$, obtained from the invariant amplitude $\bar{D}^+$ at the CD point according to Eq.~(\ref{eq:EQ60}), and $\sigma(0)$ has been the subject of considerable research.
\begin{itemize}
\item The $\Sigma$ term is first related to $\sigma (2 m_c^2)$ via the expression
\begin{equation*}
\Sigma = \sigma(2 m_c^2) + \Delta_R \, \, \, ,
\end{equation*}
where the remainder $\Delta_R$ was first treated in Ref.~\cite{gss} and found to be small (around $0.35$ MeV).
\item The next step in obtaining $\sigma(0)$ is the evaluation of the difference
\begin{equation*}
\Delta_\sigma = \sigma (2 m_c^2) - \sigma(0) \, \, \, . 
\end{equation*}
The correction $\Delta_\sigma$ was first treated in Ref.~\cite{gls1} and found to be very important ($15.2 \pm 0.4$ MeV).
\end{itemize}

A milestone in the history of the extraction of the $\Sigma$ term from $\pi N$ data was Koch's result in the early 1980s \cite{k}; employing hyperbolic dispersion relations on the $(\nu^2,t)$ plane, Koch obtained $\Sigma = 64 \pm 8$ 
MeV. At this point, two remarks are due. First, at the time when Ref.~\cite{k} appeared, the quantities $\Sigma$ and $\sigma(0)$ were not distinguished; the correction $\Delta_\sigma$ (as well as the much smaller $\Delta_R$) had 
not yet been considered. Second, the only low-energy $\pi N$ data which were available to Ref.~\cite{k} were proven in the 1990s to be inconsistent with the modern measurements comprising (most of) today's database. In any case, 
the result of Ref.~\cite{k} indicated a large strange-quark content of the proton and stimulated interest in this subject, which culminated in the introduction and evaluation of two sizeable corrections. Apart from the determination 
of $\Delta_\sigma$, corrections were also applied to the original $\Sigma$ value of Ref.~\cite{k}: using updated values of the $\pi N$ LECs, Ref.~\cite{glls} suggested a downward correction by about $8$ MeV.

The modern measurements indicate an enhanced (more positive) isoscalar component in the $\pi N$ scattering amplitude at low energies; this becomes evident after comparing Koch's isoscalar $s$-wave scattering length 
$b_0=-0.008 \, m_c^{-1}$ \cite{kk} with the result of Table \ref{tab:SLII}. Using most of today's database, Ref.~\cite{pasw} obtained $\Sigma = 79 \pm 7$ MeV.

We conclude the present subsection with one additional remark. It is not clear which the appropriate definition of the $\Sigma$ term is in a framework where the hadronic part of the $\pi N$ interaction violates isospin invariance 
\cite{mr3,glk,m}. In all probability, the effects induced by this effect are sizeable; for instance, it is long known that the position of the CD point on the $t$ axis is strongly affected by the nucleon-mass splitting (i.e., by the 
proton-neutron mass difference, see the last part of Subsection 8.1.1 of Ref.~\cite{h}, p.~562). We are not aware of any works addressing this subject.

\subsubsection{\label{sec:SigmaTermFormula}Model prediction}

Regarding the extraction of the $\Sigma$ value from the data, one may come up with two advantages of the model over schemes employing dispersion relations.
\begin{itemize}
\item The graphs of the model lead to uniquely defined $K$-matrix amplitudes at all $(\nu,t)$ points, including those in the unphysical region \cite{glmbg}. It thus appears that the extrapolation of the model amplitudes into the 
unphysical region would be straightforward. We will shortly explain why this is not entirely true.
\item The determination of the $\Sigma$ value involves the use of low-energy $\pi N$ data exclusively; the data at $T > 100$ MeV do not influence our results. The importance of this remark is revealed after recollecting that 
the energy dependence of the partial-wave amplitudes in analyses using dispersion relations is (almost entirely) determined from the high-energy data. We have already commented on the mismatch at low energies between the amplitudes 
obtained from the entire $\pi N$ database and those extracted exclusively on the basis of low-energy $\pi N$ data \cite{mworg,mr1}, in particular in the $s$ waves.
\end{itemize}

We have investigated the possibility of setting forth a $\Sigma$-evaluation scheme, also incorporating our unitarisation prescription of Eq.~(\ref{eq:EQ57}). However, a number of problems render this goal hard to achieve. To start 
with, the implementation must inevitably involve the entirety of each partial-wave amplitude. Unfortunately, as the nucleon Born-term contribution is removed from the hadronic part of the scattering amplitude, prior to its extrapolation 
into the unphysical region, the prescription of Eq.~(\ref{eq:EQ57}) is inapplicable. We have not yet found a way to circumvent this problem.

To obtain a prediction for the $\Sigma$ term within the context of our model, we are thus bound to follow one of the next two options: a) make use of the tree-level approximation of Ref.~\cite{glmbg} or b) employ a scheme featuring 
the determination of the $\Sigma$ value (predominantly) from $\pi N$ LECs. In the remaining part of the present subsection, we will investigate these two options.

In Ref.~\cite{glmbg}, a simple formula for the $\Sigma$ term had been obtained from the model amplitudes, as they have been detailed in Section \ref{sec:ModelAmplitudes}; that formula reads as:
\begin{equation} \label{eq:EQ62}
\Sigma = F_\pi^2 \left( \frac{2 G_\sigma m_\sigma^2 m_c}{m_\sigma^2 - 2 m_c^2} - \frac{g_{\pi N N}^2}{m_p} \frac{x^2}{(1+x)^2} \right) + \delta_R \, \, \, .
\end{equation}
It is easy to identify the two main contributions in Eq.~(\ref{eq:EQ62}): the first term within the brackets is the $\sigma$-exchange contribution to $A^+$ at the CD point, see Eq.~(\ref{eq:EQ29}), whereas the second term is the 
remainder in $D^+_N$ of Eq.~(\ref{eq:EQ33}) after the removal of the nucleon Born-term contribution; of course, as we currently use a pure pseudovector $\pi N$ coupling, the contribution of the $N$ graphs vanishes. Within our model, 
the remainder $\delta_R$ comprises contributions (mostly) of the $\Delta(1232)$ graphs; the contribution to $\delta_R$ of the $N(1720)$ graphs is about $600$ times smaller, whereas all other contributions from the graphs treated herein 
vanish. Using the relations of Subsections \ref{sec:PoleDelta} and \ref{sec:NonPoleDelta}, one can prove that the $\Delta(1232)$ contribution to the isoscalar amplitude $D^+$ at the CD point is independent of the parameter $Z$ and 
reads as:
\begin{equation*}
D^+_\Delta (s_{CD},\xi_{CD}) = \frac{g_{\pi N \Delta}^2 (2 m_\Delta + m_p) m_c^4}{18 m_p^2 m_\Delta^2 (m_\Delta^2 - m_p^2)} \, \, \, .
\end{equation*}
Summing up the contributions of the $\Delta(1232)$ and $N(1720)$ graphs, we finally obtain:
\begin{equation*}
\delta_R = 0.637 \pm 0.012 \, {\rm MeV} \, \, \, .
\end{equation*}
Using the results of our fits to the low-energy $\pi^\pm p$ elastic-scattering data (as explained in the beginning of Section \ref{sec:Results}), we obtain from Eq.~(\ref{eq:EQ62}):
\begin{equation} \label{eq:EQ64}
\Sigma = 72.4 \pm 3.1 \, {\rm MeV} \, \, \, .
\end{equation}

More than one decade ago, Olsson set forth a pioneering scheme for the evaluation of the $\Sigma$ term, resting on the knowledge of a few LECs of the $\pi N$ system \cite{ols}. In his method, the $\Sigma$ value may be obtained via 
the formula:
\begin{align} \label{eq:EQ65}
\Sigma = \frac{F_\pi^2}{m_c} \Bigg[ 4 \pi \Bigg( & (1+r)^2 \Big( m_c a^+_{0+} - \frac{1.047 m_c^2}{3(1+2r)} \big( 2 (a^{3/2}_{0+})^2 + (a^{1/2}_{0+})^2 \big) - \frac{1+2r}{3} m_c^3 C^+ \Big) \nonumber \\
 & + \frac{r (6 r^2 + 13 r + 6)}{1+2r} m_c^3 a^+_{1+} - \frac{r^2}{1+2r} m_c^3 a^+_{1-} \Bigg) + \delta \Bigg] \, \, \, ,
\end{align}
where $r=\frac{m_c}{2 m_p}$, $a^+_{l\pm} = \frac{1}{3} (2 a^{3/2}_{l\pm}+a^{1/2}_{l\pm})$, and $C^+=\frac{1}{3} (2 C^{3/2}_{0+}+C^{1/2}_{0+})$; the quantities $C^{3/2}_{0+}$ and $C^{1/2}_{0+}$ are associated with the effective 
ranges (coefficients of $\vec{q}^{\, 2}$ in the expansion of the real part of the two $s$-wave amplitudes around threshold~\footnote{For the definition of $C^+$, see Eq.~(3.7) and footnote 4 of Ref.~\cite{ols}.}). In Eq.~(\ref{eq:EQ65}), 
the input $s$-wave scattering lengths $a^+_{0+}$, $a^{3/2}_{0+}$, and $a^{1/2}_{0+}$ are assumed expressed in units of $m_c^{-1}$, whereas the $p$-wave scattering volumes $a^+_{1+}$ and $a^+_{1-}$, as well as the isoscalar 
effective range $C^+$, in $m_c^{-3}$. The value of $1.047$ is the result of a numerical integration \cite{ols}. Finally, $\delta$ is given by:
\begin{equation*}
\delta = 2 \, r \, g_{\pi N N}^2 \, \left( \frac{r}{1-r^2} \right)^2 - I_1 - I_2 \, \, \, ,
\end{equation*}
where the dispersion integral $I_1=0.21 \pm 0.02$ was evaluated in Ref.~\cite{ols} from SAID results and $I_2 = 0.02 \pm 0.02$ represents the contributions of the partial waves with $l>1$. Our predictions for the relevant $s$-wave 
scattering lengths and $p$-wave scattering volumes are given in Tables \ref{tab:SLII} and \ref{tab:SLSI}; our corresponding result for the isoscalar effective range $C^+$ is $-0.1092 \pm 0.0044 \, m_c^{-3}$. The prediction 
for the $\Sigma$ term, using Olsson's method and Eq.~(\ref{eq:EQ65}), is: $\Sigma = 70.4 \pm 2.5 {\rm (stat.)} \pm 1.7 {\rm (syst.)}$ MeV, i.e., a value which matches well our result of Eq.~(\ref{eq:EQ64}). The statistical uncertainties 
correspond to the results of our analysis of the low-energy $\pi^\pm p$ elastic-scattering data, whereas the systematic ones pertain to extraneous material (i.e., to the uncertainties of $I_1$ and $I_2$); the partial uncertainties have 
been combined in quadrature, to yield the quoted total uncertainties.

Regarding the main result of Ref.~\cite{ols}, a few comments are due. Therein, the extracted value of $\sigma (2 m_c^2)$ was $71 \pm 9$ MeV. However, we have found a number of inconsistencies in that paper. To start with, the input 
values of the $s$-wave scattering lengths, given in Eqs.~(4.9)-(4.11) therein, are inconsistent; the expected relation between $a^+_{0+}$, $a^{3/2}_{0+}$, and $a^{1/2}_{0+}$, see the first of our Eqs.~(\ref{eq:EQ59}), is not obeyed 
by these values. Furthermore, the origin of these values is unclear. Equally problematic is that Olsson's formula (4.7) does not contain a factor $1+2r$ in the denominators of the second and third terms within the square brackets; 
assuming the validity of Eqs.~(3.11) and (3.12) of Ref.~\cite{ols}, our Eq.~(\ref{eq:EQ65}) should be the correct expression. Unfortunately, there are indications that also Eq.~(3.12) might not be correct: the numerical value of 
$1.036$, appearing in that equation, does not represent the ratio $\frac{(1+r)^2}{1+2r} \approx 1.00482$. Our efforts notwithstanding, it has not been possible to clarify any of these issues with the author of Ref.~\cite{ols}. As a 
result, we must emphasise that the validity of our Eq.~(\ref{eq:EQ65}) rests on the correctness of a number of relations appearing in Ref.~\cite{ols}. Given the importance of the $\Sigma$ term, the re-investigation of this subject, 
in the light of our findings, would be welcome.

The reduction of the uncertainty in the estimates of the present work, compared to those of Ref.~\cite{ols}, is noticeable. The large uncertainty in Ref.~\cite{ols} originates from the treatment of the input uncertainties; 
Ref.~\cite{ols} had no other choice than to treat them as independent. Of course, these uncertainties are not independent in our scheme. The predictions for the $\pi N$ LECs involve a Monte-Carlo generation, in which the fitted 
model-parameter values and their uncertainties, as well as the Hessian matrices of the optimisation, are used as input. Therefore, the model predictions (i.e., for the low-energy observables, for the $s$-wave scattering lengths 
and $p$-wave scattering volumes, etc.) are interconnected.

It is interesting to mention a recent result for $\sigma(0)$, obtained within the framework of the ChPT from our phase-shift solution. Using our 2006 phase-shift solution \cite{mworg}, Ref.~\cite{aco1} obtained $\sigma(0)=59 \pm 2$ 
MeV. Responding promptly to our request, the authors applied their method \cite{aco1,aco2} to the phase-shift solution of the present work (Table \ref{tab:HPS}) and obtained for $\sigma(0)$ the value of $61.3 \pm 2.1$ MeV 
\cite{aco3}. The truncation of the chiral expansion in the method of Refs.~\cite{aco1,aco2} introduces an additional (systematic) uncertainty of about $7$ MeV \cite{aco3}; therefore, the ChPT result, using our current phase-shift 
solution, should rather read as: $\sigma(0)=61 \pm 2 {\rm (stat.)} \pm 7 {\rm (syst.)}$ MeV.

\section{\label{sec:Conclusions}Discussion and conclusions}

The ETH model was put forth in the early 1990s to account for the hadronic part of the pion-nucleon ($\pi N$) interaction at low energies. The model contains $t$-channel $\sigma$ and $\rho$ exchanges, as well as the $s$- and 
$u$-channel contributions with the well-established $s$ and $p$ baryon states with masses below $2$ GeV. The model amplitudes obey crossing symmetry and isospin invariance. In the past, this model was used in partial-wave analyses 
of the low-energy (pion laboratory kinetic energy $T \leq 100$ MeV) $\pi N$ data, aiming at: a) investigating the consistency and the reproduction of the available experimental information, b) extracting the values of low-energy 
constants (LECs) of the $\pi N$ system, and c) testing the isospin invariance in the $\pi N$ system.

One of the main goals in the present work was to list all the analytical expressions for the model contributions to the $K$-matrix elements up to (and including) the $f$ waves. The contributions of (only) the main Feynman graphs 
of the model to (only) the $s$- and $p$-wave $K$-matrix elements had been listed in Ref.~\cite{glmbg}. The publication of these amplitudes is expected to facilitate the use of the model in other works. To make our analysis self-contained 
(independent of extraneous information in the physical region), it is needed to include in the model the $s$- and $u$-channel contributions with the $d$ and $f$ higher baryon resonances as intermediate states; there are six such 
states, two of which could be easily included. The development of the theoretical background for the treatment of the four remaining fields is pending. The hope is that the present work will serve as motivation to advance further 
the treatment of the propagation of massive spin-$\frac{5}{2}$ (and, perhaps, spin-$\frac{7}{2}$) particles.

Our results now contain also the effects of the variation of the $\sigma$-meson mass $m_\sigma$ in the interval which is recommended by the Particle-Data Group; the current range of the $m_\sigma$ values is between $400$ and $550$ 
MeV \cite{pdg}. Our approach was modified on principle, not because of necessity; up to now, the sensitivity of our analysis to the choice of the $m_\sigma$ value has been very low. In the present paper, we applied the methodology 
of Ref.~\cite{mr1} to an enhanced database, comprising the truncated combined $\pi^\pm p$ elastic-scattering database of Ref.~\cite{mr1} and $28$ analysing-power measurements which were added to the input for the first time (see 
the beginning of Section \ref{sec:Results}). We obtained new values for the model parameters (Table \ref{tab:ModelParameters}), for the $s$- and $p$-wave phase shifts (Table \ref{tab:HPS}), and for the $s$-wave scattering lengths 
and $p$-wave scattering volumes (Tables \ref{tab:SLII} and \ref{tab:SLSI}).

There is little doubt that the bulk of the modern (meson-factory) low-energy $\pi^\pm p$ elastic-scattering data favours an enhanced (more positive) isoscalar component in the $\pi N$ dynamics at low energies, thus leading to 
results for the $\pi N$ $\Sigma$ term which exceed the canonical value of Ref.~\cite{k}. Our result for the $\Sigma$ term, obtained within the tree-level approximation of Ref.~\cite{glmbg}, is $72.4 \pm 3.1$ MeV, see Eq.~(\ref{eq:EQ64}).

In Ref.~\cite{ols}, Olsson set forth a pioneering method for the evaluation of the $\Sigma$ term, resting on the knowledge of a few LECs of the $\pi N$ system. In Subsection \ref{sec:SigmaTermFormula}, we attempted to correct some 
inconsistencies which Olsson's paper contains. Assuming the validity of a number of relations appearing in that work (which we can hardly assert), we corrected Olsson's main formula (4.7) and used the amended expression, i.e., our 
Eq.~(\ref{eq:EQ65}), along with updated information on the relevant LECs, to evaluate $\Sigma$ using Olsson's scheme; our final result was found to be in good agreement with the value obtained within the tree-level approximation of 
Ref.~\cite{glmbg}.

\begin{ack}
We would like to thank the two reviewers of the present paper for their careful reading and constructive criticism. This research programme has been shaped to its current form following the long-term interaction with our colleagues 
B.L. Birbrair, A. Gashi, P.F.A. Goudsmit, A.B. Gridnev, H.J. Leisi, G.C. Oades(deceased), and W.S. Woolcock(deceased); we are grateful to them for their contributions, suggestions, and comments. An additional attempt to single out 
individual contributions is bound to lead to a long list; we thus refrain from citing additional names and refer the interested reader to the acknowledgments in earlier papers. The Feynman graphs of the present document have been 
drawn with the software package JaxoDraw \cite{JaxoDraw}, available from http://jaxodraw.sourceforge.net/.
\end{ack}

\newpage
\begin{table}[h!]
{\bf \caption{\label{tab:Constants}}}The current values of the physical constants, used in the hadronic part of the $\pi N$ scattering amplitude obtained within the context of the ETH model; these values have been taken from the most recent 
compilation of the Particle-Data Group \cite{pdg}. Regarding the well-established $s$ and $p$ higher baryon resonances (HBRs), $M_R$ and $\Gamma_T$ denote the Breit-Wigner mass and total decay width respectively, whereas $\eta$ is the 
branching ratio for the $\pi N$ decay mode.
\vspace{0.2cm}
\begin{center}
\begin{tabular}{|l|c|}
\hline
Physical quantity (unit) & Value \\
\hline
Pion-decay constant $F_\pi$ (MeV) & $92.214$ \\
Charged-pion mass $m_c$ (MeV) & $139.57018$ \\
$\rho(770)$ mass $m_\rho$ (MeV) & $775.49$ \\
Proton mass $m_p$ (MeV) & $938.272046$ \\
$\Delta(1232)$ mass $m_{\Delta}$ (MeV) & $1232$ \\
$\Delta(1232)$ decay width $\Gamma_{\Delta}$ (MeV) & $117$ \\
\hline
\multicolumn{2}{|c|}{$s$ and $p$ HBRs} \\
\hline
$N(1440)$ $M_R$ (MeV) & $1440$ \\
$N(1440)$ $\Gamma_T$ (MeV) & $300$ \\
$N(1440)$ $\eta$ & $0.650$ \\
\hline
$N(1535)$ $M_R$ (MeV) & $1535$ \\
$N(1535)$ $\Gamma_T$ (MeV) & $150$ \\
$N(1535)$ $\eta$ & $0.450$ \\
\hline
$N(1650)$ $M_R$ (MeV) & $1655$ \\
$N(1650)$ $\Gamma_T$ (MeV) & $150$ \\
$N(1650)$ $\eta$ & $0.700$ \\
\hline
$N(1720)$ $M_R$ (MeV) & $1720$ \\
$N(1720)$ $\Gamma_T$ (MeV) & $250$ \\
$N(1720)$ $\eta$ & $0.110$ \\
\hline
$\Delta(1620)$ $M_R$ (MeV) & $1630$ \\
$\Delta(1620)$ $\Gamma_T$ (MeV) & $140$ \\
$\Delta(1620)$ $\eta$ & $0.250$ \\
\hline
$\Delta(1910)$ $M_R$ (MeV) & $1890$ \\
$\Delta(1910)$ $\Gamma_T$ (MeV) & $280$ \\
$\Delta(1910)$ $\eta$ & $0.225$ \\
\hline
\end{tabular}
\end{center}
\end{table}

\vspace{0.5cm}
\begin{table}
{\bf \caption{\label{tab:ModelParameters}}}Average values of the seven parameters of the ETH model, obtained from fits to the truncated combined $\pi^\pm p$ elastic-scattering databases (see the beginning of Section \ref{sec:Results}). 
The results correspond to a pure pseudovector coupling ($x=0$) in the contributions of the graphs of Subsection \ref{sec:Nucleon} (and in those of the graphs of Subsection \ref{sec:P11}).
\vspace{0.2cm}
\begin{center}
\begin{tabular}{|l|c|}
\hline
Parameter & Fitted value \\
\hline
$G_\sigma(GeV^{-2})$ & $24.8 \pm 1.3$ \\
$\kappa_\sigma$ & $-0.111 \pm 0.066$ \\
$G_\rho(GeV^{-2})$ & $54.52 \pm 0.64$ \\
$\kappa_\rho$ & $0.57 \pm 0.44$ \\
$g_{\pi N N}$ & $12.81 \pm 0.13$ \\
$g_{\pi N \Delta}$ & $29.81 \pm 0.27$ \\
$Z$ & $-0.565 \pm 0.056$ \\
\hline
\end{tabular}
\end{center}
\end{table}

\newpage
\begin{table}
{\bf \caption{\label{tab:HPS}}}Our current solution for the $\pi N$ $s$- and $p$-wave phase shifts (in degrees), obtained from fits to the truncated combined $\pi^\pm p$ elastic-scattering databases (see the beginning of Section \ref{sec:Results}). 
$T$ denotes the pion laboratory kinetic energy.
\vspace{0.2cm}
\begin{center}
\begin{tabular}{|c|c|c|c|c|c|c|}
\hline
$T$ (MeV) & ${\delta}_{0+}^{3/2}$ (S31) & ${\delta}_{0+}^{1/2}$ (S11) & ${\delta}_{1+}^{3/2}$ (P33) & ${\delta}_{1-}^{3/2}$ (P31) & ${\delta}_{1+}^{1/2}$ (P13) & ${\delta}_{1-}^{1/2}$ (P11) \\ 
\hline
$20$ & $-2.358(36)$ & $4.192(28)$ & $1.278(10)$ & $-0.2224(45)$ & $-0.1572(38)$ & $-0.3654(76)$ \\
$25$ & $-2.755(38)$ & $4.675(30)$ & $1.816(14)$ & $-0.3065(63)$ & $-0.2134(53)$ & $-0.482(10)$ \\
$30$ & $-3.148(39)$ & $5.106(31)$ & $2.430(17)$ & $-0.3975(84)$ & $-0.2725(69)$ & $-0.597(13)$ \\
$35$ & $-3.540(40)$ & $5.494(32)$ & $3.120(20)$ & $-0.494(11)$ & $-0.3338(88)$ & $-0.708(16)$ \\
$40$ & $-3.935(40)$ & $5.849(33)$ & $3.890(22)$ & $-0.597(13)$ & $-0.397(11)$ & $-0.813(19)$ \\
$45$ & $-4.333(40)$ & $6.175(35)$ & $4.742(24)$ & $-0.704(16)$ & $-0.460(13)$ & $-0.910(23)$ \\
$50$ & $-4.735(40)$ & $6.474(37)$ & $5.681(26)$ & $-0.815(19)$ & $-0.525(15)$ & $-0.998(26)$ \\
$55$ & $-5.142(41)$ & $6.751(40)$ & $6.712(28)$ & $-0.930(22)$ & $-0.590(18)$ & $-1.076(30)$ \\
$60$ & $-5.553(41)$ & $7.007(43)$ & $7.841(29)$ & $-1.048(25)$ & $-0.655(20)$ & $-1.142(33)$ \\
$65$ & $-5.970(43)$ & $7.243(47)$ & $9.078(30)$ & $-1.170(29)$ & $-0.719(23)$ & $-1.196(37)$ \\
$70$ & $-6.393(45)$ & $7.461(52)$ & $10.429(32)$ & $-1.295(32)$ & $-0.783(26)$ & $-1.236(41)$ \\
$75$ & $-6.820(49)$ & $7.662(57)$ & $11.904(35)$ & $-1.422(37)$ & $-0.847(29)$ & $-1.264(46)$ \\
$80$ & $-7.252(54)$ & $7.847(62)$ & $13.515(41)$ & $-1.552(41)$ & $-0.910(32)$ & $-1.277(50)$ \\
$85$ & $-7.690(60)$ & $8.016(68)$ & $15.273(49)$ & $-1.685(45)$ & $-0.973(36)$ & $-1.275(55)$ \\
$90$ & $-8.132(67)$ & $8.171(74)$ & $17.190(61)$ & $-1.820(50)$ & $-1.034(39)$ & $-1.258(60)$ \\
$95$ & $-8.580(76)$ & $8.311(81)$ & $19.279(76)$ & $-1.958(55)$ & $-1.095(43)$ & $-1.226(65)$ \\
$100$ & $-9.031(85)$ & $8.438(88)$ & $21.555(94)$ & $-2.097(61)$ & $-1.154(47)$ & $-1.177(71)$ \\
\hline
\end{tabular}
\end{center}
\end{table}

\vspace{0.5cm}
\begin{table}
{\bf \caption{\label{tab:SLII}}}The various contributions of the graphs of the ETH model to the $s$-wave scattering lengths and $p$-wave scattering volumes (isoscalar-isovector format); the $s$-wave scattering lengths ($b_0$ and $b_1$) are 
given in units of $m_c^{-1}$, the $p$-wave scattering volumes (remaining quantities) in $m_c^{-3}$. The row marked as HBRs contains the sum of the contributions of the well-established $s$ and $p$ higher baryon resonances detailed in 
Subsection \ref{sec:HR}. Uncertainties are quoted only for the sum of the contributions.
\vspace{0.2cm}
\begin{center}
\begin{tabular}{|l|c|c|c|c|c|c|}
\hline
 & $b_0 \equiv a^+_{0+}$ & $b_1$ & $c_0$ & $c_1$ & $d_0$ & $d_1$ \\
\hline
$\sigma$ & $0.0742$ & $-$ & $0.0056$ & $-$ & $-0.0004$ & $-$ \\
$\rho$ & $-$ & $-0.07357$ & $-$ & $-0.0111$ & $-$ & $-0.00950$ \\
$N$ & $-0.0094$ & $-0.00070$ & $0.0016$ & $0.1468$ & $-0.1453$ & $-0.00006$ \\
$\Delta(1232)$ & $-0.0613$ & $-0.00221$ & $0.1910$ & $0.0384$ & $-0.0386$ & $-0.05238$ \\
HBRs & $0.0007$ & $-0.00033$ & $0.0064$ & $-0.0017$ & $0.0017$ & $-0.00528$ \\
\hline
Sum & $0.0041(13)$ & $-0.07681(64)$ & $0.2046(24)$ & $0.1724(19)$ & $-0.1826(20)$ & $-0.06721(85)$ \\
\hline
\end{tabular}
\end{center}
\end{table}

\vspace{0.5cm}
\begin{table}
{\bf \caption{\label{tab:SLSI}}}The various contributions of the graphs of the ETH model to the $s$-wave scattering lengths and $p$-wave scattering volumes (spin-isospin format); the $s$-wave scattering lengths ($a_{0+}^{3/2}$ and $a_{0+}^{1/2}$) 
are given in units of $m_c^{-1}$, the $p$-wave scattering volumes (remaining quantities) in $m_c^{-3}$. The row marked as HBRs contains the sum of the contributions of the well-established $s$ and $p$ higher baryon resonances detailed in Subsection 
\ref{sec:HR}. Uncertainties are quoted only for the sum of the contributions.
\vspace{0.2cm}
\begin{center}
\begin{tabular}{|l|c|c|c|c|c|c|}
\hline
 & $a_{0+}^{3/2}$ & $a_{0+}^{1/2}$ & $a_{1+}^{3/2}$ & $a_{1-}^{3/2}$ & $a_{1+}^{1/2}$ & $a_{1-}^{1/2}$ \\
\hline
$\sigma$ & $0.0742$ & $0.0742$ & $0.0020$ & $0.00161$ & $0.00202$ & $0.0016$ \\
$\rho$ & $-0.0736$ & $0.1471$ & $-0.0005$ & $-0.01004$ & $0.00108$ & $0.0201$ \\
$N$ & $-0.0101$ & $-0.0080$ & $0.0979$ & $-0.04745$ & $-0.04896$ & $-0.1942$ \\
$\Delta(1232)$ & $-0.0635$ & $-0.0569$ & $0.1068$ & $0.01580$ & $0.01601$ & $0.0822$ \\
HBRs & $0.0003$ & $0.0013$ & $0.0028$ & $-0.00076$ & $-0.00084$ & $0.0115$ \\
\hline
Sum & $-0.0727(16)$ & $0.1577(14)$ & $0.2090(22)$ & $-0.04084(77)$ & $-0.03070(66)$ & $-0.0789(15)$ \\
\hline
\end{tabular}
\end{center}
\end{table}

\clearpage

\clearpage

\clearpage

\clearpage

\clearpage

\clearpage

\clearpage

\clearpage


\begin{thebibliography}{99}
\bibitem{glm1} P.F.A. Goudsmit, H.J. Leisi, E. Matsinos, `Pionic atoms, the relativistic mean-field theory and the pion-nucleon scattering lengths', Phys. Lett. B 271 (1991) 290-4.
\bibitem{glm2} P.F.A. Goudsmit, H.J. Leisi, E. Matsinos, `A pion-nucleon interaction model', Phys. Lett. B 299 (1993) 6-10.
\bibitem{glmbg} P.F.A. Goudsmit, H.J. Leisi, E. Matsinos, B.L. Birbrair, A.B. Gridnev, `The extended tree-level model for the pion-nucleon interaction', Nucl. Phys. A 575 (1994) 673-706.
\bibitem{mworg} E. Matsinos, W.S. Woolcock, G.C. Oades, G. Rasche, A. Gashi, `Phase-shift analysis of low-energy $\pi^\pm p$ elastic-scattering data', Nucl. Phys. A 778 (2006) 95-123.
\bibitem{mr1} E. Matsinos, G. Rasche, `Analysis of the low-energy $\pi^\pm p$ elastic-scattering data', J. Mod. Phys. 3 (2012) 1369-87.
\bibitem{mr2} E. Matsinos, G. Rasche, `Analysis of the low-energy $\pi^\pm p$ differential cross sections of the CHAOS Collaboration', Nucl. Phys. A 903 (2013) 65-80.
\bibitem{mr3} E. Matsinos, G. Rasche, `Analysis of the low-energy $\pi^- p$ charge-exchange data', Int. J. Mod. Phys. A 28 (2013) 1350039.
\bibitem{glk} W.R. Gibbs, Li Ai, W.B. Kaufmann, `Isospin breaking in low-energy pion-nucleon scattering', Phys. Rev. Lett. 74 (1995) 3740-3.
\bibitem{m} E. Matsinos, `Isospin violation in the $\pi N$ system at low energies', Phys. Rev. C 56 (1997) 3014-25.
\bibitem{gmorw1} A. Gashi, E. Matsinos, G.C. Oades, G. Rasche, W.S. Woolcock, `Electromagnetic corrections to the hadronic phase shifts in low energy $\pi^+ p$ elastic scattering', Nucl. Phys. A 686 (2001) 447-62.
\bibitem{gmorw2} A. Gashi, E. Matsinos, G.C. Oades, G. Rasche, W.S. Woolcock, `Electromagnetic corrections for the analysis of low energy $\pi^- p$ scattering data', Nucl. Phys. A 686 (2001) 463-77.
\bibitem{fm} N. Fettes, E. Matsinos, `Analysis of recent $\pi^+ p$ low-energy differential cross-section measurements', Phys. Rev. C 55 (1997) 464-73.
\bibitem{kbl} T.W.B. Kibble, `Kinematics of general scattering processes and the Mandelstam representation', Phys. Rev. 117 (1960) 1159-62.
\bibitem{h} G. H\"ohler, `Pion Nucleon Scattering. Part 2: Methods and Results of Phenomenological Analyses', Landolt-B\"ornstein, Vol. 9b2, ed. H. Schopper, Springer, Berlin, 1983.
\bibitem{cd} T.P. Cheng, R. Dashen, `Is SU(2)$\bigotimes$SU(2) a better symmetry than SU(3)?', Phys. Rev. Lett. 26 (1971) 594-7.
\bibitem{bd} J.D. Bjorken, S.D. Drell, `Relativistic Quantum Mechanics', McGraw-Hill, 1964.
\bibitem{pdg} J. Beringer \etal~(Particle Data Group), `The Review of Particle Physics', Phys. Rev. D 86 (2012) 010001.
\bibitem{ew} T.E.O. Ericson, W. Weise, `Pions and Nuclei', Clarendon Press, Oxford, 1988.
\bibitem{oo} J.A. Oller, E. Oset, `Chiral symmetry amplitudes in the $s$-wave isoscalar and isovector channels and the $\sigma$, $f_0(980)$, $a_0(980)$ scalar mesons', Nucl. Phys. A 620 (1997) 438-56; see erratum in Nucl. Phys. A 652 (1999) 407-9.
\bibitem{dp} A. Dobado and J.R. Pel\'aez, `Inverse amplitude method in chiral perturbation theory', Phys. Rev. D 56 (1997) 3057-73.
\bibitem{ao} M. Albaladejo, J.A. Oller, `Size of the $\sigma$ meson and its nature', Phys. Rev. D 86 (2012) 034003.
\bibitem{au} J.E. Augustin \etal~(DM2 Collaboration), `Study of the J/$\psi$ decay into five pions', Nucl. Phys. B 320 (1989) 1-19.
\bibitem{torn} N.A. T\"ornqvist, M. Roos, `Confirmation of the sigma meson', Phys. Rev. Lett. 76 (1996) 1575-8.
\bibitem{gs} F. Gross, Y. Surya, `Unitary, relativistic resonance model for $\pi N$ scattering', Phys. Rev. C 47 (1993) 703-23.
\bibitem{fp} M. Fierz, W. Pauli, `On relativistic wave equations for particles of arbitrary spin in an electromagnetic field', Proc. Roy. Soc. (London) A 173 (1939) 211-32.
\bibitem{rs} W. Rarita, J. Schwinger, `On a theory of particles with half-integral spin', Phys. Rev. 60 (1941) 61.
\bibitem{wl} H.T. Williams, `Misconceptions regarding spin $\frac{3}{2}$', Phys. Rev. C 31 (1985) 2297-9.
\bibitem{bdm} M. Benmerrouche, R.M. Davidson, N.C. Mukhopadhyay, `Problems of describing spin-$\frac{3}{2}$ baryon resonances in the effective Lagrangian theory', Phys. Rev. C 39 (1989) 2339-48.
\bibitem{pasca1} V. Pascalutsa, `Quantization of an interacting spin-$\frac{3}{2}$ field and the $\Delta$ isobar', Phys. Rev. D 58 (1998) 096002.
\bibitem{pasca2} V. Pascalutsa, `Correspondence of consistent and inconsistent spin-$\frac{3}{2}$ couplings via the equivalence theorem', Phys. Lett. B 503 (2001) 85-90.
\bibitem{haber} H. Haberzettl, `Propagation of a massive spin-$\frac{3}{2}$ particle', arXiv:nucl-th/9812043.
\bibitem{te} H.-B. Tang, P.J. Ellis, `Redundance of $\Delta$-isobar parameters in effective field theories', Phys. Lett. B 387 (1996) 9-13.
\bibitem{kem} H. Krebs, E. Epelbaum, U.-G. Mei\ss ner, `Redundancy of the off-shell parameters in chiral effective field theory with explicit spin-$\frac{3}{2}$ degrees of freedom', Phys. Lett. B 683 (2010) 222-8.
\bibitem{mr4} E. Matsinos, R. Rasche, `The propagation of a massive spin-$\frac{3}{2}$ field, with application to $\pi N$ scattering' (in preparation).
\bibitem{pe} R.D. Peccei, `Chiral Lagrangian calculation of pion-nucleon scattering lengths', Phys. Rev. 176 (1968) 1812-21.
\bibitem{nek} L.M. Nath, B. Etemadi, J.D. Kimel, `Uniqueness of the interaction involving spin-$\frac{3}{2}$ particles', Phys. Rev. D 3 (1971) 2153-61.
\bibitem{slm} V. Shklyar, H. Lenske, U. Mosel, `Spin-$\frac{5}{2}$ fields in hadron physics', Phys. Rev. C 82 (2010) 015203.
\bibitem{abws} R.A. Arndt, W.J. Briscoe, I.I. Strakovsky, R.L. Workman, `Extended partial-wave analysis of $\pi N$ scattering data', Phys. Rev. C 74 (2006) 045205; SAID PSA Tool: http://gwdac.phys.gwu.edu.
\bibitem{two1} B. Tromborg, S. Waldenstr{\o}m, I. {\O}verb{\o}, `Electromagnetic corrections to $\pi^+ p$ scattering', Ann. Phys. 100 (1976) 1-36.
\bibitem{two2} B. Tromborg, S. Waldenstr{\o}m, I. {\O}verb{\o}, `Electromagnetic corrections to $\pi N$ scattering', Phys. Rev. D 15 (1977) 725-9.
\bibitem{two3} B. Tromborg, S. Waldenstr{\o}m, I. {\O}verb{\o}, `Electromagnetic corrections in hadron scattering, with application to $\pi N \rightarrow \pi N$', Helv. Phys. Acta 51 (1978) 584-607.
\bibitem{orwmg} G.C. Oades, G. Rasche, W.S. Woolcock, E. Matsinos, A. Gashi, `Determination of the $s$-wave pion-nucleon threshold scattering parameters from the results of experiments on pionic hydrogen', Nucl. Phys. A 794 (2007) 73-86.
\bibitem{ss} H.-Ch. Schr{\"o}der \etal, `The pion-nucleon scattering lengths from pionic hydrogen and deuterium', Eur. Phys. J. C 21 (2001) 473-88.
\bibitem{bhhknp} V. Baru, C. Hanhart, M. Hoferichter, B. Kubis, A. Nogga, D.R. Phillips, `Precision calculation of the $\pi^- d$ scattering length and its impact on threshold $\pi N$ scattering', Phys. Lett. B 694 (2011) 473-7.
\bibitem{gott} D. Gotta, private communication.
\bibitem{la} A.D. Lahiff, I.R. Afnan, `Solution of the Bethe-Salpeter equation for pion-nucleon scattering', Phys. Rev. C 60 (1999) 024608.
\bibitem{pt} V. Pascalutsa, J.A. Tjon, `Pion-nucleon interaction in a covariant hadron-exchange model', Phys. Rev. C 61 (2000) 054003.
\bibitem{mo} U.-G. Mei\ss ner, J.A. Oller, `Chiral unitary meson-baryon dynamics in the presence of resonances: elastic pion-nucleon scattering', Nucl. Phys. A 673 (2000) 311-34.
\bibitem{dhkm} C. Ditsche, M. Hoferichter, B. Kubis, U.-G. Mei\ss ner,`Roy-Steiner equations for pion-nucleon scattering', J. High Energy Phys. 06 (2012) 043.
\bibitem{gls1} J. Gasser, H. Leutwyler, M.E. Sainio, `Form factor of the $\sigma$-term', Phys. Lett. B 253 (1991) 260-4.
\bibitem{bali} G.S. Bali \etal~(QCDSF Collaboration), `Nucleon mass and sigma term from lattice QCD with two light fermion flavors', Nucl. Phys. B 866 (2013) 1-25.
\bibitem{g} J. Gasser, `Hadron masses and the sigma commutator in light of chiral perturbation theory', Ann. Phys. 136 (1981) 62-112.
\bibitem{gl} J. Gasser, H. Leutwyler, `Quark masses', Phys. Rep. 87 (1982) 77-169.
\bibitem{bor} B. Borasoy, `Sigma-terms in heavy baryon chiral perturbation theory revisited', Eur. Phys. J. C 8 (1999) 121-30.
\bibitem{agco} J.M. Alarc\'on, L.S. Geng, J. Martin Camalich, J.A. Oller, `The strangeness content of the nucleon from effective field theory and phenomenology', Phys. Lett. B 730 (2014) 342-6.
\bibitem{gss} J. Gasser, M.E. Sainio, A. $\rm{\check{S}}$varc, `Nucleons with chiral loops', Nucl. Phys. B 307 (1988) 779-853.
\bibitem{k} R. Koch, `A new determination of the $\pi N$ Sigma term using hyperbolic dispersion relations in the ($\nu^2$,$t$) plane', Z. Phys. C 15 (1982) 161-8.
\bibitem{glls} J. Gasser, H. Leutwyler, M.P. Locher, M.E. Sainio, `Extracting the pion-nucleon sigma-term from data', Phys. Lett. B 213 (1988) 85-90.
\bibitem{kk} R. Koch, `Improved $\pi N$ partial waves, consistent with analyticity and unitarity', Z. Phys. C 29 (1985) 597-609.
\bibitem{pasw} M.M. Pavan, R.A. Arndt, I.I. Strakovsky, R.L. Workman, `The pion-nucleon $\Sigma$ term is definitely large: results from a G.W.U. analysis of $\pi N$ scattering data', $\pi N$ Newslett. 16 (2002) 110-5; arXiv:hep-ph/0111066.
\bibitem{ols} M.G. Olsson, `The nucleon sigma term from threshold parameters', Phys. Lett. B 482 (2000) 50-6.
\bibitem{aco1} J.M. Alarc{\'o}n, J. Martin Camalich, J.A. Oller, `Chiral representation of the $\pi N$ scattering amplitude and the pion-nucleon sigma term', Phys. Rev. D 85 (2012) 051503.
\bibitem{aco2} J.M. Alarc{\'o}n, J. Martin Camalich, J.A. Oller, `Improved description of the $\pi N$-scattering phenomenology at low energies in covariant baryon chiral perturbation theory', Ann. Phys. 336 (2013) 413-61.
\bibitem{aco3} J.M. Alarc{\'o}n, J. Martin Camalich, J.A. Oller, private communication.
\bibitem{JaxoDraw} D. Binosi, L. Theu\ss{}l, `JaxoDraw: A graphical user interface for drawing Feynman diagrams', Comput. Phys. Commun. 161 (2004) 76-86.
\end{thebibliography}
\end{document}